\documentclass[longauth]{aa}
\usepackage{amsmath}
\usepackage{mathrsfs}
\usepackage{arydshln}
\usepackage{float}
\usepackage{stfloats}
\usepackage{color}
\usepackage{setspace}
\usepackage{graphicx}
\usepackage{fixltx2e}
\usepackage{booktabs, makecell, multirow}
\usepackage{txfonts}
\usepackage{placeins}
\usepackage{array, makecell} 
\usepackage{gensymb}
\usepackage{booktabs}
\usepackage{url}
\usepackage{subfloat}
\usepackage[version=3]{mhchem}
\usepackage{threeparttable}
\usepackage{multirow}
\usepackage{CJK}
\usepackage{longtable}
\usepackage{pdflscape}
\usepackage[export]{adjustbox}
\usepackage{subfig}
\usepackage{array}
\usepackage{bm}
\usepackage[singlelinecheck=false, labelfont=bf]{caption}

\usepackage{natbib,twoopt}
\usepackage[hyphenbreaks]{breakurl}
\usepackage[breaklinks]{hyperref}      
\bibpunct{(}{)}{;}{a}{}{,}             
\definecolor{cobalt}{rgb}{0.06, 0.2, 0.65}
\hypersetup{
  colorlinks,
  citecolor=cobalt,
  linkcolor=[rgb]{0.8, 0.2, 1.0},
  urlcolor=cobalt,
}
\makeatletter
  \newcommandtwoopt{\citeads}[3][][]{\href{http://adsabs.harvard.edu/abs/#3}%
    {\def\hyper@linkstart##1##2{}%
     \let\hyper@linkend\@empty\citealp[#1][#2]{#3}}}
  \newcommandtwoopt{\citepads}[3][][]{\href{http://adsabs.harvard.edu/abs/#3}%
    {\def\hyper@linkstart##1##2{}%
     \let\hyper@linkend\@empty\citep[#1][#2]{#3}}}
  \newcommandtwoopt{\citetads}[3][][]{\href{http://adsabs.harvard.edu/abs/#3}%
    {\def\hyper@linkstart##1##2{}%
     \let\hyper@linkend\@empty\citet[#1][#2]{#3}}}
  \newcommandtwoopt{\citeyearads}[3][][]%
    {\href{http://adsabs.harvard.edu/abs/#3}
    {\def\hyper@linkstart##1##2{}%
     \let\hyper@linkend\@empty\citeyear[#1][#2]{#3}}}
\makeatother

\newcommand{\myemail}{\email{\href{mailto:chentao.yang@chalmers.se}{chentao.yang@chalmers.se}}}

\newcommand{\kms}{{\hbox {\,km\,s$^{-1}$}}}
\newcommand{\mum}{{\hbox {\,$\mu$m}}}
\newcommand{\kkmspc}{{\hbox {\,K\,km\,s$^{-1}$\,pc$^{2}$}}} 
\newcommand{\lsun}{{\hbox {$L_\odot$}}}
\newcommand{\msun}{{\hbox {$M_\odot$}}}

\newcommand{\hto}{{\hbox {H\textsubscript{2}O}}}

\newcommand{\htop}{{\hbox {H$_2$O$^+$}}}
\newcommand{\httop}{{\hbox {H$_{3}$O$^{+}$}}}
\newcommand{\ci}{{\hbox {[C{\scriptsize \,I}]}}}

\newcommand{\lir}{\hbox {$L_{\mathrm{IR}}$}}
\newcommand{\lfir}{\hbox {$L_{\mathrm{FIR}}$}}
\label{key}

\def\co#1#2{{\hbox {${\mathrm{CO}}(#1\text{--}#2)$}}}

\def\htot#1#2#3#4#5#6{\hbox {\hto(\t#1#2#3#4#5#6)}}
\def\tp#1#2#3#4#5#6{{\hbox {$#1^#3_{#2}\text{--}#4^#6_{#5}$}}}
\def\t#1#2#3#4#5#6{{\hbox {$#1_{#2#3}\text{--}#4_{#5#6}$}}}

\def\cil#1#2{{\hbox {[C{\scriptsize{\,I}}](#1\text{--}#2)}}}

\newcommand{\apm}{\hbox {APM\,08279+5255}}

\begin{document}
\begin{CJK*}{UTF8}{gbsn}


\title{
{\em SUNRISE}:  The rich molecular inventory of high-redshift dusty galaxies 
revealed by broadband spectral line surveys
\thanks{
We dedicate this paper to the memory of our coauthor and friend, Yu Gao, who passed away in May 2022.
}
\thanks{The final data products of the tables derived from \texttt{UVFIT} will be available in electronic form at the CDS via anonymous ftp to \url{cdsarc.cds.unistra.fr} (\url{130.79.128.5}) or via \url{https://cdsarc.cds.unistra.fr/cgi-bin/qcat?J/A+A/}.
}
}

\author
{
Chentao Yang (杨辰涛)$^{1}$                  \and     
Alain Omont$^{2}$                           \and 
Sergio Mart\'in$^{3,4}$                     \and
Thomas G. Bisbas$^{5}$                      \and 
Pierre Cox$^{2}$                            \and \\
Alexandre Beelen$^{6}$                      \and
Eduardo Gonz{\'a}lez-Alfonso$^{7}$          \and 
Rapha{\"e}l Gavazzi$^{6,2}$                 \and \\
Susanne Aalto$^{1}$                         \and 
Paola Andreani$^{8,9,10}$                   \and 
Cecilia Ceccarelli$^{11}$                   \and   
Yu Gao (高煜)$^{12}$                         \and
Mark Gorski$^{1,13}$                        \and
Michel Gu{\'e}lin$^{14}$                    \and \\
Hai Fu (傅海)$^{15}$                         \and 
R. J. Ivison$^{8,14,17,18}$                 \and 
Kirsten K. Knudsen$^{1}$                    \and
Matthew Lehnert$^{19}$                      \and 
Hugo Messias$^{3,4}$                        \and  
Sebastien Muller$^{1}$                      \and \\
Roberto Neri$^{14}$                         \and 
Dominik Riechers$^{20}$                     \and 
Paul van der Werf$^{21}$                    \and
Zhi-Yu Zhang (张智昱)$^{22, 23}$                          
}

\institute{
Department of Space, Earth \& Environment, Chalmers University of Technology, 
SE-412 96 Gothenburg, Sweden. \myemail 
\and 
CNRS and Sorbonne Universit{\'e}, UMR 7095, Institut d'Astrophysique de Paris, 98bis boulevard Arago, 75014 Paris, France
\and
European Southern Observatory, Alonso de C{\'o}rdova 3107, Casilla 19001, 
Vitacura, Santiago, Chile. 
\and
Joint ALMA Observatory, Alonso de C{\'o}rdova, 3107, Vitacura, 
Santiago 763-0355, Chile.
\and
Research Center for Intelligent Computing Platforms, Zhejiang Laboratory, 
Hangzhou 311100, People's Republic of China
\and
Aix Marseille Univ, CNRS, CNES, LAM, Marseille, France.
\and
Universidad de Alcal{\'a}, Departamento de F\'{\i}sica y Matem{\'a}ticas, 
Campus Universitario, 28871 Alcal{\'a} de Henares, Madrid, Spain.
\and
European Southern Observatory, Karl-Schwarzschild-Strasse 2, D-85748 Garching, Germany
\and
Aristotle University of Thessaloniki, GR-54124 Thessaloniki, Greece
\and
Institute of Theoretical Astrophysics, University of Oslo, PO Box 1029, Blindern 0315, Oslo, Norway
\and
Univ. Grenoble Alpes, CNRS, IPAG, 38000 Grenoble, France
\and
Department of Astronomy, Xiamen University, Xiamen, Fujian 361005, People's Republic of China
\and
Center for Interdisciplinary Exploration and Research in Astrophysics (CIERA) and Department of Physics and Astronomy, Northwestern University, Evanston, IL 60208, USA
\and
Institut de Radioastronomie Millim{\'e}trique, 300 rue de la Piscine, 38406 Saint-Martin-d'H{\`e}res, France
\and
Department of Physics \& Astronomy, The University of Iowa, 203 Van Allen Hall, Iowa City, IA 52242, USA
\and
Department of Physics and Astronomy, Macquarie University, North Ryde, New South Wales, Australia
\and
School of Cosmic Physics, Dublin Institute for Advanced Studies, 31 Fitzwilliam Place, Dublin D02 XF86, Ireland
\and
Institute for Astronomy, University of Edinburgh, Royal Observatory, Edinburgh EH9 3HJ, UK
\and
Centre de Recherche Astrophysique de Lyon—CRAL, CNRS UMR 5574, UCBL1, ENSLyon, 9 avenue Charles Andr{\'e}, F-69230 Saint-Genis-Laval, France
\and
I. Physikalisches Institut, Universit{\"a}t zu K{\"o}ln, Z{\"u}lpicher Strasse 77, D-50937, K{\"o}ln, Germany
\and
Leiden Observatory, Leiden University, PO Box 9513, NL-2300 RA Leiden, The Netherlands
\and
School of Astronomy and Space Science, Nanjing University, Nanjing 210023, People's Republic of China
\and
Key Laboratory of Modern Astronomy and Astrophysics (Nanjing University), Ministry of Education, Nanjing 210023, People’s Republic of China
}

\date {Received .../ Accepted ...}

\abstract
{
Understanding the nature of high-redshift dusty galaxies requires a comprehensive view of their interstellar medium (ISM) and molecular complexity. However, the molecular ISM at high redshifts is commonly studied using only a few species beyond $^{12}$C$^{16}$O, limiting our understanding. In this paper, we present the results of deep 3\,mm spectral line surveys using the NOrthern Extended Millimeter Array (NOEMA) targeting two strongly lensed dusty galaxies observed when the Universe was less than 1.8\,Gyr old: \apm, a quasar at redshift $z$\,=\,3.911, and NCv1.143 ({\it H}-ATLAS J125632.7+233625), a $z$\,=\,3.565 starburst galaxy. The spectral line surveys cover rest-frame frequencies from about 330 to 550\,GHz for both galaxies. We report the detection of 38 and 25 emission lines in \apm\ and NCv1.143, respectively. These lines originate from 17 species, namely CO, $^{13}$CO, C$^{18}$O, CN, CCH, HCN, HCO$^+$, HNC, CS, C$^{34}$S, H$_2$O, H$_3$O$^+$, NO, N$_2$H$^+$, CH, c-C$_3$H$_2$, and the vibrationally excited HCN and neutral carbon. The spectra reveal the chemical richness and the complexity of the physical properties of the ISM. By comparing the spectra of the two sources and combining the analysis of the molecular gas excitation, we find that the physical properties and the chemical imprints of the ISM are different: the molecular gas is more excited in \apm, which exhibits higher molecular gas temperatures and densities compared to NCv1.143; the molecular abundances in \apm\ are akin to the values of local active galactic nuclei (AGN), showing boosted relative abundances of the dense gas tracers that might be related to high-temperature chemistry and/or the X-ray-dominated regions, while NCv1.143 more closely resembles local starburst galaxies. The most significant differences between the two sources are found in H$_2$O: the 448\,GHz ortho-\htot423330\ line is significantly brighter in \apm, which is likely linked to the intense far-infrared radiation from the dust powered by AGN. Our astrochemical model suggests that, at such high column densities, far-ultraviolet radiation is less important in regulating the ISM, while cosmic rays (and/or X-rays and shocks) are the key players in shaping the molecular abundances and the initial conditions of star formation. Both our observed CO isotopologs line ratios and the derived extreme ISM conditions (high gas temperatures, densities, and cosmic-ray ionization rates) suggest the presence of a top-heavy stellar initial mass function. From the $\sim$\,330--550\,GHz continuum, we also find evidence of nonthermal millimeter flux excess in \apm\ that might be related to the central supermassive black hole. Such deep spectral line surveys open a new window into the physics and chemistry of the ISM and the radiation field of galaxies in the early Universe.
}


\authorrunning{C. Yang et al.}

\titlerunning{SUNRISE: 
Submillimeter molecUlar liNe suRveys in dIstant duSty galaxiEs}

\maketitle


\section{Introduction}
\label{section:intro}

The interstellar medium (ISM), comprising interstellar gas, dust, and cosmic rays (CRs), is a fundamental part of the ecosystem of galaxies. The cold ISM (gas and dust with $T<10^3$\,K), which we simply refer to as the ISM hereafter unless otherwise specified, exhibits complex physical structures across different scales and plays a crucial role in a range of important physical processes, including star formation \citep{1978ppim.book.....S}. Energetic phenomena in the ISM, including shocks and jets, along with interactions with the electromagnetic radiation field and CRs emitted from various sources such as stars, supernovae, and active galactic nuclei (AGN), foster conditions that enable a vast number of chemical reactions \citep{2013RvMP...85.1021T,tielens_2021}. Consequently, a comprehensive understanding of the ISM in galaxies requires observational studies that delve into its chemical complexity \citep[e.g.,][]{2007RPPh...70.1099O, 2016IAUS..315...17V}.

Astrochemical studies of nearby galaxies have flourished over the past two decades, driven by advancements in radio/(sub)millimeter instruments. Deep spectral line surveys covering large frequency ranges (see Table\,1 of \citealt{2021A&A...656A..46M} for a list of extragalactic line surveys and references) allow us to study a wide variety of ISM tracers without preselecting the canonical species, such as $^{12}$C$^{16}$O (CO hereafter), thereby offering an unbiased perspective of the ISM by appreciating its chemical complexity. These line surveys of nearby galaxies have unveiled a complicated picture of how different ISM phases interplay with radiation in the process of star formation and AGN activities and have greatly enriched our understanding of the evolution and ecosystem of galaxies. Interferometers such as the NOrthen Extended Millimeter Array (NOEMA) and the Atacama Large Millimeter/submillimeter Array (ALMA) now enable different gas tracers in nearby galaxies to be mapped down to the scales of giant molecular clouds, unveiling significant spatial variations in the ISM chemistry that are distinct from those found in the Milky Way, potentially arising from super star clusters and AGN \citep[e.g.,][]{2015ApJ...801...63M, 2021A&A...656A..46M, 2021ApJ...923..240S}.

However, in contrast to local galaxies, high-redshift galaxies have remained largely unexplored (though there have been several detections in molecular absorbers against bright quasars; \citealt{2011A&A...535A.103M,2014A&A...566A.112M,2020A&A...636L...7T}), primarily due to sensitivity limitations. The lack of high-sensitivity line surveys at high redshifts prevents us from attaining a comprehensive understanding of the ISM in the early Universe, which often hosts the most extreme ISM conditions and 
is a key laboratory for testing physical and chemical theories of the ISM.

A breakthrough came with the discovery of large samples of strongly gravitationally lensed dusty star-forming galaxies at high redshifts \citep[e.g.,][]{2010Sci...330..800N, 2013Natur.495..344V, 2015A&A...581A.105C}. Lensing magnification enables us to achieve sensitivities that allow for the detection of more complex molecules beyond CO. These strongly lensed infrared-luminous galaxies are among the most active dusty starbursts in the early Universe \citep[e.g.,][]{2002PhR...369..111B, 2014PhR...541...45C, 2020RSOS....700556H}. They are rich in molecular gas and dust and feature bright emission and/or absorption lines, which are often highly excited by intense radiation fields, such as infrared (IR), ultraviolet (UV), and X-ray. Additionally, CRs, stellar winds, and shocks can also serve as excitation sources. The discovery of these galaxies has paved the way for in-depth spectral line surveys, particularly in galaxies with extreme ISM conditions, facilitating an unparalleled insight into their ISM. The spectral richness of such high-redshift dusty galaxies has been demonstrated in recent dedicated observations of the lensed Cloverleaf quasar at $z$\,=\,2.56 where a variety of molecules have been reported, including HCN, HNC, HCO$^+$, CN, and CO isotopologs \citep[see][and references therein]{2022EPJWC.26500024G}. Stacking together the spectra of high-redshift dusty star-forming galaxies from $\sim$\,200 to 800\,GHz in the rest frame \citep{2014ApJ...785..149S, 2023ApJ...948...44R, 2023MNRAS.521.5508H}, several molecular lines in addition to CO have also been detected, including \hto, \htop, $\rm ^{13}CO$, HCN, HCO$^+$, CH and CN. However, stacked spectra only reflect the averaged quantities of the sample, and results can vary with different weighted average methods, making it difficult to interpret the physical properties. 

Therefore, to accurately depict the ISM properties, it is crucial to conduct in-depth spectral line surveys in individual sources. However, individual detections of molecular gas tracers beyond CO lines remain scarce. Apart from the Cloverleaf \citep{2022EPJWC.26500024G}, only a limited number of high-redshift sources have been reported with detections of a few species, predominantly only single transitions per molecule \citep[e.g.,][]{2003Natur.426..636S, 2004ApJ...614L..97V, 2005ApJ...618..586C, 2007ApJ...660L..93G, 2007A&A...462L..45G, 2006ApJ...645L..13R, 2007ApJ...671L..13R, 2010ApJ...725.1032R, 2011ApJ...726...50R, 2011MNRAS.410.1687D, 2017ApJ...850..170O, 2018A&A...620A.115B, 2021A&A...645A..45C, 2022A&A...667A..70R, 2023arXiv230802886R}. As a result, our understanding of the high-redshift ISM remains limited.

In this paper, we report the first results of NOEMA line surveys toward two strongly lensed high-redshift galaxies of contrasting extremes, \apm\ and NCv1.143, from the {\em SUNRISE} (Submillimeter molecUlar liNe suRveys in dIstant duSty galaxiEs) project. This survey project includes multiple deep broadband millimeter spectral line surveys toward four high-redshift targets --- \apm, NCv1.143, G09v1.97, and BR\,1202+0725 --- using NOEMA (this paper) and ALMA (Yang et al., in prep.). The first results presented in this work highlight the detection of a rich set of emission lines in the rest-frame frequency from about 330 to 550\,GHz in two sources with very different ISM environments --- the dusty quasar \apm\ and the starburst galaxy NCv1.143 (throughout this work, we use the term ``ISM environments'' to refer either AGN- or starburst-dominated ISM). These comprehensive and unbiased line surveys, spanning a continuous bandwidth, are essential for detecting multiple transitions of molecules beyond CO. They allow us to study the ISM by probing a large range of ISM conditions, provide rich and robust ISM diagnostic tool sets, improve our current potentially biased understanding of the physics and chemistry of the ISM, and reveal how various extreme ISM environments influence its properties.

Throughout this work, we assume a \citet{2003PASP..115..763C} stellar initial mass function (IMF) and a definition of infrared luminosity integrated over 8--1000\,$\mu$m. Based on the starburst evolutionary synthesis models in \cite{1995ApJS...96....9L} with a \cite{1955ApJ...121..161S} IMF, \citet{1998ApJ...498..541K} derived the star formation rate (SFR) to infrared luminosity, SFR-$L_\mathrm{IR}$, calibration for starburst galaxies (with continuous bursts of age $\sim$\,100\,Myr) as $\mathit{SFR}$\,=\,$1.7\times 10^{-10}$\,(\lir/\lsun)\,\msun\,yr$^{-1}$. After translating this calibration to the \citet{2003PASP..115..763C} IMF using a factor of $\,\sim$\,0.63 for the SFR calibration \citep{2014ARA&A..52..415M}, we derive $\mathit{SFR}$\,=\,$1.1 \times 10^{-10}$\,(\lir/\lsun)\,\msun\,yr$^{-1}$ \citep[with $\approx$\,30\% uncertainty due to variations in star formation histories within individual sources;][]{1998ApJ...498..541K}. We note that using a top-heavy stellar IMF, which is found in dusty starburst galaxies, may reduce the SFR/\lir\ ratio by a factor of at least $\sim$\,2 \citep{2020ApJ...891...74C} and up to $\sim$\,5 \citep{2017A&A...607A.126Y, 2018Natur.558..260Z}. Additionally, if the burst time is shorter by one order of magnitude, the calibration factor will increase by a factor of 3 \citep{2014ARA&A..52..415M}, introducing additional uncertainties to the SFR-$L_\mathrm{IR}$ calibration. 

We adopt a spatially flat Lambda cold dark matter ($\Lambda$CDM) cosmology model with $H_{0}=(67.4\pm0.5)\,{\rm km\,s^{-1}\,Mpc^{-1}}$ and $\Omega_\mathrm{M}=0.315 \pm 0.007$ \citep{2020A&A...641A...6P}. This translates to luminosity distances and scales of $35700\pm370$\,Mpc and 7.44\,kpc/\arcsec\ for \apm\ and $31970\pm320$\,Mpc and 7.17\,kpc/\arcsec\ for NCv1.143.

This paper is organized as follows: Sect.\,\ref{section:sample} introduces the targets of the surveys, Sect.\,\ref{section:obs_redu} describes the NOEMA observations and the process of data reduction, and Sect.\,\ref{section:results} presents a detailed analysis of the data, including line identification. Section\,\ref{subsection:ctm} discusses the continuum emission covered by our line survey. Section\,\ref{section:line-atlas-diff} delves into a detailed discussion about the line survey results and a comparison between the two sources, which is followed by chemical modeling in Sect.\,\ref{subsection:Chemical-model}. The impact of differential lensing is addressed in Sect.\,\ref{section:lensing}. We conclude in Sect.\,\ref{section:conclusion}.

\section{Targets --- APM\,08279+5255 and NCv.1.143}
\label{section:sample}

APM\,08279+5255 is a $z$\,=\,3.911 strongly lensed broad absorption line quasar with complex multiple absorption features mainly seen in the C\,{\small IV} line \citep{1998ApJ...505..529I} and a prodigious apparent bolometric luminosity of $5\times10^{15}$\,\lsun\ \citep{2000ApJ...535..561E}. It has been suggested that radiation pressure plays an important role in producing quasar outflow \citep{2011ApJ...737...91S}. Observations of H$\beta$ and Mg\,{\small II} suggest a mass of the central supermassive black hole (SMBH) of $\sim$\,$10^{10}$\,\msun\ \citep{2018A&A...617A.118S}. \apm\ is one of the brightest dust- and gas-rich quasars in the far-infrared and submillimeter \citep[e.g.,][]{1999ApJ...513L...1D, 2000ApJ...535..561E, 2007A&A...467..955W}. It is extremely dusty with a total infrared luminosity \lir\,=\,($3.4\pm0.4$)$\times10^{14} \mu^{-1}$\,\lsun\ (\citealt{{2019ApJ...876...48L}}, note the different definition of \lir\textsuperscript{\ref{note1}}). The molecular gas is found to be highly excited with extreme conditions by examining the CO spectral line energy distribution (SLED). The CO excitation can be attributed to two prominent excitation components --- a cold one with $T_\mathrm{kin}$\,$\sim$\,$65$\,K and $n_\mathrm{H_2}$\,$\sim$\,$10^{5}$\,cm$^{-3}$, and a warm one with $T_\mathrm{kin}$\,$\sim$\,$220$\,K and $n_\mathrm{H_2}$\,$\sim$\,$10^{4}$\,cm$^{-3}$ which is probably heated by the central AGN \citep{2007A&A...467..955W}. Observations of multiple \hto\ lines in \apm\ suggest that the submillimeter \hto\ emission originates in similar regions as CO with a dust temperature $T_\mathrm{d}$\,$\sim$\,$220$\,K and the optical depth at 100\,$\mu$m, $\tau_{100}\approx0.9$, where the X-ray radiation of the quasar nucleus penetrates deeply into the buried regions and contributes to the ISM heating \citep{2010A&A...518L..42V, 2011ApJ...738L...6L}. From the early optical and near/mid-infrared observations, \apm\ shows a triple-image structure, likely composed of the lensed images of the quasar \citep{2013PASJ...65....9O}. Based on these observations, \apm\ is believed to exhibit a very high lensing magnification ($\mu$\,$\sim$\,90--120). This magnification is notably sensitive to the size of the source within the source plane, showing a differential lensing factor of up to 10 \citep{1998A&A...339L..77L, 2000ApJ...535..561E}. However, later analysis of the 0.3\arcsec\ resolution \co10\ Very Large Array (VLA) image suggests a naked cusp lens model with a significantly lower magnification of $\sim$\,7 \citep{2002MNRAS.330L..15L}. Moreover, a magnification of $\mu$\,=\,2--4 was derived based on a different lens model based on a deeper 0.3\arcsec\ \co10\ image \citep{2009ApJ...690..463R}, suggesting that the CO disk has an apparent (magnified) radius of about 1.3\,kpc, consistent with the CO size derived from excitation analysis \citep{2007A&A...467..955W}. In this lens model, differential lensing is insignificant (variations $\lesssim$\,20\% for source radius smaller than 5\,kpc) and is at most a factor of 2 for the most extreme intrinsic source radius that exceeds 20\,kpc. Throughout this work, we adopt this latest lens model.

The dusty starburst galaxy NCv1.143 is among the brightest strongly lensed sources discovered from the {\it Herschel}-ATLAS survey \citep[or {\it H}-ATLAS for short;][]{2010PASP..122..499E}; its IAU name is {\it H}-ATLAS J125632.7+233625 (or HerBS-5, \citealt{2018MNRAS.473.1751B}). At $z$\,=\,3.565, NCv1.143 is an intrinsically hyper-luminous infrared galaxy with an apparent total infrared luminosity\footnote
{\label{note1}
For the values of far-infrared luminosity (rest frame 40--120\,\mum) \lfir, we converted \lfir\ to \lir\ (without AGN contribution) using a factor of 1.91  \citep{2013ApJ...779...25B}. We performed a similar conversion for \apm, taking the star-forming \lfir, by adopting the \lfir\,=\,($1.8\pm0.2$)$\times10^{14} \mu^{-1}$\,\lsun\ \citep{2019ApJ...876...48L}. We define hyper-luminous infrared galaxies with a luminosity criterion of \lir\,$\geq$\,10$^{13}$\,\lsun.
} of \lir\,=\,($1.4\pm0.2$)$\times10^{14} \mu^{-1}$\,\lsun. The magnification is derived to be $\mu$\,=\,$11.3\pm0.7$ based on the 880\,$\mu$m continuum from \citet{2013ApJ...779...25B} and $\mu$\,=\,$12.2\pm1.2$ based on the 2\,mm continuum from \citet{2017PhDT........21Y}. Throughout this work, we have adopted the latter value derived from the lens model described in Appendix\,\ref{appen:lens-model}. Thus, the intrinsic infrared luminosity is $\gtrsim$\,$10^{13}$\,\lsun\ with a molecular gas reservoir of ($6.2\pm1.6$)$\times$10$^{10}$\,\msun\ \citep{2017A&A...608A.144Y}. The source has one of the highest \lir\ surface densities in the {\it H}-ATLAS sample  \citep{2013ApJ...779...25B}, reaching $\Sigma_{L_{\rm IR}}$\,$\sim$\,$\times10^{13}\,\lsun$\,kpc$^{-2}$. A series of extensive follow-up studies \citep[e.g.,][]{2016A&A...595A..80Y, 2017A&A...608A.144Y} found no clear AGN signatures, indicating that this galaxy is likely dominated by starburst activity with an estimated $\mathit{SFR}$\,=\,$1.3 \times 10^{3}$\,\msun\,yr$^{-1}$, although a deeply buried AGN cannot be completely ruled out. Studies of the molecular gas conditions using the CO SLED in NCv1.143 found two prominent excitation components of molecular gas --- a lower-excitation one with a kinematic temperature $T_\mathrm{kin}$\,$\sim$\,20\,K and a molecular gas density $n_\mathrm{H_2}$\,$\sim$\,$10^{4.1}$\,cm$^{-3}$, and a high-excitation molecular gas component with $T_\mathrm{kin}$\,$\sim$\,63\,K and $n_\mathrm{H_2}$\,$\sim$\,10$^{4.2}$\,cm$^{-3}$ \citep{2017A&A...608A.144Y}. Detection of the $J_{\rm up}$\,=\,2, 3 and 4 \hto\ lines in NCv1.143 suggests the existence of an optically thick ($\tau_{100}>1$) dusty nucleus with a dust temperature $T_{\rm{d}}$\,$\sim$\,75\,K, surrounded by an extended cooler disk with $T_{\rm{d}}$\,$\sim$\,35\,K \citep{2016A&A...595A..80Y, 2017PhDT........21Y}. The similarities between the line profiles of the high-$J$ CO and \hto\ lines indicate that they both arise from similarly dense warm gas in star-forming regions across the galaxy.

In addition to CO and \hto, detections of dense gas tracers such as HCN, HNC, and HCO$^+$ in \apm\ \citep{2005ApJ...634L..13W, 2006ApJ...645L..17G, 2007A&A...467..955W, 2010ApJ...725.1032R} and NCv1.143 \citep{2017PhDT........21Y} have also been reported. These detections, however, lean toward the brightest lines, focusing on specific frequency windows and often capturing only one transition per molecule. This limited scope presents a challenge for conducting a thorough examination of the ISM and its interaction with radiation fields, as such an analysis necessitates the consideration of complex astrochemistry. 

\section{Observations and data reduction}
\label{section:obs_redu}

\setlength{\tabcolsep}{0.33em}
\renewcommand{\arraystretch}{1.0}
\begin{table*}[!htbp]
\scriptsize
\centering
\caption{NOEMA observation log of \apm\ and NCv1.143.}
\vspace{-0.1cm}
\begin{tabular}{rccccccccccc}
\toprule
Source                           &  Redshift              &        \multicolumn{2}{c}{Tuning}                    &      Date\,($N_\mathrm{ant})$                              &   $t_\mathrm{on}$    &         Baseline         &          \multicolumn{2}{c}{Sideband}              & RMS$_\mathrm{/2\,MHz}$ &   \multicolumn{2}{c}{Synthesized Beam}    \\\cmidrule{3-4}\cmidrule{8-9}\cmidrule{11-12} 
                                 &                        &          Name              &        LO Frequency     &                                                            &                      &                          &    Name     &            Frequency Range           &                        &        Size         &       PA            \\
                                 &                        &                            &         (GHz)           &                                                            &      (h)             &           (m)            &             &               (GHz)                  &       (mJy/beam)            &        (\arcsec)    &      ($^\circ$)     \\
\midrule                                                                                                                                                                                                                                                                                                                                                              
                                 &                        & \multirow{2}{*}{w18eb001}  & \multirow{2}{*}{76.359} & \multirow{2}{*}{2019: Mar30\,(10)}                         &\multirow{2}{*}{4.9}  &\multirow{2}{*}{24--176}  &     LSB     & \llap{70.788\;}--\rlap{\;\;\,78.724} &         1.0            &   5.6$\times$4.2    &    \,$-$70          \\
                                 &                        &                            &                         &                                                            &                      &                          &     USB     & \llap{86.267\;}--\rlap{\;\;\,94.202} &         0.6            &   4.7$\times$3.6    &        112          \\
\multirow{2}{*}{\apm} & \multirow{2}{*}{3.911} & \multirow{2}{*}{w18eb002}  & \multirow{2}{*}{84.103} & \multirow{2}{*}{2019: Mar23\,(10), Mar27\,(10)}            &\multirow{2}{*}{6.6}  &\multirow{2}{*}{24--176}  &     LSB     & \llap{78.532\;}--\rlap{\;\;\,86.468} &         0.7            &   4.9$\times$3.8    &        105          \\
                                 &                        &                            &                         &                                                            &                      &                          &     USB     & \llap{94.011\;}--\rlap{\,101.946}    &         0.7            &   4.2$\times$3.3    &    \,$-$75          \\
                                 &                        & \multirow{2}{*}{w18eb003}  & \multirow{2}{*}{105.273}& \multirow{2}{*}{2019: Mar19\,(10)}                         &\multirow{2}{*}{4.9}  &\multirow{2}{*}{24--176}  &     LSB     & \llap{87.796\;}--\rlap{\;\;\,95.731} &         0.6            &   4.2$\times$3.3    &    \,$-$81          \\
                                 &                        &                            &                         &                                                            &                      &                          &     USB     & \llap{103.274\,}--\rlap{\,111.209}   &         0.8            &   3.5$\times$3.0    &    \,$-$82          \\      
\midrule                                                                                                                                                                                                                                                                                                                                                  
                                 &                        & \multirow{2}{*}{s18dc001}  & \multirow{2}{*}{76.670} & 2018: Aug28\,(7), Oct18\,(9), Oct26\,(9),                  &\multirow{2}{*}{12.6} &\multirow{2}{*}{24--368}  &     LSB     & \llap{70.789\;}--\rlap{\;\;\,78.724} &         0.9            &   3.5$\times$2.0    & \llap{$-$}162       \\
                                 &                        &                            &                         & Nov22\,(10), Dec13\,(10); 2019: Feb23\,(10)                &                      &                          &     USB     & \llap{86.267\;}--\rlap{\;\;\,94.201} &         0.6            &   3.0$\times$1.7    & \llap{$-$}163       \\
\multirow{2}{*}{NCv1.143}        & \multirow{2}{*}{3.565} & \multirow{2}{*}{s18dc002}  & \multirow{2}{*}{84.414} & \multirow{2}{*}{2018: Aug28\,(6), Oct20\,(9), Nov18\,(10)} &\multirow{2}{*}{7.0}  &\multirow{2}{*}{24--368}  &     LSB     & \llap{78.534\;}--\rlap{\;\;\,86.467} &         1.1            &   4.1$\times$2.3    &      \;\;18         \\
                                 &                        &                            &                         &                                                            &                      &                          &     USB     & \llap{94.009\;}--\rlap{\,101.943}    &         1.4            &   3.5$\times$1.9    & \llap{$-$}162       \\
                                 &                        & \multirow{2}{*}{s18dc003}  & \multirow{2}{*}{92.158} & \multirow{2}{*}{2018: Jun23\,(7)}                          &\multirow{2}{*}{2.9}  &\multirow{2}{*}{24--176}  &     LSB     & \llap{86.277\;}--\rlap{\;\;\,94.212} &         1.3            &   5.5$\times$3.0    &      \;\;10         \\
                                 &                        &                            &                         &                                                            &                      &                          &     USB     & \llap{101.755\,}--\rlap{\,109.690}   &         1.6            &   4.7$\times$2.6    & \llap{$-$}170       \\
\bottomrule
\end{tabular}
   \begin{tablenotes}[flushleft]
   \small
	 \item\textbf{Note:} 
        The J2000 coordinates (pointing/phase centers) for \apm\ are RA\,08h31m40.70s DEC\,$+$52$^\circ$45$'$17.60$''$ (note the peak position of \apm\ is at RA\,08h31m41.70s DEC\,$+$52$^\circ$45$'$17.46$''$), while for NCv1.143 they are RA\,12h56m32.544s DEC\,$+$23$^\circ$36$'$27.630$''$. The coordinates and redshifts of \apm\ and NCv1.143 are taken from \citet{2017A&A...608A.144Y} and \citet{2007A&A...467..955W}, respectively. The tuning names also correspond to the \texttt{Prog} name that can be found on the Science Data Archive of IRAM (\url{http://vizier.u-strasbg.fr/viz-bin/VizieR-3?-source=B/iram/noema}). The synthesized beam sizes correspond to natural weighting. The RMS$_\mathrm{/2\,MHz}$ is the root mean square noise per 2\,MHz bandwidth of the spectra for each sideband of each tuning.
   \end{tablenotes}
   \label{table:obs-log}
\end{table*}
\normalsize 

\begin{figure}[!htbp] 
\;\;\includegraphics[scale=0.395]{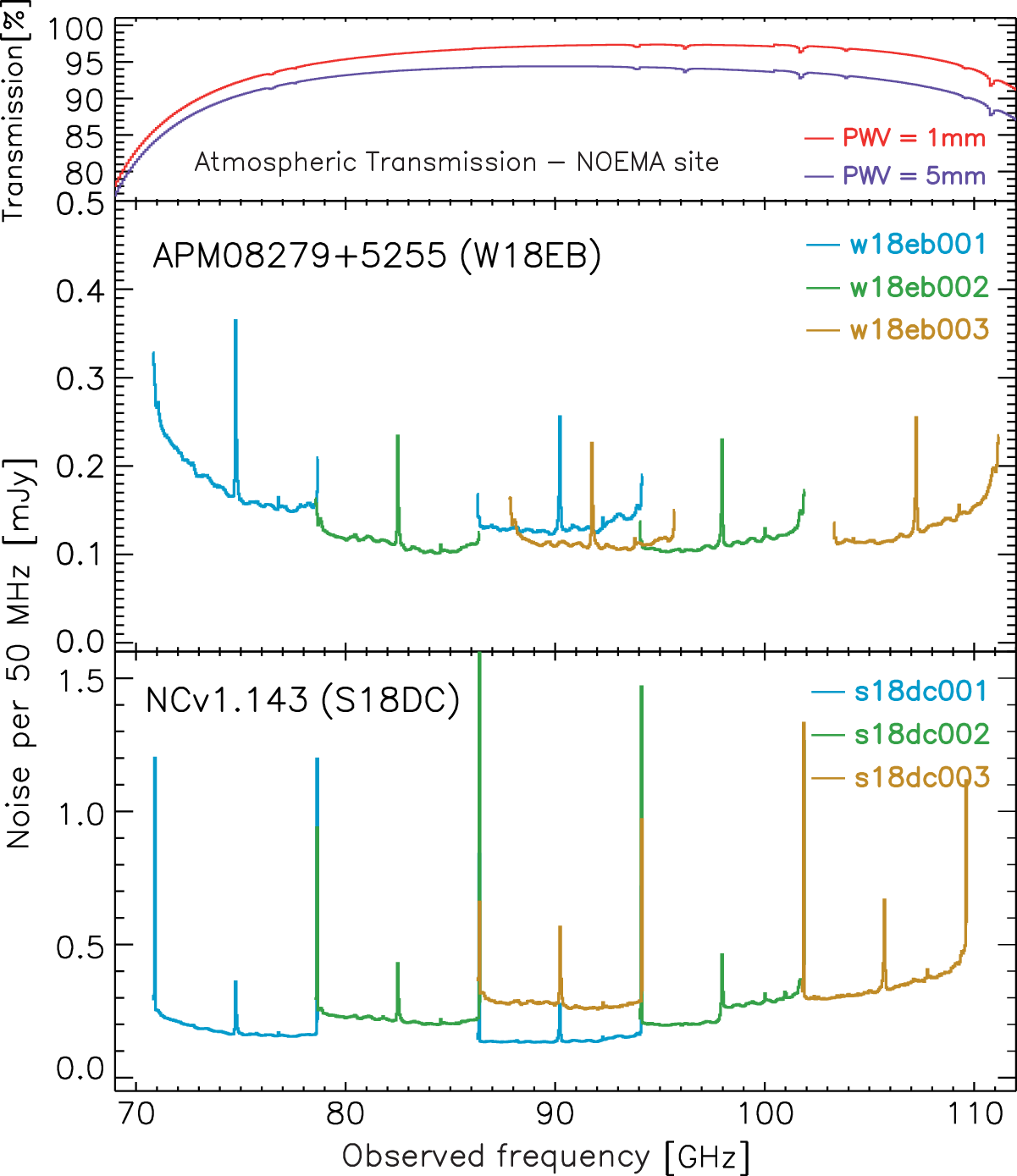}
\caption
{Tuning setups of the NOEMA observations. 
{\em Top panel:} Atmospheric transmission curve at the zenith of the NOEMA site for the frequency range covered by our line survey. The red and blue lines show the transmission curves for PWV\,=\,1\,mm and PWV\,=\,5\,mm, respectively. {\em Middle and bottom panels:} Tuning setups, their sky (observed) frequency coverage, and the RMS noise reached in each spectral window for \apm\ and NCv1.143. For display purposes, we smoothed the spectral resolution to 50\,MHz, resulting in an RMS that is five times smaller than the noise level reported in Table\,\ref{table:obs-log} for the 2\,MHz bins. As expected, the edges of the sidebands and the conjunction channels of the two spectral windows within each sideband exhibit large RMS values.
}
\vspace{-0.25cm}
\label{fig:tuning_noise}
\end{figure}

The deep 3\,mm spectral line surveys of \apm\ and NCv1.143 were conducted using NOEMA with its {\em PolyFix} correlator \citep{2019adw..confE..16G}. The observations were carried out under projects W18EB and S18DC (PIs: C. Yang \& A. Omont). The {\em PolyFix} correlator can process a total instantaneous bandwidth of 7.744\,GHz for up to 12 antennas per polarization (dual polarization) per sideband -- namely the upper sideband (USB) and the lower sideband (LSB) -- which provides a frequency coverage of 15.488 GHz per tuning in dual-polarization, with a fixed channel spacing of 2\,MHz across LSB and USB, separated by a gap of 7.744\,GHz. As shown in Fig.\,\ref{fig:tuning_noise} and Table\,\ref{table:obs-log}, we designed the tunings to allow for a continuous frequency coverage from 71 to 109.5\,GHz for NCv1.143 and from 71 to 111\,GHz for \apm\ (except for a gap between 101.7\,GHz and 103.3\,GHz, intentionally placed to cover some crucial lines at the high-frequency end, close to 111\,GHz). In the rest frame, the frequency coverage for \apm\ corresponds to 347.71--500.76\,GHz and 507.28--546.26\,GHz, and for NCv1.143, it corresponds to 323.15--500.73\,GHz. Therefore, the total frequency coverage for both sources extends to approximately 200\,GHz in the rest frame. This extensive coverage, as intended, enables us to detect multiple transitions of some key molecular emitters.

\begin{figure}[!htbp]
\centering
\includegraphics[scale=0.25]{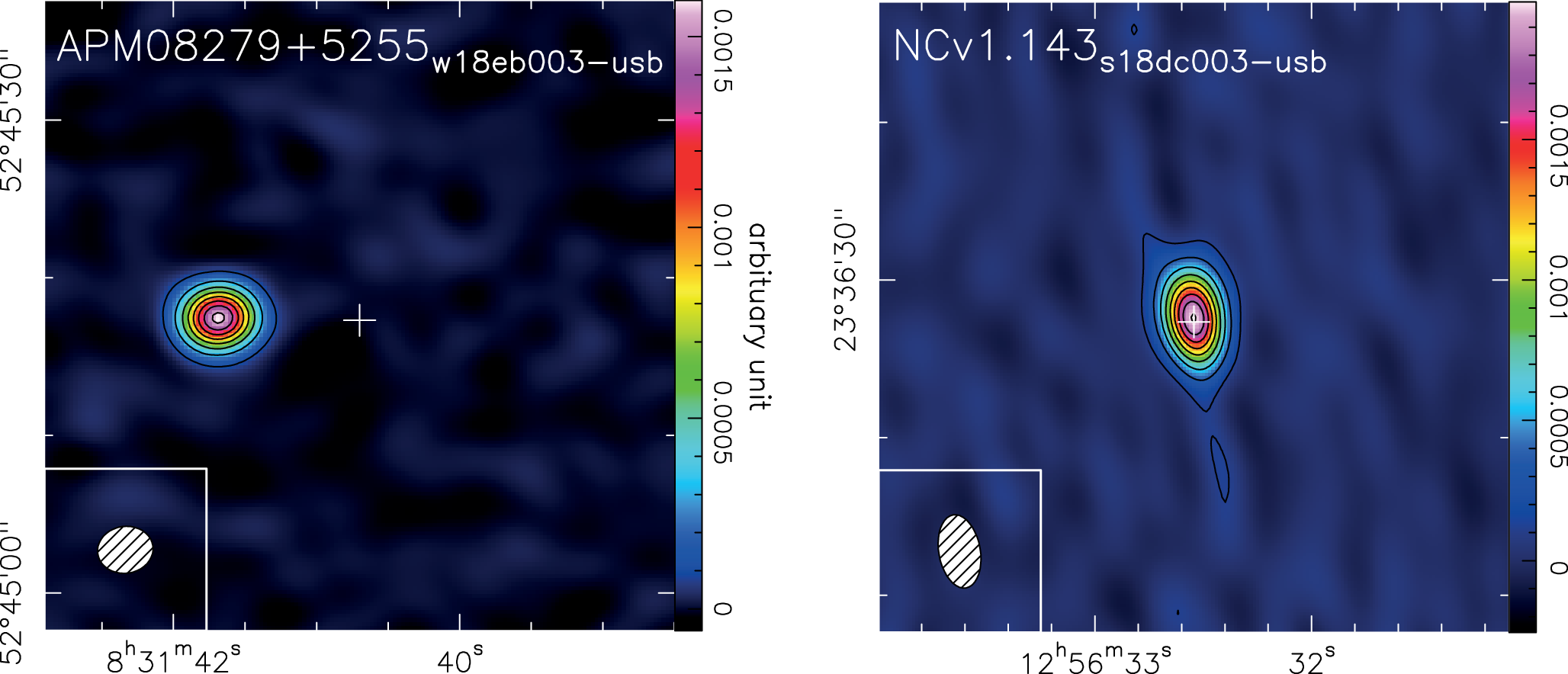}
\caption
{
NOEMA \texttt{CLEAN}ed images of \apm\ and NCv1.143, made from the visibilities obtained by averaging all the USB channels from the highest-frequency tunings (w18eb003 and s18dc003; Table~\ref{table:obs-log}). These tunings were chosen since they are expected to offer the highest spatial resolution among all the spectral windows in our tunings. The contour levels begin at 5\,$\sigma$ and increase in increments of 10\,$\sigma$. The unit of the color bar and contours is arbitrary, with values representing fluxes integrated over the entire USB of the tuning. The primary objective here is to illustrate that the sources are unresolved rather than provide precise flux measurements. The offset for \apm\ is explained in Sect.\,\ref{section:obs_redu}.
}
 \label{fig:003_sub_maps}
 \end{figure}

As summarized in Table\,\ref{table:obs-log}, the observations spanned two semesters from June 2018 to March 2019. Since our aim is to detect weak spectral features, we conducted our observations in compact C or D configurations to maximize sensitivity. The number of antennas used ranged from six to ten, with ten being used most of the time (Table\,\ref{table:obs-log}). The baselines remain compact, with lengths from 24 to 176\,m (D configuration) for most of the observations, while for a small portion of the observations, we also used baselines up to 368\,m (C configuration). The resulting synthesized beam sizes with natural weightings have modest/low resolutions of $\sim$\,1.7\arcsec$\times$3.0\arcsec\ to 4.2\arcsec$\times$5.6\,\arcsec (about 12$\times$22\,kpc to 31$\times$41\,kpc). The pointing positions are given in Table\,\ref{table:obs-log}. We note that, for \apm, there is an offset of about 9\arcsec\ between the phase center and the peak flux position (RA\,08h31m41.7s DEC\,$+$52$^\circ$45$'$17.5$''$; see Fig.\,\ref{fig:003_sub_maps}), which does not have any significant impact in our observation, as the position shift can be neglected considering the size of the primary beam (FWHM $\sim$\,50\arcsec\ at 3\,mm). Most observations were carried out in good or excellent weather conditions, except for those on November 22 and December 13, 2018, when the weather conditions were moderate. The observations of NCv1.143 were performed mostly in the summer semester and partially in the winter semester, with a mean phase RMS of $\sim$15$^\circ$ and precipitable water vapor (PWV) below 7$\mathrm{mm}$, while \apm\ was observed in the winter semester with a mean phase RMS of $\sim$\,5$^\circ$ and PWV below 5$\mathrm{mm}$. Consequently, the total data loss due to flagging is below 20\%. The total on-source time for all the tunings was $\sim$\,22.5 hours for NCv1.143 and $\sim$\,16.4 hours for \apm. In the end, the flux RMS levels in the native 2\,MHz spectral resolution reach $\sim$\,0.6--1.6\,mJy/beam and $\sim$\,0.6--1.0\,mJy/beam for NCv.143 and \apm, respectively. The RMS is also relatively stable across all the frequency coverage, except for the edge of the spectral windows where the values are high (Fig.\,\ref{fig:tuning_noise}). Details of the setups and conditions of the observations are listed in Table\,\ref{table:obs-log} and Fig.\,\ref{fig:tuning_noise}. The phase and bandpass were calibrated by measuring standard calibrators that are regularly monitored at NOEMA, including 3C279, 3C273, MWC349, and 0923+392. The accuracy of the absolute flux calibration is estimated to be about $\sim$10\% in the 3\,mm band. The final $uv$ tables containing the visibilities were produced after calibration using the \texttt{GILDAS}\footnote{See \url{http://www.iram.fr/IRAMFR/GILDAS} for more information about the \texttt{GILDAS} software.} packages \texttt{CLIC}, with the native 2\,MHz spectral resolution. Then imaging, \texttt{CLEAN}ing, $uv$ fitting, and spectra extraction were performed with \texttt{GILDAS}'s \texttt{MAPPING} on the $uv$ tables.

\section{Data analysis}
\label{section:results}

Upon acquiring the calibrated $uv$ tables, we collapsed all the channels from each sideband per tuning. Subsequently, we 
imaged and \texttt{CLEAN}ed those collapsed $uv$ tables to validate that the sources are not spatially resolved by the synthesized beams. As an example, we show the \texttt{CLEAN}ed images from the USB of the highest frequency tunings (s18dc003 and w18eb003, which correspond to the tunings with the highest spatial resolution) for NCv1.143 and \apm\ in Fig.\,\ref{fig:003_sub_maps}. From the perspective of the all-channel-collapsed images, it is clear that both sources are not significantly resolved by the synthesis beams with dimensions of approximately 3.0\arcsec{$\times$}1.7{\arcsec}--5.6\arcsec{$\times$}4.2{\arcsec} (using natural-weighting) for the signal surpassing the 5\,$\sigma$ levels. 

However, since the all-channel-collapsed images are dominated by the dust continuum, it is important to check if any of our observed line emissions can be significantly more extended than the dust continuum. Notably, some molecular line maps  (e.g., low-$J$ CO) show a more extended spatial distribution than the dust continuum in dusty star-forming galaxies \citep[e.g.,][]{2011MNRAS.412.1913I, 2018ApJ...863...56C, 2023MNRAS.523.4654T}. After further examining specific channels where the emission is dominated by lines (e.g., CO), we find no significant difference in the size of the source between the all-channel-collapsed images and the line-dominated channel images (all the differences are below 3\,$\sigma$). We thus conclude that all the line emissions and the dust continuum are not resolved by our beams. Additionally, we do not expect any significant missing flux, given the compactness of the emission with regard to the beam sizes.

\setlength{\tabcolsep}{0.86em}
\renewcommand{\arraystretch}{1.0}
\begin{table}[t]
\small
\centering
\caption{Parameter of the elliptical Gaussian model used in fitting the $uv$ data.}
\begin{tabular}{rccc}
\toprule
Source                            &  tuning       &        $\delta_\mathrm{RA}$  &  $\delta_\mathrm{DEC}$    \\
                                  &               &            \arcsec           &      \arcsec              \\
\midrule                                                                                                     
\multirow{3}{*}{\apm}             &   w18eb001    &           $9.14\pm0.02$      &  \llap{$-$}$0.24\pm0.01$  \\
                                  &   w18eb002    &           $9.09\pm0.01$      &  \llap{$-$}$0.11\pm0.01$  \\
                                  &   w18eb003    &           $8.96\pm0.01$      &  \llap{$-$}$0.12\pm0.01$  \\
\multirow{3}{*}{NCv1.143}         &   s18dc001    &           $0.11\pm0.01$      &  \llap{$-$}$0.03\pm0.01$  \\
                                  &   s18dc002    &           $0.08\pm0.02$      &  \llap{$-$}$0.02\pm0.02$  \\
                                  &   s18dc003    & \llap{$-$}$0.01\pm0.02$      &  \llap{$-$}$0.04\pm0.03$  \\
\bottomrule  
\end{tabular}
   \begin{tablenotes}[flushleft]
   \small
	 \item\textbf{Note:} 
        The phase centers for \apm\ and NCv1.143 are the pointing positions (Table\,\ref{table:obs-log}). The $\sim$\,9\arcsec\ shift in $\delta_\mathrm{RA}$ of \apm\ was due to a shifted observational phase center. The $\delta_\mathrm{RA}$ and $\delta_\mathrm{DEC}$ positions are the center of the Gaussian profiles relative to the phase center. 
        The rest of the fitted parameters --- the major and minor axis of the ellipse,  $r_\mathrm{major}$ and $r_\mathrm{minor}$ and the position angle PA are all consistent across different spectral windows within the uncertainties. As they are irrelevant to the purpose of this work, we do not list them here.
   \end{tablenotes}
   \label{table:uvfit}
\end{table}
\normalsize

Since the sources are not resolved and remain compact for both the emission from the continuum and the lines, with the purpose of extracting the total fluxes and maximizing the signal-to-noise ratios, the emission of our sources can thus be approximated using simple elliptical Gaussian models in the $uv$ plane across all the channels. Such an assumption is also consistent with other long-baseline interferometric data of the sources, namely the 0.5\arcsec{$\times$}0.3{\arcsec} resolution 1.2\,mm dust continuum, CO and \hto\ line images of NCv1.143 (see \citealt{2017PhDT........21Y} nad our Appendix\,\ref{appen:lens-model}) and the 0.3\arcsec\ resolution 2.6\,mm dust continuum and CO images of \apm\ \citep{2009ApJ...690..463R}. These high-angular-resolution data show that the largest structures of the ISM emission are $\lesssim$\,1{\arcsec} for \apm\ and $\lesssim$\,2\arcsec\ for NCv1.143. These structures are characterized by the lensing-produced multiple image components distributed along the Einstein rings. The sizes of the Einstein rings are slightly smaller than the synthesized beam sizes as listed in Table\,\ref{table:obs-log}. The individual image components are about  $\lesssim$\,0.2{\arcsec} and 0.7{\arcsec} for \apm\ and NCv1.143, respectively.

\subsection{Extracting spectra}
\label{subsection:spec-extraction}

For flux extraction, we assume a model of a single two-dimensional elliptical Gaussians and directly fit the visibilities in the $uv$ plane. We verified our fitting results by checking the residual images and did not find any significant residual fluxes. Nevertheless, we caution that our purpose is only to extract the total fluxes in each channel, and the study of the ISM morphology is beyond the scope of this work. Therefore, while the two-dimensional elliptical Gaussian models are accurate enough as approximations to extract source flux and reduce the ambiguity in \texttt{CLEAN}ing and aperture selection in the traditional method of spectra extraction, they do not reflect the source morphology (see Appendix\,\ref{appen:lens-model} for discussion of the morphology of the sources with high spatial resolution data). Considering the spatial resolution of our line survey data and the main goal of maximizing spectral sensitivity and extracting fluxes, two-dimensional elliptical Gaussian is thus a good approximation.

The two-dimensional elliptical Gaussian profiles are characterized by the central positions relative to the phase centers, $\delta_\mathrm{RA}$ and $\delta_\mathrm{DEC}$, the major and minor axes, $r_\mathrm{major}$ and $r_\mathrm{minor}$, the position angle PA, and the total flux. We used the \texttt{UVFIT} package of the \texttt{GILDAS} to fit the visibilities in the $uv$-plane directly with the model. We fixed the parameter of $\delta_\mathrm{RA}$, $\delta_\mathrm{DEC}$, $r_\mathrm{major}$, $r_\mathrm{minor}$, and PA to the values derived from the all-channel-collapsed $uv$ tables for each tuning (assuming that the parameters, except for the flux, do not vary significantly between the LSB and USB within the same tuning), and only allow the flux to change across all the channels. Accordingly, the fitted fluxes, along with the errors across different channels, produce the final spectra containing fluxes and their errors. 

The central positions, which are part of the fixed parameters derived in each tuning, are listed in Table\,\ref{table:uvfit}. For NCv1.143, we find an elliptical Gaussian size of $\sim$\,(1.3--1.6)\arcsec$\times$(1.2--1.4)\arcsec, depending on the frequencies. For \apm, the values of $r_\mathrm{major}$/$r_\mathrm{minor}$ are from $1.41\pm0.38$ to $2.60\pm1.60$, indicating its elongated morphology, with similar PA values derived from $40\pm10$ to $46\pm21$ deg across three tunings. This is consistent with the elongated distribution seen at high-spatial-resolution images \citep{2009ApJ...690..463R}. For NCv1.143, the aspect ratio ($r_\mathrm{major}$/$r_\mathrm{minor}$) is from $1.03\pm0.06$ to $1.40\pm0.09$, showing little degree of deviation from circular symmetry, consistent with high-angular-resolution images (Appendix\,\ref{appen:lens-model}). Both $uv$-plane models are robust; the fitted parameters have a good agreement across all three tunings for each source. 

The positions of \apm\ are particularly well constrained because of the stable observation conditions and the compactness of the source compared with the synthesis beams (Table\,\ref{table:uvfit}). Nevertheless, we also find that the values of $r_\mathrm{major}$ and $r_\mathrm{minor}$ of \apm\ are decreasing with increasing frequencies, suggesting that the true source size is significantly smaller than the synthesis beam. While for NCv1.143, similar values of $r_\mathrm{major}$ and $r_\mathrm{minor}$ across three tunings indicate that the source is only slightly smaller than the synthesis beam.

\begin{landscape}
\begin{figure}[!htbp]
\centering
\includegraphics[scale=1.325]{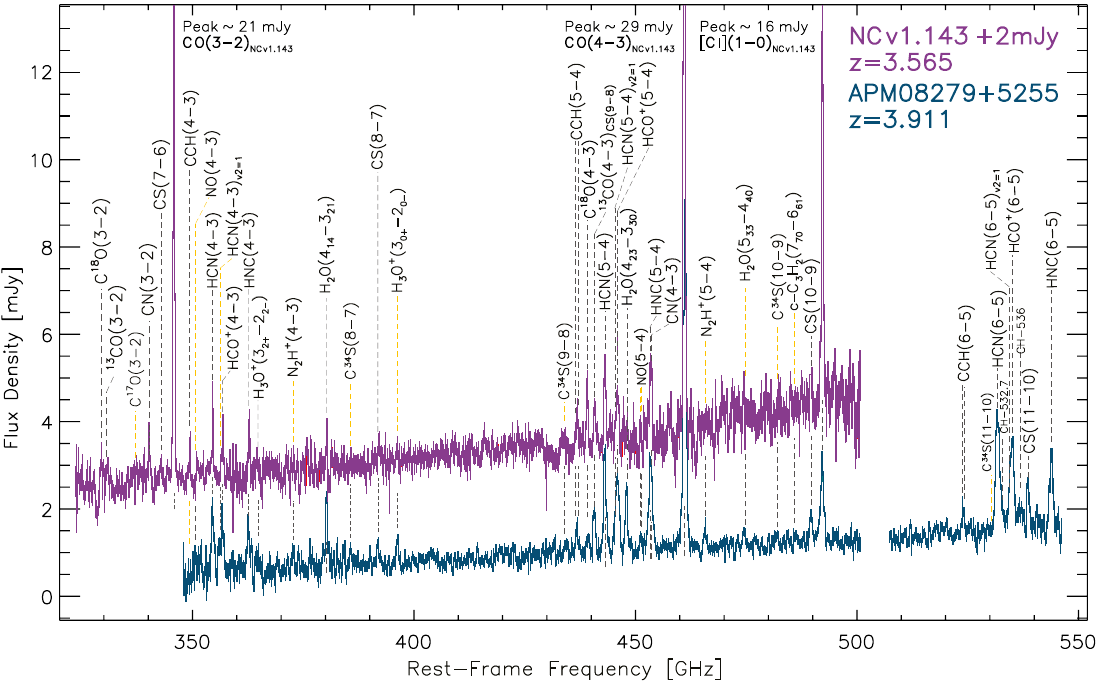}
\vspace{-0.25cm}
\caption
{
NOEMA 3\,mm band spectral surveys of \apm\ and NCv1.143, shown as the blue and violet spectra, respectively. For visualization purposes, the spectrum of NCv1.143 has been shifted up by 2\,mJy. Both spectra are binned to 50\,MHz (about 150\,\kms) to highlight the line detection. Error bars are indicated by thin lines (the same color as the spectra) overlaid on the spectra. Dashed black lines identify the lines detected in both sources, while dashed orange lines highlight the lines that are only detected in one of the two sources. The line names are labeled above the dashed lines. The peak values indicated for \co32, \co43, and \ci(1-0)\ emission lines in NCv1.143 are the true peak values before the 2\,mJy shift, and they are outside the box of the figure. A zoomed-in view is provided in Fig.\,\ref{fig:zoom-in_full-spec}.
}
\label{fig:spec}
\end{figure}
\end{landscape}
\newpage

From the aforementioned method, we obtained six spectra for each source corresponding to all the USB and LSB of the three tunings per source. The spectra are then combined through a simple linear re-grid along the frequency axis, taking the errors of each channel into account. The final combined spectra are displayed in Fig.\,\ref{fig:spec} (a zoom-in view is provided in Fig.\,\ref{fig:zoom-in_full-spec}). The following analyses throughout the work are then based on these final combined spectra. 
 
\subsection{Line identification}
\label{subsection:line-id}

\setlength{\tabcolsep}{2.97em}
\renewcommand{\arraystretch}{1.1}
\begin{table*}[htbp]
\centering
\caption{Detected species and transitions in the two sources from our NOEMA surveys.}
\begin{tabular}{lcc}
\toprule
Species            & \apm                         & NCv1.143                              \\ 
\midrule         
\ci                & 1--0                                    & 1--0                                  \\ 
$^{12}$C$^{16}$O   & 4--3                                    & 3--2, 4--3                            \\
$^{13}$C$^{16}$O   & 4--3                                    & 3--2, 4--3                            \\
$^{12}$C$^{18}$O   & 4--3                                    & 3--2, 4--3                            \\
CN                 & 4--3                                    & 3--2, 4--3                            \\
CCH                & 5--4, 6--5                              & 4--3, 5--4                            \\
HCN                & 4--3, 5--4, 6--5                        & 4--3, 5--4                            \\
HCN(${v_2=1f}$)    & 4--3, 5--4, 6--5                        & --                                    \\
HCO$^+$            & 4--3, 5--4, 6--5                        & 4--3, 5--4                            \\
HNC                & 4--3, 5--4, 6--5                        & 4--3, 5--4                            \\
CS                 & 8--7, 9--8, 10--9, 11--10               & 7--6, 8--7, 9--8, 10--9               \\
\hto               & \t414321, \t423330, \t533440            &\t414321, \t423330                     \\
\httop             & 3$^+_{2}$--2$^-_{2}$, 3$^+_{0}$--2$^-_{0}$& 3$^+_{2}$--2$^-_{2}$, 3$^+_{0}$--2$^-_{0}$ \\
NO                 & 4--3, 5--4                              & --                                    \\
N$_2$H$^+$         & 4--3, 5--4                              & --                                    \\
C$^{34}$S          & 8--7, 9--8, 10--9                       & --                                    \\
CH                 & $N$\,=\,4--3($J=3/2^+\text{--}1/2^-$), $N$\,=\,4--3($J=3/2^-\text{--}1/2^+$)& --\\
c-C$_3$H$_2$       & \t770611                                & --                                    \\
\bottomrule  
\end{tabular}
   \begin{tablenotes}[flushleft]
   \small
	 \item\textbf{Note:} 
        Some of the lines are not detected because of the difference in the rest-frame frequency coverage for the two sources. See Fig.\,\ref{fig:spec} for details.
   \end{tablenotes}
   \label{table:detection-summary}
\end{table*}
\normalsize 

To maximize the accuracy of the line identification, we adopt two complementary methods that cross-verify each other to produce the final line catalogs. First, we utilized  \texttt{MADCUBA}\footnote{MADCUBA VERSION 6.0 (07/05/2018). \url{https://www.cab.inta-csic.es/madcuba/index.html}} (MAdrid Data CUBe Analysis; \citealt{2019A&A...631A.159M}) to identify molecular species by simultaneously fitting all their transitions within our frequency coverage, also accounting the brightness of the lines as predicted by the local thermal equilibrium (LTE) assumption. Second, we input the identified line detection list, as priors, into our matched filter identification code to cross-check and ensure that no line features are overlooked. In the following sections, we provide the details of these processes.

\subsubsection{Local thermal equilibrium analysis and line identification}
\label{subsection:MADCUBA}

As an initial step to identify the spectral line detections in our observations, we deployed \texttt{MADCUBA} to conduct line fittings of all the lines across the entire spectra, assuming LTE conditions. Fitting the emission of all transitions for all the possible species, rather than identifying individual spectral features by their central observed frequency, allows for a robust identification when multiple transitions are detected. Additionally, this method accounts for line blending based on identified species and their LTE-predicted fluxes. We included a total of 29 species in the model, drawing from previously published extragalactic line surveys \citep{2021A&A...656A..46M}. A list of the molecules and transitions detected is compiled in Table\,\ref{table:detection-summary}. Within the 29-species list, undetected species included in the fit are CH$_3$CN, CH$_3$OH, H$_2$CO, H$_2$S, N$_2$D$^+$, the isotopologs C$^{17}$O, DNC, C$^{15}$N,  HC$^{15}$N, H$^{15}$NC, and the vibrational states of HNC. We do not identify any other feature outside this list in our spectra (which is cross-checked in Sect.\,\ref{subsection:blind-id}).

The model adopts a linewidth of 520 and 330\,\kms\ for \apm\ and NCv1.143, respectively, as derived from CO emission from our data. These values are consistent with previous measurements \citep{2007A&A...467..955W, 2017A&A...608A.144Y}. Here, we assume an apparent source size of 0.13 arcsec$^2$, which should be within a factor of $\lesssim$\,3 from the sizes of both sources (with significant uncertainties; see our Appendix\,\ref{appen:lens-model} and \citealt{2009ApJ...690..463R}). The effect of the source size in the model partially degenerates with the column density, where a smaller source size would require higher column densities to model the observed spectra. Given that the bulk of our analysis is rooted in comparing relative column densities (to that of C$^{18}$O, see Sect.~\ref{section:line-atlas-diff}), the uncertainties associated with size estimations are likely canceled out if the line-emitting regions are similar. If this is the case, the choice of source size will have minimal influence on our overall conclusions. We also acknowledge that some of the lines might have different sizes than that of the C$^{18}$O lines, such as more extended low-$J$ CO. However, because of the limited spatial resolution of our data, future high-angular-resolution observations are needed to disentangle the uncertainties in the sizes of the emitting regions for a more accurate accounting of the size variations across all the lines. With the assumed source size, the models are still within the optically thin regime for almost all the lines. A much smaller source size would result in line saturation, mostly on the CO transitions toward NCv1.143, while the rest of the species is still in the optically thin regime. For those species for which the excitation temperature could not be constrained with the observations, mainly those with a single spectral line feature, a $T_{\rm ex}$ of 36 and 25\,K has been derived from HCN, HCO$^+$, and HNC and assumed for all other species (except for \hto\ and the vibration excited HCN), for \apm\ and NCv1.143, respectively. It is important to note that \hto\ lines are mainly powered by radiative pumping in both sources \citep{2010A&A...518L..42V, 2016A&A...595A..80Y}; therefore, the fitting here (assuming a molecular-gas temperature of 230\,K for both sources) serves only to identify the \hto\ features. We assume an ad hoc high temperature of 300\,K, which is the typical value found in Arp\,220 \citep{2011A&A...527A..36M} and NGC\,4418 \citep{2010ApJ...725L.228S}, to fit the vibrational transitions of HCN given its high energy levels. Table\,\ref{table:madcuba-fit} lists the fitted column density ($N$) and temperature of the molecular species under LTE for both sources. We note that the excitation temperature ($T_{\rm ex}$) should be treated as a lower limit of the kinetic temperature ($T_{\rm kin}$) of the molecular gas \citep{1999ApJ...517..209G}. Nevertheless, the higher $T_{\rm ex}$ values found in \apm\ as compared to NCv1.143 reflect that the molecular gas conditions are generally more extreme in the former.

\subsubsection{Matched-filtering line identification and the final fitting results}
\label{subsection:blind-id}
After identifying the spectral lines using \texttt{MADCUBA} under the LTE assumptions as described in the previous section, we fed in the list of identified lines as priors and performed a second search based on a matched-filtering algorithm, where we convolve the spectra with Gaussian kernels based on the best fit CO linewidths, utilizing the \texttt{specutils} package of \texttt{AstroPy} \citep[similar approach as in, e.g.,][]{2023ApJ...945..111B}. This process leveraged the well-detected bright CO-line profile to enhance detection and filter out noise for better detecting weak emission features.

First, we conducted a ``blind'' matched-filter detection using only a 5-$\sigma$ detection threshold. Then, we further lowered our detection threshold based on the prior information about the line detection from \texttt{MADCUBA}, especially for the case of the blended lines, and performed a second line search. Accordingly, we identified all the emission and absorption lines after this search. Statistical tests are performed to choose the adequate threshold to make sure more than 99.7\% of the detected lines are real. In the end, we also verified all the line identifications visually as a triple-check. We do not see any significant unidentified lines beyond the initial \texttt{MADCUBA} list. Thus, our initial list of 29 species is sufficient for line identification. 

The final cross-verified detections are summarized in Table\,\ref{table:detection-summary}. During this process, we simultaneously fitted all the line features, along with the underlying continuum emission (Figs.\,\ref{fig:flux-fitting-1} and Fig.\,\ref{fig:flux-fitting-2}). The discussion on the physical aspects of the continuum for both sources is presented in Sect.\,\ref{subsection:ctm}. In the fitting process, we also deployed a Markov chain Monte Carlo (MCMC) technique, where we used samplers that are twice the number of the free parameters with 2000 interactions after 1000 burn-in steps. The derived posterior distribution of the fitted fluxes can better account for the overall uncertainties. The resulting line fluxes are listed in Table\,\ref{table:line-flux}.

{\centering
\setlength{\tabcolsep}{0.73em}
\begin{table}[hbpt]
\renewcommand{\arraystretch}{1.25}
\small
\caption{\texttt{MADCUBA}-derived physical parameters.}
\begin{tabular}{lcccc}
\toprule
Species             &  \multicolumn{2}{c}{Column density}                       &  \multicolumn{2}{c}{Excitation temperature}  \\
                    &  \multicolumn{2}{c}{log\,$N$ (cm$^{-2}$)~\tablefootmark{a}}  &  \multicolumn{2}{c}{$T_{\rm ex}$\,(K)}    \\
                    &          APM                 &   NC                       &         APM        &         NC              \\
\midrule                                                                                                   
C                   &          18.2 (16.9)         &  18.7 (17.6)               &         36         &         25              \\
$^{12}$C$^{16}$O    &          17.5 (15.7)         &  17.9 (16.7)               &         40         &         25              \\
$^{13}$C$^{16}$O    &          16.1 (15.2)         &  16.3 (15.5)               &         36         &         25              \\
$^{12}$C$^{18}$O    &          16.2 (15.3)         &  16.7 (15.7)               &         36         &         25              \\
CN                  &          14.0 (12.8)         &  14.4 (13.5)               &         36         &         25              \\
CCH                 &          14.7 (13.8)         &  14.9 (14.0)               &         36         &         25              \\
HCN                 &          14.2 (12.9)         &  14.1 (13.4)               &         37 (2)     &         26 (4)          \\
HCO$^+$             &          13.9 (12.7)         &  13.7 (13.2)               &         36 (2)     &         24 (4)          \\
HNC                 &          14.0 (13.0)         &  14.0 (13.6)               &         32 (2)     &         19 (4)          \\
CS                  &          14.6 (13.4)         &  14.3 (13.7)               &         36         &         25              \\
\hto                &          16.4 (14.9)         &  15.6 (15.2)               &         230 (20)   &         230             \\
\httop              &          15.5 (14.5)         &  --                        &         36         &         --              \\
NO                  &          16.0 (15.3)         &  --                        &         36         &         --              \\
HCN$_{v_2=1f}$      &          15.1 (14.0)         &  --                        &         300        &         --              \\
N$_2$H$^+$          &          13.1 (12.1)         &  --                        &         36         &         --              \\
C$^{34}$S           &          14.0 (13.4)         &  14.1 (13.7)               &         36         &         25              \\
CH                  &          14.1 (13.0)         &  --                        &         36         &         --              \\
c-C$_3$H$_2$        &          14.0 (13.2)         &  --                        &         36         &         --              \\
\bottomrule  
\end{tabular}
   \begin{tablenotes}[flushleft]
   \small
	 \item\textbf{Note:} 
        LTE-derived column densities and excitation temperatures of \apm\ (APM for short) and NCv1.143 (NC for short). We note that the values are not corrected for lensing magnification. Temperature values without errors indicate species for which this parameter could not be constrained and was fixed to the value derived from HCN and HCO$^+$. Although most of the lines are well modeled with a single temperature, for APM\,0879+5255, the vibrationally excited HCN (${v_2=1f}$) requires significantly higher kinetic temperatures. 
        The same applies to the \hto\ lines, where a pure collisional excitation cannot explain the fluxes well. Values in brackets are errors. \tablefoottext{a}{The reported column density $\log_{10} N=A(B)$ corresponds to $N=10^A\pm 10^B$.}
   \end{tablenotes}
   \label{table:madcuba-fit}
\end{table}
\normalsize}

\setlength{\tabcolsep}{1.125em}
\begin{table*}[htbp]
\renewcommand{\arraystretch}{1.05}
\scriptsize
\centering
\caption{Integrated line fluxes in \apm\ and NCv1.143.}
\begin{tabular}{lrrccccc}
\toprule
Line                                                                   & $E_\mathrm{up}/k_\mathrm{B}$ & $\nu_\mathrm{rest-frame}$ &  $A_\mathrm{UL}$   &       \multicolumn{2}{c}{Integrated Flux}        &   \multicolumn{2}{c}{Reference}       \\
                                                                       &              [K]             &      [GHz]                &      [s$^{-1}$]    &       \multicolumn{2}{c}{[Jy\,\kms]}             &                                       \\
                                                                       &                              &                           &                    &  \apm       & NCv1.143                &   \apm &   NCv1.143        \\
\midrule                                                                                                                                                                                                   
\ci(1--0)                                                              &             23.62            &      492.161              &      7.88e-8       & $1.02\pm0.04$          & $3.38\pm0.11$           &   This work       &   This work       \\
\ci(2--1)                                                              &             62.46            &      809.342              &      2.65e-7       &      --                & $4.4\pm0.9$             &   --              &   Y17a            \\
\co10                                                                  &              5.53            &      115.271              &      7.20e-8       & $0.168\pm0.015$        &       --                &   R09             &   --              \\
\co21                                                                  &             16.60            &      230.538              &      6.91e-7       &   $0.81\pm0.18$        &       --                &   R09             &   --              \\
\co32                                                                  &             33.19            &      345.795              &      2.50e-6       &      --                & $5.27\pm0.04$           &   This work       &   This work       \\
\co43                                                                  &             55.32            &      461.041              &      6.13e-6       & $4.54\pm0.03$          & $7.37\pm0.07$           &   This work       &   This work       \\ 
\co54                                                                  &             82.97            &      576.268              &      1.22e-5       &      --                &  $10.7\pm0.9$           &   --              &   Y17a            \\
\co65                                                                  &            116.16            &      691.473              &      2.14e-5       &      7.3               &  $ 9.6\pm1.2$           &   W07             &   Y17a            \\
\co76                                                                  &            154.87            &      806.652              &      3.42e-5       &      --                &  $11.5\pm0.9$           &   --              &   Y17a            \\
\co98                                                                  &            248.88            &     1036.912              &      7.33e-5       &     12.5               &       --                &   W07             &   --              \\
CO(10--9)                                                              &            304.16            &     1151.985              &      1.01e-4       &     11.9               &  $ 4.2\pm1.3$           &   W07             &   Y17b            \\
CO(11--10)                                                             &            364.97            &     1267.014              &      1.34e-4       &     10.4               &       --                &   W07             &   --              \\
\llap{$^{13}$}${\mathrm{CO}}(3\text{--}2)$                             &             31.73            &      330.588              &      2.18e-6       & --                     & $0.19\pm0.06$           &   --              &    This work      \\ 
\llap{$^{13}$}${\mathrm{CO}}(4\text{--}3)$ $_\mathrm{ blended\,with\, CS(9\text{--}8),\,deblended\,flux}$ &             52.89            &      440.765              &      5.35e-6       & $\mathbf{0.27\pm0.14}$ & $\mathbf{0.14\pm0.05}$  &   This work       &    This work      \\
${\mathrm{C^{18}O}}(3\text{--}2)$                                      &             31.61            &      329.331              &      2.17e-6       & --                     & $0.29\pm0.06$           &   This work       &    This work      \\
${\mathrm{C^{18}O}}(4\text{--}3)$                                      &             52.68            &      439.089              &      5.33e-6       & $0.24\pm0.03$          & $0.42\pm0.06$           &   This work       &    This work      \\
CN($N=3\text{--}2$)($J=7/2\text{--}5/2$)                               &             32.66            &      340.248              &      4.13e-4       & --                     & $0.29\pm0.04$           &   --              &    This work      \\
CN($N=4\text{--}3$)($J=9/2\text{--}7/2$)                               &             54.43            &      453.607              &      1.02e-3       & $0.33\pm0.05$          & $0.41\pm0.08$           &   This work       &    This work      \\
CCH($N=4\text{--}3$)                                                   &             41.91            &      349.338              &      1.28e-4       & --                     & $0.26\pm0.05$           &   --              &    This work      \\
CCH($N=5\text{--}4$)                                                   &             62.87            &      436.661              &      2.57e-4       & $0.38\pm0.03$          & $0.18\pm0.06$           &   This work       &    This work      \\
CCH($N=6\text{--}5$)                                                   &             88.02            &      523.972              &      4.53e-4       & $0.24\pm0.04$          & --                      &   This work       &    --             \\
HCN(4--3)                                                              &             42.53            &      354.505              &      2.05e-3       & $0.80\pm0.08$          & $0.61\pm0.04$           &   This work       &    This work      \\
HCN(5--4)                                                              &             63.80            &      443.116              &      4.10e-3       & $1.29\pm0.03$          & $0.62\pm0.05$           &   This work       &    This work      \\
HCN(6--5)                                                              &             89.32            &      531.716              &      7.20e-3       & $1.49\pm0.04$          & --                      &   This work       &    --             \\
HCO$^+$(4--3)                                                          &             42.80            &      356.734              &      3.63e-3       & $0.59\pm0.08$          & $0.37\pm0.04$           &   This work       &    This work      \\
HCO$^+$(5--4)                                                          &             64.20            &      445.903              &      7.25e-3       & $1.02\pm0.04$          & $0.39\pm0.05$           &   This work       &    This work      \\
HCO$^+$(6--5)                                                          &             89.88            &      535.062              &      1.27e-2       & $1.05\pm0.06$          & --                      &   This work       &    --             \\
HNC(4--3)                                                              &             43.51            &      362.630              &      2.30e-3       & $0.54\pm0.08$          & $0.43\pm0.08$           &   This work       &    This work      \\
HNC(5--4)                                                              &             65.26            &      453.270              &      4.58e-3       & $1.07\pm0.04$          & $0.37\pm0.07$           &   This work       &    This work      \\
HNC(6--5)                                                              &             91.37            &      543.898              &      8.04e-3       & $0.82\pm0.07$          & --                      &   This work       &    --             \\
CS(7--6)                                                               &             65.83            &      342.883              &      8.40e-4       & --                     & $0.09\pm0.04$           &   --              &    This work      \\
CS(8--7)                                                               &             84.63            &      391.847              &      1.26e-3       & $0.24\pm0.04$          & $0.11\pm0.06$           &   This work       &    This work      \\
CS(9--8) $_\mathrm{blended\,with\, ^{13}CO(4\text{--}3),\,deblended\,flux}$                          &            105.79            &      440.803              &      1.81e-3       & $\mathbf{0.27\pm0.14}$ & $\mathbf{0.14\pm0.07}$  &   This work       &    This work      \\
CS(10--9)                                                              &            129.30            &      489.751              &      2.50e-3       & $0.30\pm0.04$          & $0.16\pm0.09$           &   This work       &    This work      \\
CS(11--10)                                                             &            155.15            &      538.689              &      3.34e-3       & $0.39\pm0.05$          & --                      &   This work       &    --             \\ 
p-\htot202111                                                          &            100.85            &      987.927              &      5.84e-3       &   $9.1\pm0.9$          & --                      &   V11             &    --             \\
p-\htot211202                                                          &            136.94            &      752.033              &      7.06e-3       &   $4.3\pm0.6$          &   $5.8\pm0.3$           &   V11             &    Y16            \\
p-\htot220211                                                          &            195.91            &     1228.789              &      1.87e-2       &   $6.7\pm0.8$          & --                      &   L11             &    --             \\
p-\htot422413                                                          &            454.34            &     1207.639              &      2.86e-2       &   $7.5\pm2.1$          &   $3.2\pm0.4$           &   V11             &    Y17b           \\
p-\htot533440                                                          &            725.11            &      474.689              &      4.53e-5       & $0.14\pm0.04$          & --                      &   This work       &    --             \\ 
o-\htot312221                                                          &            249.44            &     1153.127              &      2.63e-3       & --                     &   $2.5\pm1.4$           &   --              &    Y17b           \\
o-\htot321312                                                          &            305.25            &     1162.912              &      2.29e-2       &   $8.0\pm0.9$          &   $9.3\pm0.2$           &   V11             &    Y17b           \\ 
o-\htot414321                                                          &            323.50            &      380.197              &      2.99e-5       & $0.69\pm0.06$          & $0.24\pm0.06$           &   This work       &    This work      \\
o-\htot423330                                                          &            432.16            &      448.001              &      5.26e-5       & $0.85\pm0.03$          & $0.09\pm0.06$           &   This work       &    This work      \\
p-\httop(\tp32+22-)                                                     &            139.34            &      364.797              &      2.79e-4       & $0.14\pm0.06$          & $0.07\pm0.06$           &   This work       &    This work      \\
o-\httop(\tp30+20-)                                                     &            169.14            &      396.272              &      6.44e-4       & $0.30\pm0.05$          & $0.04\pm0.03$           &   This work       &    This work      \\
NO($N=4\text{--}3$)($J=9/2^-\text{--}7/2^+$)                           &             36.06            &      350.689              &      5.43e-6       & --                     & --                      &   --              &    --             \\ 
NO($N=4\text{--}3$)($J=9/2^+\text{--}7/2^-$)                           &             36.13            &      351.044              &      5.43e-6       & $0.11\pm0.10$          & --                      &   This work       &    --             \\ 
NO($N=5\text{--}4$)($J=11/2^+\text{--}9/2^-$)                          &             57.70            &      450.940              &      1.19e-5       & $0.13\pm0.05$          & --                      &   This work       &    --             \\ 
NO($N=5\text{--}4$)($J=11/2^-\text{--}9/2^+$)                          &             57.79            &      451.289              &      1.20e-5       & $0.06\pm0.05$          & --                      &   This work       &    --             \\ 
HCN(4--3)$_{v2=1f}$                                                    &           1067.14            &      356.256              &      1.78e-3       & $0.26\pm0.10$          & --                      &   This work       &    --             \\ 
HCN(5--4)$_{v2=1f}$                                                    &           1088.51            &      445.303              &      1.24e-2       & $0.18\pm0.04$          & $<0.15$                 &   This work       &    This work      \\
HCN(6--5)$_{v2=1f}$                                                    &           1114.16            &      534.340              &      2.21e-2       & $0.22\pm0.05$          & --                      &   This work       &    --             \\ 
N$^2$H$^+$(4--3)                                                       &             44.71            &      372.673              &      3.09e-3       & $0.18\pm0.06$          & --                      &   This work       &    --             \\ 
N$^2$H$^+$(5--4)                                                       &             67.07            &      465.825              &      6.18e-3       & $0.27\pm0.03$          & --                      &   This work       &    --             \\ 
C$^{34}$S(8--7)                                                        &             83.28            &      385.577              &      1.14e-3       & $0.14\pm0.06$          & --                      &   This work       &    --             \\ 
C$^{34}$S(9--8)                                                        &            104.10            &      433.751              &      1.60e-3       & $0.10\pm0.03$          & --                      &   This work       &    --             \\ 
C$^{34}$S(10--9)                                                       &            127.23            &      481.916              &      2.14e-3       & $0.11\pm0.04$          & --                      &   This work       &    --             \\ 
CH($N=4\text{--}3$)($J=3/2^+\text{--}1/2^-$)                           &             25.73            &      532.746              &      4.14e-2       & $0.24\pm0.05$          & --                      &   This work       &    --             \\ 
CH($N=4\text{--}3$)($J=3/2^-\text{--}1/2^+$)                           &             25.76            &      536.779              &      6.38e-2       & $0.24\pm0.06$          & --                      &   This work       &    --             \\ 
c-C$_3$H$_2$(\t661550)                                                 &             68.67            &      410.296              &      3.35e-3       & --                      & --                      &   This work       &    --             \\ 
c-C$_3$H$_2$(\t770661)                                                 &             91.99            &      485.732              &      5.80e-3       & $0.06\pm0.04$          & --                      &   This work       &    --             \\ 
\bottomrule  
\end{tabular}
   \begin{tablenotes}[flushleft]
   \small
	 \item\textbf{Note:} 
	      The upper energy levels, frequencies, and Einstein $A$ coefficients of the molecules are taken from \citet{2005A&A...432..369S}. To complete the table, we included line detections reported in previous studies. The references for the line fluxes are: L11 = \citet{2011ApJ...738L...6L}, R09 = \citet{2009ApJ...690..463R}, V11 = \citet{2011ApJ...741L..38V}, W07 = \citet{2007A&A...467..955W}, Y16 = \citet{2016A&A...595A..80Y}, Y17a = \citet{2017A&A...608A.144Y}, Y17b = \citet{2017PhDT........21Y}. For the CN lines, we do not distinguish between the hyper-fine splitting caused by the nuclear spin of nitrogen (described by quantum number F) due to limited spectral resolution. For the CCH lines, because the frequencies of the fine and hyper-fine splitting within the same $N+1{\rightarrow}N$ are very close (considering the four brightest transitions for a given $N_{J, F}$, we have the brightest four lines that are within 0.06\,GHz), we do not distinguish the hyper-fine lines. For the NO and CH, we only list the fine structure and $\Lambda$-doublet because the hyper-fine lines are too close to be separated. For the aforementioned molecules, we list the $E_\mathrm{up}/k_\mathrm{B}$ and $\nu_\mathrm{rest-frame}$ with the highest Einstein $A$ coefficient as a representative (for the CH lines, we list $3/2,\,1^+ {\rightarrow} 1/2,\,0^-$ and $3/2,\,2^- {\rightarrow} 1/2,\,1^+$). The data of C$^{34}$S, HCN-VIB (${v_2=1f}$) and c-C$_3$H$_2$ are taken from CDMS \citep{2016JMoSp.327...95E}. For the HCN and the N$^2$H$^+$ lines, we do not consider their hyper-fine structure lines considering the spectral resolution and the intrinsic linewidths of the galaxies.  
   \end{tablenotes}
   \label{table:line-flux}
\end{table*}
\normalsize

\section{Continuum: Thermal dust and free-free emission}
\label{subsection:ctm}
Due to the exquisite capability of NOEMA to calibrate the continuum over large frequency ranges, we can constrain the continuum emission of the two sources using the full coverage of the continuum flux densities across the rest-frame frequency range of $\sim$\,330--550\,GHz. 

For both galaxies, we assume two components that contribute to the continuum at this frequency range: the Rayleigh-Jeans tail of the modified black body emission from the dust and the free-free (thermal Bremsstrahlung) emission produced by free electrons scattering off ions commonly in HII regions. Our assumption aligns with the global spectral energy distribution (SED) analysis of these two sources \citep{2019ApJ...876...48L, 2017A&A...608A.144Y}, where the contribution from the nonthermal part is negligible around $\sim$\,330--550\,GHz. The full SEDs of the two galaxies align with the typical synchrotron emission in normal galaxies \citep{1992ARA&A..30..575C}, contrasting with radio-loud sources where nonthermal emissions can have significant contributions \citep[e.g.,][]{2019A&A...621A..27F}.

\begin{figure}[htbp]
\;\includegraphics[scale=0.63]{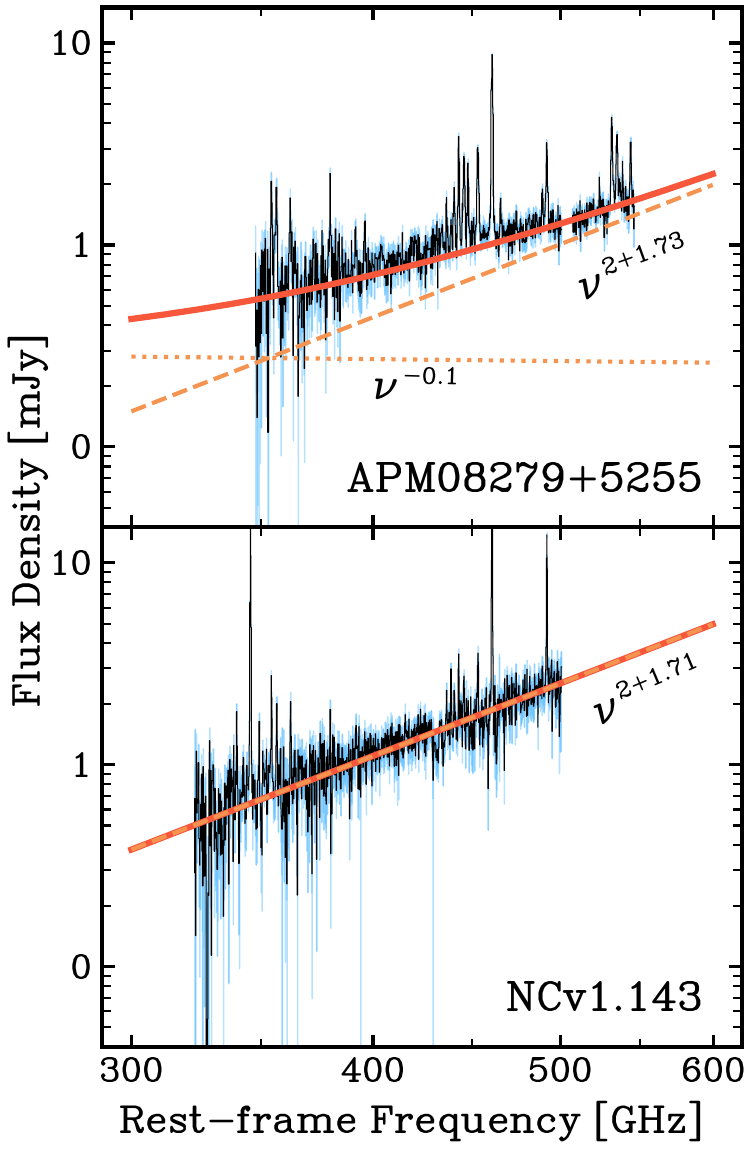}
\vspace{-0.2cm}
\caption
{Fitting of the continuum of APM\,08279+5255 and NCv1.143. Fluxes and errors are in black and blue, respectively. The solid red lines represent the total flux from the model. The dotted and dashed orange lines are the free-free and thermal dust components from the model, respectively. It is clear from the fitting that APM\,08279+5255 has a flatter spectrum, which can be explained by contributions from free-free and/or corona emission from the SMBH, while the continuum of NCv1.143 is dominated by thermal dust.}
\label{fig:cont-only}
\end{figure}

As shown in Fig.\,\ref{fig:cont-only}, to simplify the fitting, we used the Rayleigh-Jeans approximation for the thermal dust emission while adopting a $\alpha=-0.1$ spectral index for the free-free; thus, we have the following expression of the total flux, which consists of two power-law components:
\begin{equation}
    C^{\rm R-J}(\nu_{\rm rest}/\nu_{0})^{(2 + \beta)} + C^{\rm free-free}(\nu_{\rm rest}/\nu_{0})^{-0.1}
\end{equation}
where $C^{\rm R-J}$ and $C^{\rm free-free}$ are the scaling factors of the fluxes that reflect relative contributions from each component, $\nu$ is the rest-frame frequency ($\nu_{0}$\,=\,300\,GHz) and $\beta$ is the dust emissivity index. The fitting results are shown in Fig.\,\ref{fig:cont-only}. The fitted values for $C^{\rm R-J}$ and $C^{\rm free-free}$ are $0.38\pm0.03$\,mJy and $0.00\pm0.02$\,mJy for NCv1.143 and $C^{\rm R-J}$\,=\,$0.15\pm0.01$\,mJy and $C^{\rm free-free}$\,=\,$0.38\pm0.01$\,mJy for \apm. Therefore, for NCv1.143, we find the continuum covered by our line survey is totally dominated by dust emission, with $\beta$\,=\,$1.71\pm0.02$, while for \apm, the free-free contribution is more prominent with a contribution from about 57\% at 350\,GHz to 27\% at 500\,GHz, and the dust continuum contributing the rest 43\% to 73\% with $\beta$\,=\,$1.73\pm0.03$. While the thermal dust flux contribution of 0.27\,mJy is in excellent agreement with \citet{2019ApJ...876...48L} for \apm, our free-free component (0.37\,mJy) is at least three times larger than previous studies of the global SED \citep{2018MNRAS.476.5075S, 2019ApJ...876...48L}. We note that the cosmic microwave background can affect the fitted spectral index in the Rayleigh-Jeans part depending on redshift and dust temperature. Following \citet{2013ApJ...766...13D}, assuming the dust temperatures of $\gtrsim$\,40\,K \citep{2017A&A...608A.144Y, 2019ApJ...876...48L}, we estimate that the value of $\beta$ may be slightly underestimated by about 9\%, while this value can be even smaller in \apm\ due to a higher dust temperature \citep[see also][]{2016RSOS....360025Z}. This will further increase the fraction of the free-free contribution toward the low-frequency end.

This raises the question of what causes such a difference in the ratio between free-free and thermal dust emission. Comparing the two sources, while the dust emission remains similar, the free-free emission from \apm\ is significantly higher. Both sources have high apparent SFRs around $1.2\text{--}1.5\times10^{4}$\,$\mu^{-1}$\,\msun\,yr$^{-1}$. Using the correlation between free-free luminosity and the SFR \citep{2022ApJ...924...76A}, assuming the typical electron temperature of the H{\small II} regions of 10$^{4}$\,K and a thermal fraction of 1 (neglecting synchrotron emission around 300--500\,GHz), we derive a luminosity of free-free emission of about $1.7\times10^{31}$\,erg\,s$^{-1}$\,Hz$^{-1}$ at 350\,GHz, which translates to fluxes of about 0.06\,mJy, about six times lower than the observed value. But this value is in agreement with the free-free contribution from the global SED fitting of \apm\ \citep{2018MNRAS.476.5075S, 2019ApJ...876...48L}. One possibility is that the contribution from the synchrotron emission in \apm\ is significant because of its AGN activities. However, longer wavelength data of \apm\ showed that the synchrotron emission is almost two orders of magnitude smaller than thermal dust and free-free, which rules out this scenario \citep{2018MNRAS.476.5075S, 2019ApJ...876...48L}. Another possibility is an additional contribution from very cold dust ($T$\,<\,15\,K) emission (typically found in nearby star-forming galaxies, \citealt{2018ARA&A..56..673G}) in \apm. However, considering the similarity of the bulk of the dust emission in NCv1.143 and \apm, it will be difficult to explain why such very cold dust is not present in NCv1.143. The third possibility is an additional millimeter flux contribution from the non-self-absorbed part of the synchrotron radiation from the hot corona around the SMBH \citep[e.g.,][]{2008MNRAS.390..847L, 2014PASJ...66L...8I, 2022ApJ...938...87K}. Accordingly, using the 2--10\,keV X-ray luminosity of \apm\ \citep[$\mu L_{\rm X}=3\times10^{46}$\,erg\,s$^{-1}$;][]{2022A&A...662A..98B} and the correlation between millimeter and X-ray corona emission \citep{2022ApJ...938...87K}, we find a corresponding millimeter flux of about 0.05\,mJy at 350\,GHz (assuming the millimeter corona emission peaks around 300\,GHz), with significant uncertainties. Such a millimeter emission from the corona component may explain the observed elevated fluxes of \apm\ at lower frequencies. If this is the case, the value of $\beta$ could be even higher to account for the steeper increase with less contribution from the free-free after including this corona component. In the global SED of \apm\ shown in \citet{2018MNRAS.476.5075S}, it is evident that there is flux excess, which is about 0.1--0.2\,mJy depending on the frequency, that cannot be explained by free-free combining thermal dust around rest frame 100--400\,GHz. The excess value is in broad agreement with a 50\% contribution from the millimeter corona flux inferred from the X-ray luminosity. However, we will need to further combine data at the rest-frame frequency range of 50--300\,GHz, where the corona emission usually peaks, to test this scenario. This is beyond the scope of this work and will be presented in another paper (del Palacio et al. ~in prep.).

\section{A detailed look at the detected molecules and a comparison of 
the two sources}
\label{section:line-atlas-diff}
As displayed in Fig.\,\ref{fig:spec} and Table\,\ref{table:detection-summary}, we have detected a total of 17 species, including two isotopologs of CO and one isotopolog of CS, plus the vibrational line of the HCN. Except for the \ci\ line, all of the species have at least two transitions detected in at least one of the sources. Among these lines, the CH, NO, HCN-VIB (rotational transitions within the vibrationally excited state of HCN\,${v_2=1f}$), 380\,GHz \hto\ maser, and H$_3$O$^+$ lines are the first high-redshift detections in individual sources reported.

\begin{figure*}[htbp]
\centering
\includegraphics[scale=0.224]{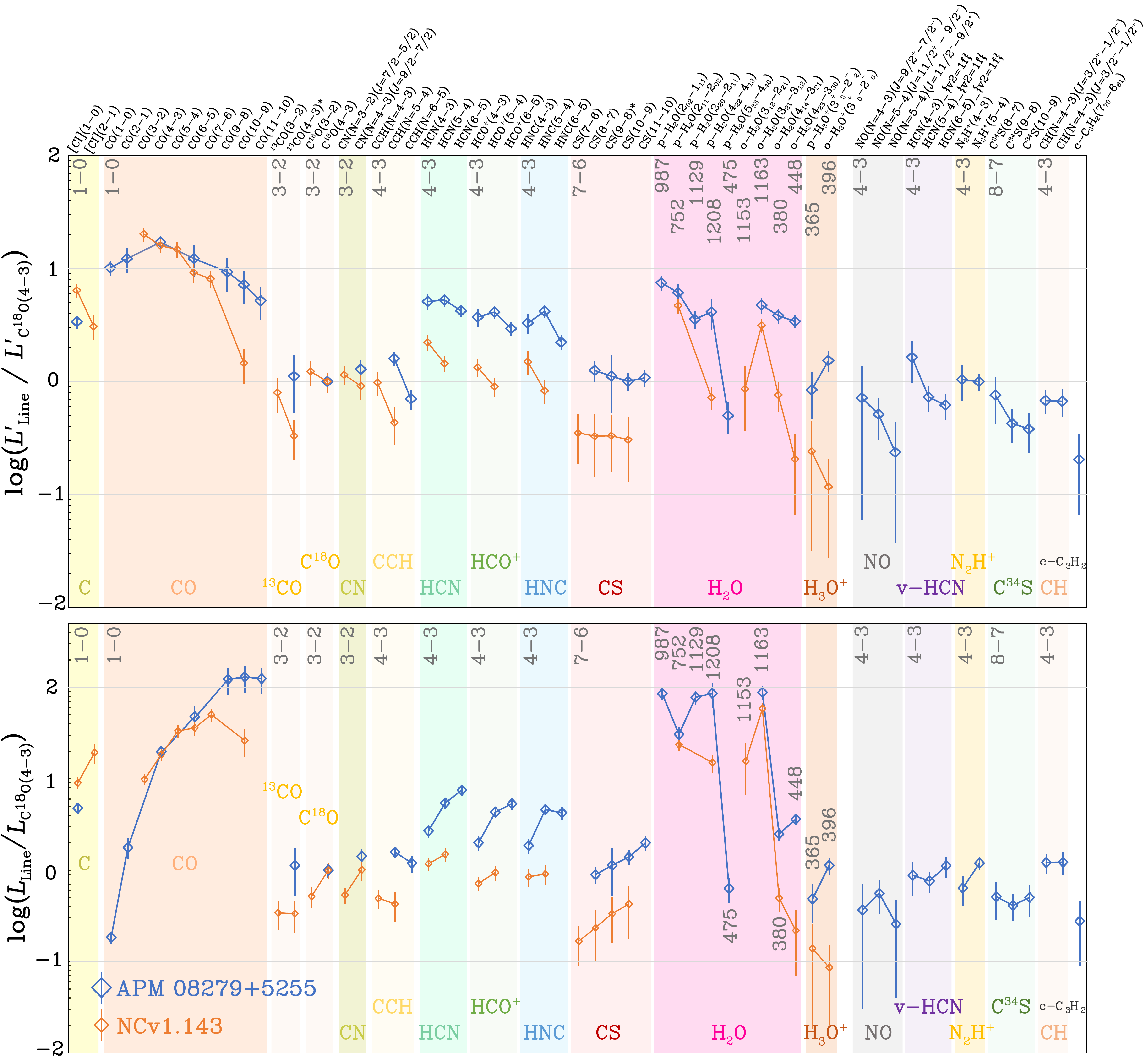}
\vspace{-0.2cm}
\caption
{Line luminosities of \apm\ and NCv1.143 (based on the line fluxes listed in Table\,\ref{table:line-flux}, where we also include several detections from the literature). Error bars are propagated from the measured flux errors. The upper panel shows the line luminosity in K\,km\,s$^{-1}$\,pc$^2$, $L^{\prime}_\mathrm{line}$, normalized by the value of C$^{18}$O(4--3). The lower panel shows the line luminosity in $L_\odot$, $L_\mathrm{line}$, normalized by the C$^{18}$O(4--3) line. The lowest levels of the transitions of each molecule detected are labeled at the top of each panel, except for \hto\ and \httop, where the rest-frame frequencies in GHz are given. Each transition is identified in the upper part of the figure. Compared with NCv1.143, the excitation condition is more extreme in \apm\ as seen from the SLEDs from several molecules.}
\label{fig:flux-compare}
\end{figure*}

Given the existing extensive discussion of the CO SLED in \apm\ and NCv1.143 \citep{2007A&A...467..955W, 2017A&A...608A.144Y}, we do not delve into the discussions of the CO lines in this work. Here, we only highlight that the CO SLED of \apm\ shows a more elevated shape toward the high-$J$ CO lines in Fig.\,\ref{fig:flux-compare}, indicating an overall more excited molecular gas condition \citep{2015ApJ...801...72R}.

From the observed integrated fluxes in Table\,\ref{table:detection-summary}, we derive the apparent line luminosities, $\mu L_\mathrm{line}$ (in units of \lsun) and $\mu L^{\prime}_\mathrm{line}$ (in units of \kkmspc) following \citet{1992ApJ...387L..55S} using \mbox{$L_\mathrm{line}=1.04 \times 10 ^{-3} I_\mathrm{line} \nu_\mathrm{rest} (1+z)^{-1}D_\mathrm{L}^2$} and \mbox{$L^{\prime}_\mathrm{line}=3.25 \times 10 ^{7} I_\mathrm{line} \nu_\mathrm{obs}^{-2} (1+z)^{-3}D_\mathrm{L}^2$}, where $I_\mathrm{line}$, $\nu$ and  $D_\mathrm{L}$ are in the units of Jy\,\kms, GHz and Mpc, respectively. We note here that because the $^{13}$CO(4--3) and CS(9--8) lines are blended due to their close frequencies, we extrapolate the integrated flux of CS(9--8) by combining the data of CS(7--6), CS(8--7) and CS(10--9), and we subtracted this flux from the blended line flux for obtaining the integrated flux of $^{13}$CO(4--3). Considering the smooth CS SLED in both sources, we do not expect a strong overestimation of the CS(9--8) line flux, which can cause an underestimation of the integrated flux of the $^{13}$CO(4--3) line in \apm. 

Assuming that the magnification factors do not vary significantly among the emissions (Sect.\,\ref{section:lensing}), we remove the magnification factor by normalizing the luminosities using the C$^{18}$O(4--3) line (given that the C$^{18}$O line is likely optically thin as we argue in Appendix\,\ref{app:taus} and the emitting size of the C$^{18}$O is very similar to most of the other dense gas tracers; see, e.g., \citealt{2021A&A...656A..46M}). In Fig.\,\ref{fig:flux-compare}, we show the comparison between the line luminosities normalized by C$^{18}$O(4--3) ($L_\mathrm{line}$ and $L^{\prime}_\mathrm{line}$) of \apm\ and NCv1.143. We caution that the ratio in Fig.\,\ref{fig:flux-compare} depends on the normalization factor, which, however, does not affect the overall shapes of the SLEDs.

As shown in Fig.\,\ref{fig:flux-compare}, the luminosity of CO(4--3) and C$^{18}$O(4--3) are similar for both sources, despite \apm\ having an overall elevated CO SLED. \apm\ has a significantly low luminosity ratio of \cil10/C$^{18}$O(4--3) compared to NCv1.143, while the $^{13}$\co43/C$^{18}$O(4--3) ratio is higher than NCv1.143. We also see a similarly high ratio of the luminosities of other dense gas tracers, such as CCH, HCN, HCO$^+$, HNC, and CS over the value of C$^{18}$O(4--3). If there is no deficit of C$^{18}$O in \apm, then the lines of dense gas tracers are at least a factor of 2 brighter in \apm, indicating that the gas conditions in the quasar host are more extreme. This is consistent with molecular gas conditions derived from the CO SLEDs of \apm, where the thermal pressure reaches $2\times10^{6}$\,K\,cm$^{-3}$ \citep{2007A&A...467..955W}. While in NCv1.143, the thermal pressure is $\lesssim$\,$1\times10^{6}$\,K\,cm$^{-3}$ \citep{2017A&A...608A.144Y}. This is also in line with our non-LTE excitation analysis of the dense gas tracers presented in Sect.\,\ref{sect:lvg}.

\begin{figure*}[hbtp]
\centering
\includegraphics[scale=0.32]{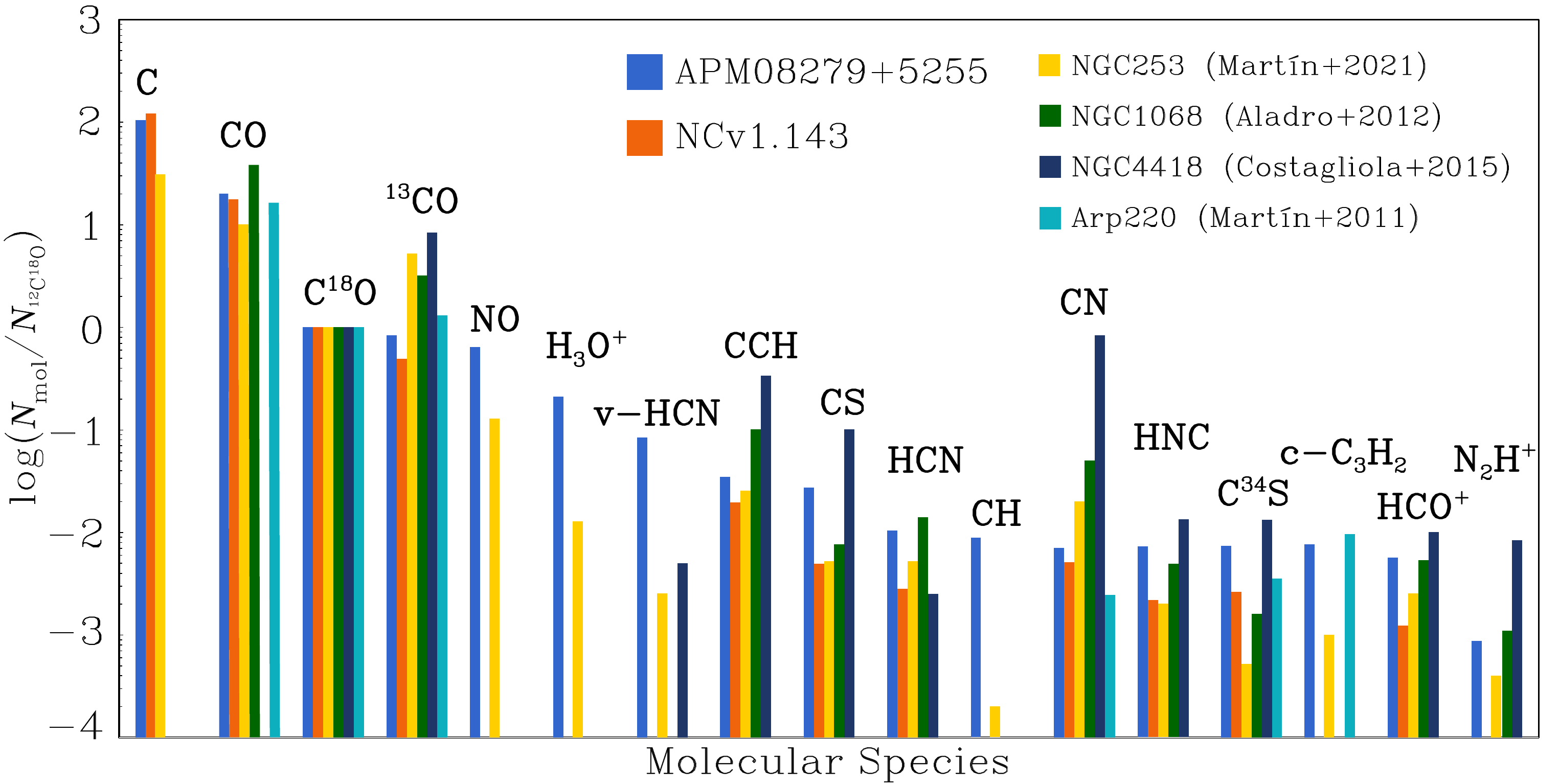}
\caption
{
Column density ratios to C$^{18}$O in \apm\ and NCv1.143 derived by \texttt{MADCUBA} (Table\,\ref{table:madcuba-fit}), which can be treated as the relative abundances. The plot is ordered, from left to right, according to the relative abundance in \apm\ from high to low. Some local sources from the literature are also shown: NGC\,253 \citep[CMZ;][]{2021A&A...656A..46M, 2014ApJ...788..147R}, NGC\,1068 \citep[inner $\sim$\,2\,kpc;][]{2015A&A...579A.101A}, NGC\,4418 \citep{2015A&A...582A..91C}, and Arp\,220 \citep{2011A&A...527A..36M}. 
} 
\label{fig:abundance-compare}
\end{figure*}

Before applying any gas excitation and radiative transfer model to the observed SLEDs, it is useful to inspect their shapes, where different heating mechanisms can leave different imprints \citep{2015ApJ...801...72R}. In Fig.\,\ref{fig:flux-compare}, the CO SLEDs reveal a clear difference at the high energy levels, where \apm\ shows almost constant $L^{\prime}_\mathrm{line}$ (thermalized), while the luminosity drops by a large factor for higher energy level CO lines in NCv1.143. This suggests that the gas conditions are more extreme in \apm\ than in NCv1.143. For the dense gas tracer lines of HCN, HCO$^+$, HNC, and CS, \apm\ has slightly more elevated SLEDs than NCv1.143, showing a more extreme heating mechanism in the quasar host. The most striking and complex difference is observed in the \hto\ lines, where we find slightly elevated \htot211202\ and \htot321312\ luminosity in NCv1.143, while the 448\,GHz \htot423330\ is about nine times brighter in \apm. Such a difference suggests that the $J_\mathrm{up}$\,=\,2 and 3 \hto\ lines might arise from very different conditions compared to the region that the \htot423330\ traces. The latter might be tightly related to AGN-powered radiative pumping. Thus, this line is much more excited in \apm\ than in NCv1.143. This is further explained by the detailed analysis of \hto\ excitation in Sect.\,\ref{sect:h2o-lines}. Besides \hto, the \httop\ line in \apm\ is also much enhanced. Additionally, we identify the first high-redshift detection of the HCN-VIB line -- three HCN-VIB lines with $J_\mathrm{up}$ ranging from 4 to 6 are detected in \apm, while they remain undetected in NCv1.143.

\subsection{Comparison of the LTE-derived abundances}
\label{section:abundance-LTE}

To understand the astrochemical differences between the two sources, we compare their relative abundances derived using \texttt{MADCUBA}, together with some local prototypical galaxies: the central molecular zone (CMZ) region of the starburst galaxy NGC\,253 \citep{2021A&A...656A..46M, 2014ApJ...788..147R}, the central $\sim$\,2\,kpc nuclear region of the Seyfert 2 galaxy NGC\,1068 \citep{2015A&A...579A.101A}, globally integrated values of the luminous infrared galaxy with a deeply buried nuclei NGC\,4418 \citep{2015A&A...582A..91C} and the archetypal ultra-luminous infrared galaxy Arp\,220 \citep{2011A&A...527A..36M}. We also note that all the line surveys mentioned above are performed at scales $\sim$\,1--2\,kpc that is comparable with the size of our galaxies, and this can potentially minimize the chemical variations due to differences of spatial scales \citep{2022A&A...667A.131B}.

\begin{figure*}[htbp]
\centering
\includegraphics[scale=0.54]{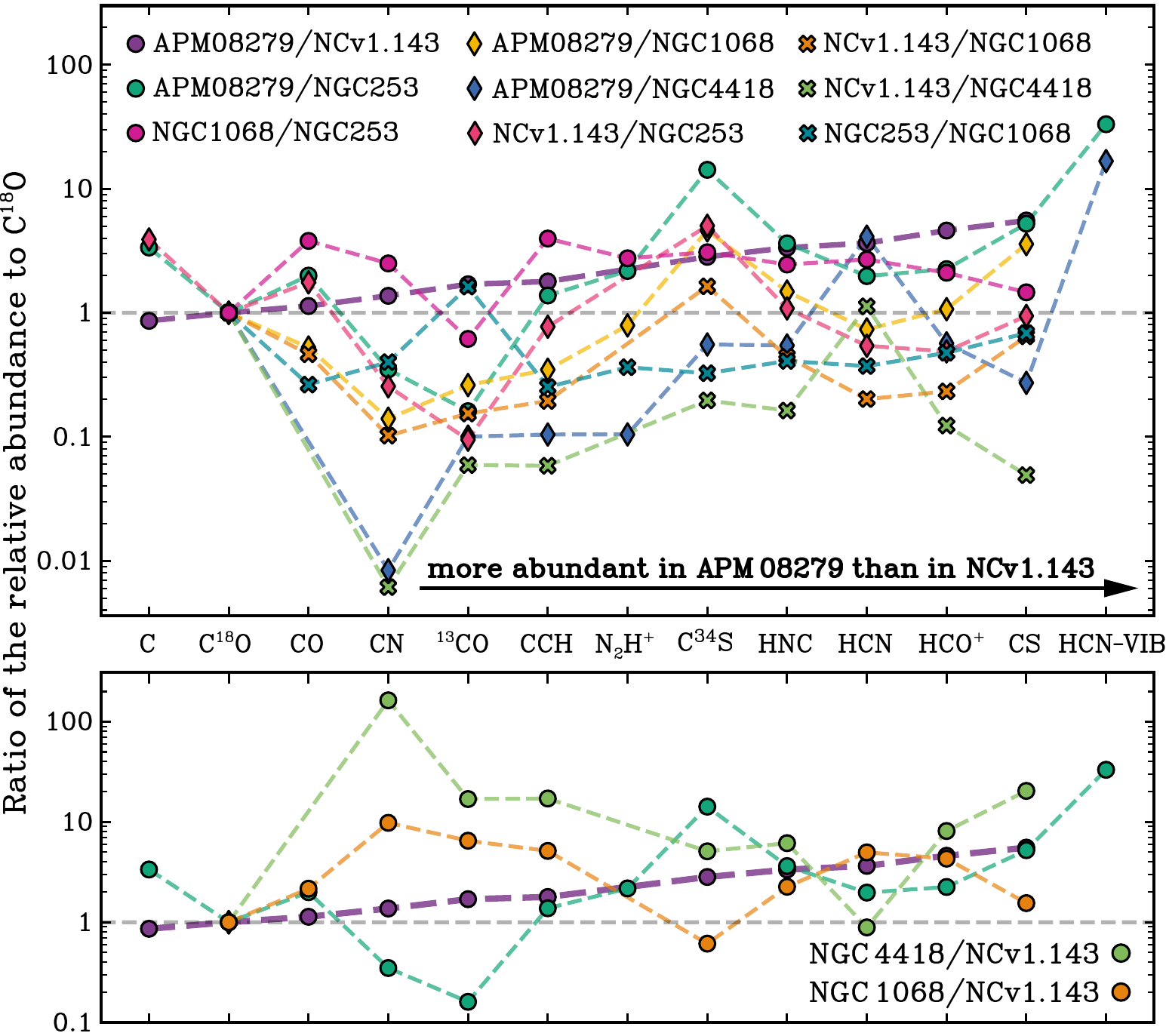}
\caption{
Comparison of the abundance ratios of \apm\ (APM\,08279 for short), NCv1.143, and three archetype local galaxies -- the CMZ of NGC\,253 \citep{2021A&A...656A..46M, 2014ApJ...788..147R}, $\sim$\,2\,kpc nuclear region of NGC\,1068 \citep{2015A&A...579A.101A}, and NGC\,4418 \citep{2015A&A...582A..91C} -- for the species detected in both \apm\ and NCv1.143. The abundance ratios of the galaxies are computed by dividing the column densities of each molecule by that of C$^{18}$O, and these ratios are then compared among the galaxies. The horizontal axis is ordered by the increasing value of the ratio APM\,08279/NCv1.143 (in the case of C$^{34}$S and HCN-VIB,  where the value of NCv1.143 is missing, we use the ratio of APM\,08279/NGC\,253 for ordering). Circles are AGN-dominated (APM\,08279, NGC\,1068, and NGC\,4418) over starburst-dominated (NCv1.143 and NGC\,253) ratios, diamonds are comparisons between similar types, and crosses are starburst-dominated over AGN-dominated ratios. The upper panel shows the ratio of the two high-redshift galaxies over local galaxies, while the lower panel shows the ratio of AGN-dominated galaxies over starburst-dominated galaxies only.
}
\label{fig:abundances-compare-2}
\end{figure*}

Given that we do not have a direct estimate of the absolute abundance of each molecule, we take the column densities listed in Table\,\ref{table:madcuba-fit}, normalize all the values with that of the C$^{18}$O (as a proxy of total H$_2$, it is optically thin as shown in Appendix\,\ref{app:taus}; besides, the size effect mentioned in Sect.\,\ref{subsection:MADCUBA} is minimum because the size of the C$^{18}$O emission is close to other dense gas tracers as found in, e.g., \citealt{2021A&A...656A..46M}), as a proxy of the relative abundance ratio (Fig.\,\ref{fig:abundance-compare}). Such a ratio also removes the uncertainty of the unknown source size (which includes the lensing magnification) if we assume that all the molecules are well-mixed. Nevertheless, we acknowledge the uncertainties as the molecular abundances can vary by a large factor within a galaxy \citep[e.g.,][]{2014A&A...567A.125G, 2022A&A...666A.102H}. However, given that our observation is spatially unresolved, such an assumption can be used as a first-order approximation until high-spatial-resolution data are available. 

The comparison of the relative abundance (with respect to C$^{18}$O) of all the molecules between the two sources, plus four nearby galaxies, is shown in Fig.\,\ref{fig:abundance-compare}. The molecules are ranked, from left to right, with the decreasing order of the relative molecular abundances of \apm. Except for the neutral carbon, almost all the molecules show higher relative abundances in the quasar \apm\ than in NCv1.143, where the difference becomes most significant in the dense gas tracers like HCN, HCO$^+$, HNC, CS, and C$^{34}$S. Notably, the CN abundance of NGC\,4418 reported by \citet{2015A&A...582A..91C} is exceptionally high compared with all other galaxies. We also note here that the relative abundances are quite uncertain, given many sources of the uncertainties involved in the LTE analysis. 

To better understand the differences in the relative abundances between the two sources and compare these abundances with some local sources, we present the ratios of relative abundances for \apm\ and NCv1.143 in Fig.\,\ref{fig:abundances-compare-2}. In addition to the ratio between \apm\ and NCv1.143, we include the ratios for our high-redshift targets when compared with -- the CMZ of NGC\,253, which is driven by pure starburst activity (\citealt{2021A&A...656A..46M}); NGC\,4418, characterized by AGN activity (\citealt{2015A&A...582A..91C}); and NGC\,1068, the inner $\sim$\,2\,kpc nuclear region, where the abundances are expected to be dominated by AGN (\citealt{2015A&A...579A.101A}). In Fig.\,\ref{fig:abundances-compare-2} we group the ratios between each two of these galaxies into three categories: (1) AGN/starburst (solid circles), which are ratios of \apm/NCv1.143, \apm/NGC\,253, and NGC\,1068/NCv1.143; 
(2) AGN/AGN and starburst/starburst (solid diamonds), which are ratios of \apm/NGC\,1068, \apm/NGC\,4418, and NCv1.143/NGC\,253; and 
(3) starburst/AGN (crosses), which are NCv1.143/NGC\,1068, NCv1.143/NGC\,4418, and NGC\,253/NGC\,1068.

From left to right in Fig.\,\ref{fig:abundances-compare-2}, we rank the molecules in the increasing order of the relative abundance ratio of \apm/NCv1.143. The figure highlights the fact that the dense gas tracers like HNC, HCN, HCO$^+$, and CS are considerably more abundant in \apm\ than in NCv1.143, while the difference is less prominent in CN, $^{13}$CO and CCH. It is worth noting that in high-temperature environments, where AGN can heat up the dust and gas, providing adequate conditions for high-temperature chemistry, CN can be converted into HCN effectively \citep{2010ApJ...721.1570H}, leading to a decrease in CN and an increase in HCN. This is fully consistent with the fact that the relative abundance ratio of HCN/CN is significantly higher in \apm\ than in NCv1.143. The relative abundance ratio \apm/NGC\,253 shows a similar trend as  \apm/NCv1.143 -- abundances of the dense gas tracers are enhanced, with the most extreme enhancement seen in C$^{34}$S. However, when we examine the ratios of \apm/NGC\,1068 and \apm/NGC\,4418 (AGN/AGN ratios), we do not see significantly enhanced relative abundances of the dense gas. The values of the ratios of \apm/NGC\,1068 and \apm/NGC\,4418 (AGN/AGN) are smaller than \apm/NCv1.143 and \apm/NGC\,253 (AGN/starburst), and they are close to or below unity. As opposed to the ratios of \apm/NCv1.143 and \apm/NGC\,253, we do not see systematic trends of these AGN/AGN ratios.

Conversely, when examining the abundance of NCv1.143 compared to local AGN-dominated sources (starburst/AGN), values lower than unity are consistently observed, which aligns with the higher values noted in the ratios of \apm\ over NCv1.143 and NGC\,253 (AGN/starburst). This is not surprising, as the ratios of NCv1.143/NGC\,1068 and NCv1.143/NGC\,4418 are the inverse of the AGN/starburst ratios. When comparing NCv1.143 with NGC\,253, we see a similar trend as seen in the AGN/AGN ratios (\apm/NGC\,1068 and \apm/NGC\,4418), where the values are found distributed around unity, and they are also often higher than the starburst/AGN ratios.

While the absolute abundance ratios for all the molecules in both our targets remain uncertain, distinct trends can be found in Fig.\,\ref{fig:abundances-compare-2}. First, when examining each species, the AGN/starburst ratios (\apm/NCv1.143 and \apm/NGC\,253) generally exhibit higher values that are mostly above unity, whereas the starburst/AGN ratios (NCv1.143/NGC\,4418 and NCv1.143/NGC\,1068) often have lower values that are mostly below unity; the similar pairs, AGN/AGN or starburst/starburst (\apm/NGC\,4418, \apm/NGC\,1068, and NCv1.143/NGC\,253) ratios,  fall in between. This indicates that the molecular abundances in \apm\ behave like those found in the AGN-dominated sources, while the abundances in NCv1.143 resemble the starburst chemistry. Second, the dense gas tracers such as HCN, HCO$^+$, HNC, CS, N$_2$H$^+$, and C$^{34}$S are generally more abundant among the molecules studied in AGN-dominated sources compared to other molecules as shown in the lower panel of Fig.\,\ref{fig:abundances-compare-2}. Similar trends of an enhanced abundance of these dense gas tracers relative to C$^{18}$O have also been found in the AGN-dominated CMZ in comparison to the starburst regions in NGC\,1068 \citep{2023ApJ...955...27N}.

The parallel between \apm\ and local AGN prototypes, coupled with the similarity in behavior of NCv1.143 to the starburst CMZ of NGC\,253, suggests that the ISM chemistry in \apm\ is influenced by the central AGN, through either the high-temperature chemistry proposed by \citet{2010ApJ...721.1570H} or X-ray irradiate chemistry \citep{2007A&A...461..793M}. Correspondingly, the chemistry of the ISM in NCv1.143 likely bears a resemblance to the CMZ of NGC\,253, and we do not see any significant evidence of any influence from a buried AGN in this source from the molecular abundance.

\subsection{The \ci\ line}
\label{section:carbon} 

Both galaxies are detected in the neutral carbon \ci\ $^3$P$_1$--$^3$P$_0$ fine structure line at 492\,GHz (\cil10\ hereafter), with luminosities $\mu L'_{\rm [C{\small I}]}$ of 3.55$\times$$10^{10}$\,\kkmspc\ and 10.1$\times$$10^{10}$\,\kkmspc\ for \apm\ and NCv1.143, respectively. The luminosity of the \ci\ line is weaker in \apm\ compared with NCv1.143, which can be explained by the X-ray-dominated region (XDR) model when the column density of gas $N_\mathrm{H}$ is around 10$^{24}$\,cm$^{-2}$ \citep{2005A&A...436..397M}. 

The \ci\ emission lines allow us to derive the properties of the atomic carbon gas in these systems and compare the H$_2$ masses derived independently from the CO lines. We note that the reported values here are not corrected for magnification. We estimated the neutral carbon masses using Eq.~1 in \cite{2005A&A...429L..25W}, assuming a \ci\ excitation temperature equal to $T_{\rm ex}$\,=\,30\,K, which is close to the value in \cite{2011ApJ...730...18W}, $<T_{\rm ex}>$\,=\,$29.1 \pm 6.3$\,K, and the mean temperature of $<T_{\rm ex}>$\,=\,$25.6 \pm 1.0$\,K found by \cite{2020ApJ...890...24V}: 
\begin{equation}
M_{\rm [C{\small I}]} = 5.706 \times 10^{-4} \, Q(T_{\rm ex}) \frac{1}{3} e^{23.6/T_{\rm ex}} \, L^{\prime}_{\rm [C{\small I}](1-0)} \textrm{,}
\end{equation}
\noindent
where $Q(T_{\rm ex}) = 1 + 3e^{-23.6 \, K/T_{\rm ex}} + 5e^{-62.5 \, K/T_{\rm ex}}$ is the partition function of \ci\, and the result is expressed in units of \msun. The derived masses are $\mu M_{\rm [C{\small I}]} = (4.40\pm0.17) \times 10^{7}$\,\msun\ and  $(1.25\pm0.05) \times 10^{8}$\,\msun\ for \apm\ and NCv1.143, respectively. 

In the case of NCv1.143, \citet{2017A&A...608A.144Y} reported the detection of the \cil21\ emission line with a line flux of $4.4\pm0.9$\,Jy\,\kms. Assuming LTE and, under the condition that both \ci\ lines are optically thin, the excitation temperature equals the kinetic temperature:
\begin{equation}
\label{eq:CI-Texc}
{T_{\rm ex}}/{\rm K} = 38.8/{\rm ln}(2.11/[L'_{\rm [C{\small I}](2\text{--}1)}/L'_{\rm [C{\small I}](1\text{--}0)}]).
\end{equation}  

Using the measured flux of the \cil21\ emission line yielding $\mu L'_{\rm [C{\small I}](2\text{--}1)} = (4.9\pm1.0)\times \, 10^{10}$\,\kkmspc\  \citep{2017A&A...608A.144Y}, we derive an excitation temperature of 26.3~K for NCv1.143, close to the above-adopted value and comparable to the excitation temperatures found in previous studies of high-redshift galaxies \citep[e.g.,][]{2019A&A...624A..23N,2020ApJ...890...24V}. As pointed out in Sect.\,\ref{subsection:MADCUBA}, this value of $T_{\rm ex}$ is likely a lower limit of $T_{\rm kin}$. This is consistent with the $T_{\rm kin}$ of 20--63\,K derived from the large velocity gradient (LVG) model of the CO SLED, while the \ci\ excitation temperature is lower than the dust temperature $T_{\rm dust}$\,=\,40\,K in NCv1.143 \citep{2017A&A...608A.144Y}. This effect has been pointed out by \citet{2022MNRAS.510..725P} via a sample of 106 galaxies, that the excitation condition of the \ci\ lines are strongly sub-thermal.

Adopting an atomic carbon abundance of $8.4 \times 10^{-5}$ \citep{2011ApJ...730...18W} and an excitation temperature of 30~K, we derive total molecular masses of $\mu M_\textrm{mol} = (8.7\pm0.3)\times 10^{10}$\,\msun\ and $(2.5\pm0.1)\times 10^{11}$\,\msun\ for \apm\ and NCv1.143, respectively. In the case of NCv1.143, the value is significantly smaller than the apparent molecular gas mass $(5.2\pm1.1)\times 10^{11}$\,\msun\ derived by \citet{2017A&A...608A.144Y} and for \apm\ the molecular gas mass estimated by \citet{2009ApJ...690..463R} of $5.3\times 10^{11}$\,\msun\ is higher by a factor of $\sim 2$. 

Adopting the average $L^\prime_{\rm [C{\small I}]}/M_\textrm{mol}$ conversion factor by \citet{2022MNRAS.517..962D}, $\alpha_{\rm [C{\small I}]} = 17.0$\,\msun\,(\kkmspc)$^{-1}$, the molecular gas masses would be $\mu M_\textrm{mol}$\,=\,$(6.03\pm0.24) \times 10^{11}$\msun\ and $(8.3\pm1.0) \times 10^{11}$\msun\ for \apm\ and NCv1.143, respectively, which are more comparable with the CO-estimated total molecular gas masses \citep{2007A&A...467..955W, 2017A&A...608A.144Y}.

\begin{figure*}[bhpt]
\centering
\includegraphics[scale=0.58]{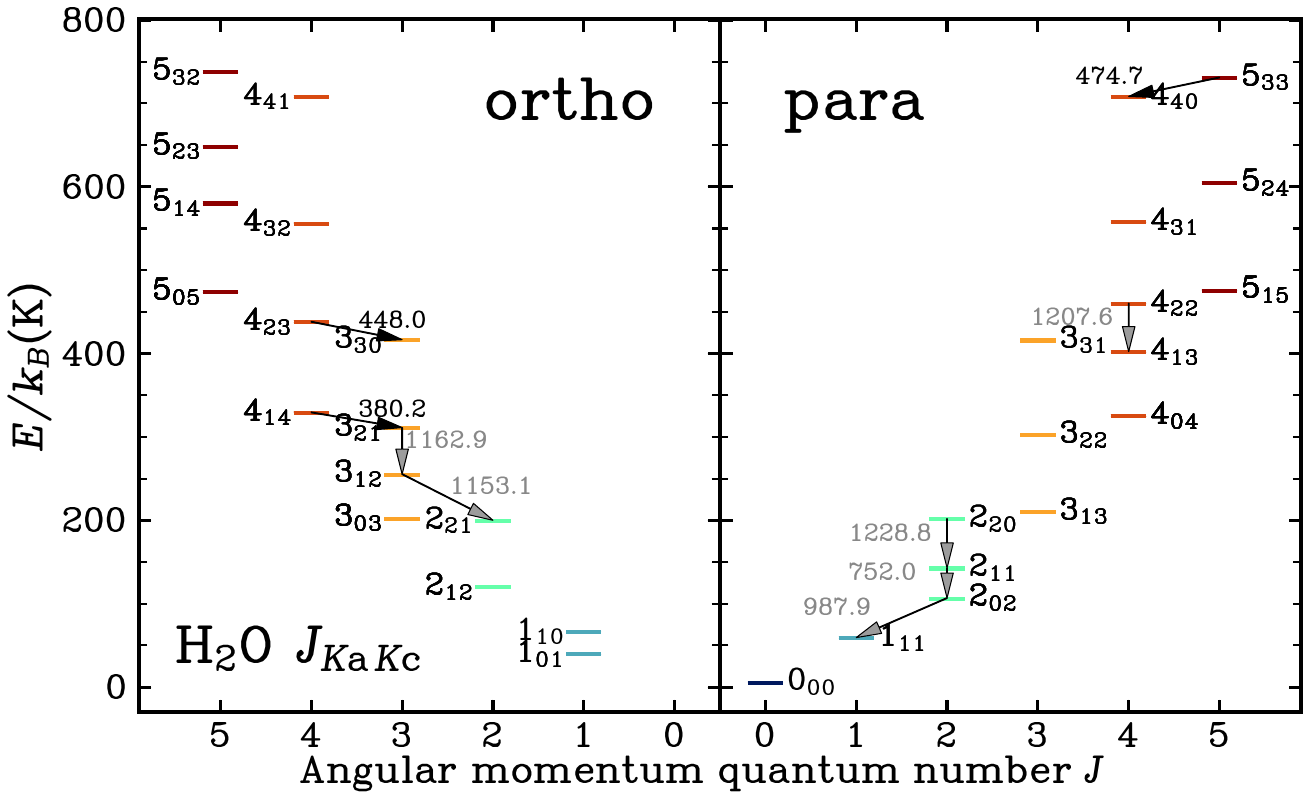}
\caption
{
Energy diagram of an \hto\ molecule. Black lines indicate the \hto\ lines detected from our line survey, while gray lines indicate the observed water transitions from other works (see Table\,\ref{table:line-flux} for references), with their frequencies indicated (in GHz). The \hto\ data are taken from the JPL Molecular Spectroscopy website  \url{https://spec.jpl.nasa.gov}. 
}
 \label{fig:h2o-diag}
 \end{figure*}

\subsection{Submillimeter H\textsubscript{2}O lines}
\label{sect:h2o-lines}

Thanks to the observations of extragalactic \hto\ lines done by {\it Herschel} \citep{2010A&A...518L...1P} and ground-based telescopes \citep[e.g.,][]{2011A&A...530L...3O, 2013A&A...551A.115O, 2013ApJ...771L..24Y, 2016A&A...595A..80Y}, it is now established that \hto\ is one of the most important interstellar molecules after CO. It is because the submillimeter \hto\ lines are bright, capable of several key diagnostics of the ISM, and provide critical complementary information to the CO emission lines \citep[e.g.,][]{2010A&A...518L..43G, 2011ApJ...741L..38V, 2019ApJ...880...92J, 2020A&A...634L...3Y,2021A&A...646A.178S, 2022A&A...667A...9P,2023A&A...673A.157D}. This is due to the combination of the high abundance and large dipole of \hto, as well as its rich energy-level structure and rotational excitation processes (Fig.\,\ref{fig:h2o-diag}), including infrared pumping in dusty starburst regions \citep{2022A&A...666L...3G}, and collisional excitation where the heating source is from post-shock gas and/or shock-front radiation heating in young stellar objects \citep{2021A&A...648A..24V, 2023A&A...674A..95D}, and the warm dense gas regions powered by warm dust heated by massive stars and/or AGN \citep{2017ApJ...846....5L}. \hto\ has a central role in the oxygen chemical networks in the ISM, where \hto\ chemistry can be dominated by grain desorption, ionic reactions in typical conditions, and neutral-neutral reactions in high-temperature conditions \citep[see the review of][and references therein]{2013ChRv..113.9043V}.

\begin{figure}[htbp]
\centering
\includegraphics[scale=0.32]{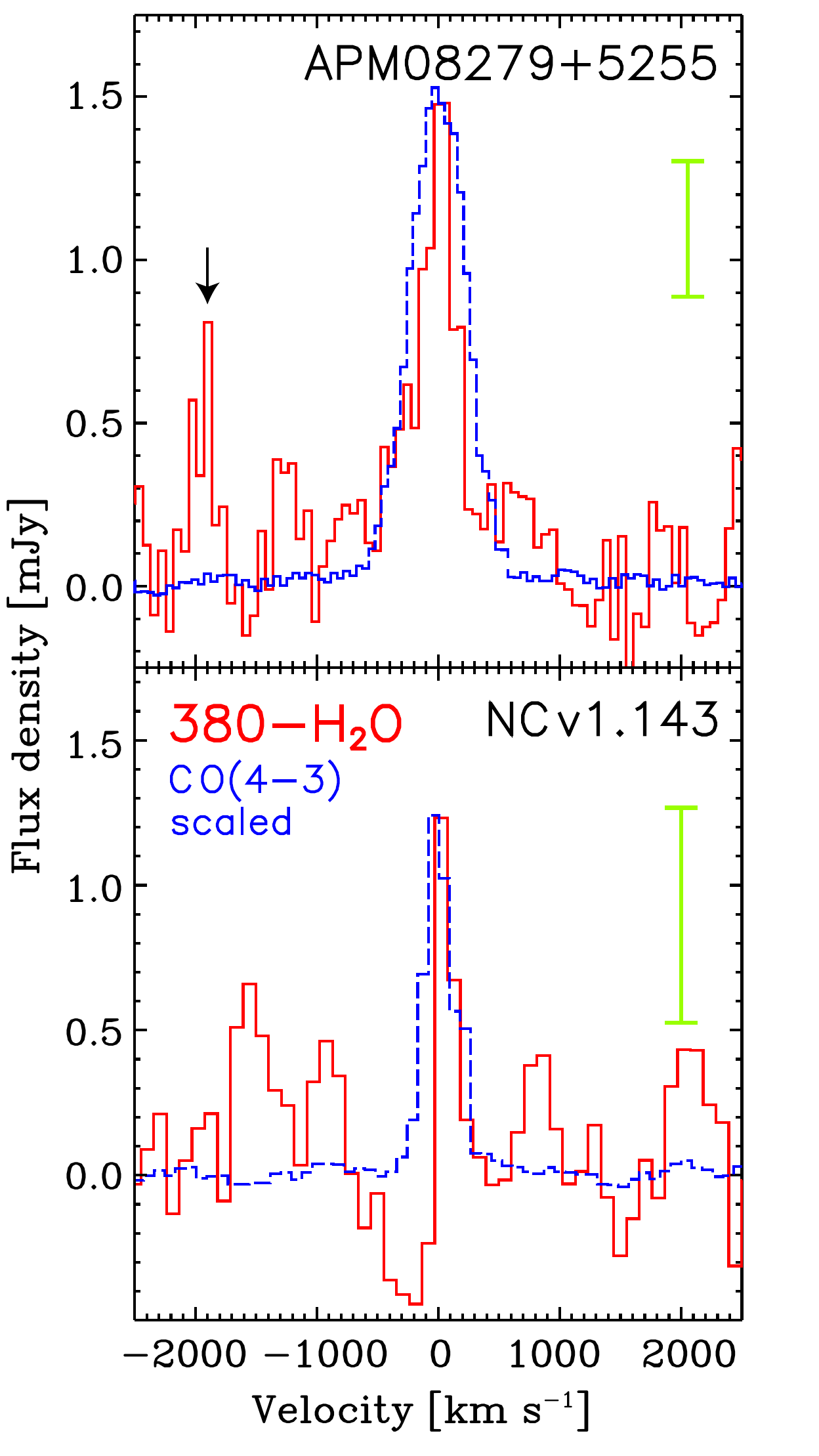}
\caption
{
Spectra of the 380\,GHz \htot414321\ line in \apm\ and NCv1.143 in red (the green error bar indicates the averaged noise level around the spectra per bin), overlaid with the scaled-down profile of the \co43\ line. The 380\,GHz \hto\ line appears narrower by a factor of a few compared with \co43, indicating that it is likely associated with maser origins. Black arrows indicate a $\sim$\,2\,$\sigma$ satellite line detection, possibly from the \hto\ maser, which could originate from accretion disks around the central SMBH.
}
\label{fig:380-h2o}
\end{figure}

For \apm\ and NCv1.143, we explored their \hto\ spectra at rest-frame frequencies below 550\,GHz, where there are only three relatively strong \hto\ lines available -- \htot414321, \htot423330\ and \htot533440: all three are detected in \apm\ and two of them, \htot414312 and \htot423330, in NCv1.143 (Tables\,\ref{table:detection-summary} and \ref{table:line-flux}). This number may appear marginal compared to the typical strong submillimeter \hto\ lines that have been observed at rest frequencies between 557 and 1410\,GHz in local starburst galaxies such as Mrk\,231 and Arp\,220 \citep[e.g.,][]{2010A&A...518L...1P, 2011ApJ...743..159H, 2013ApJ...771L..24Y}. However, these water lines with $E_\mathrm{up}/k_\mathrm{B}$\,$\sim$\,320--730\,K in \apm\ and NCv1.143 can provide great insights into the ISM conditions, as demonstrated in the following subsections.

\subsubsection{The 380\,GHz H$_2$O(4$_{14}-$3$_{21}$) maser line}

Extragalactic water masers are powerful tools for measuring masses of SMBHs and can be used to derive angular distances to the host galaxies with high precision, thus constraining critical cosmological parameters such as the Hubble constant \citep{1999Natur.400..539H, 2005ARA&A..43..625L,2016ApJ...817..128G,2020ApJ...891L...1P}. Despite many searches for \hto\ megamasers at high redshifts, to date, only one has been reported. The 22\,GHz \hto\ megamaser was identified in the $z$\,=\,2.64 lensed quasar MG\,J0414+0534 \citep{2008Natur.456..927I}. The \htot414321\ line has also been detected in the same source \citep{2019PASJ...71...57K, 2020MNRAS.493.5290S}. However, due to the complex broad line profile, it is not definitively confirmed whether this is a maser line or not.  

This ortho-\htot414321\ line at 380.197\,GHz (380\,GHz hereafter), with its metastable backbone level $J_{K_a, K_c}$\,=\,$4_{14}$ ($E_\mathrm{up}$\,=\,324\,K), is one of the most promising \hto\ maser lines predicted by radiative transfer models \citep[e.g.,][]{1991ApJ...368..215N, 2016MNRAS.456..374G}. The detection of the 380\,GHz \hto\ in \apm\ and NCv1.143 (with signal/noise ratio, S/N\,$\sim$\,11 and 6, respectively -- see Table\,\ref{table:line-flux}) is one of the most important results of our NOEMA line survey. In both sources, unlike MG\,J0414+0534 \citep{2019PASJ...71...57K}, the profiles of this line have about half the width of the CO and all the other emission lines, clearly indicating a maser emission line (Fig.\,\ref{fig:380-h2o}). This is the highest-redshift maser detection to date. We note that the 380\,GHz \hto\ is the only submillimeter \hto\ maser detected in both sources in the rest-frame frequency range of 330--550\,GHz. Moreover, the 380\,GHz \hto\ maser is at least four times stronger than its 22\,GHz \htot616523\ counterpart in \apm\ \citep[undetected by][]{2006MNRAS.370..495I}. The flux is significantly bright when juxtaposed with other possible submillimeter masers (at 355, 380, 437, 439, and 471\,GHz; see \citealt{2016MNRAS.456..374G}, with the exception of the 475\,GHz line as described later in this section), as well as the undetected 22\,GHz canonical \hto\ maser line. It is thus one of the best extragalactic \hto\ masers found so far and should be systematically searched for in high-redshift strongly lensed submillimeter galaxies.

Besides the narrow feature at the central velocity, we also identify tentative ($\sim$\,2.0-$\sigma$) detections of narrow emission features close to the 380\,GHz maser line in \apm\ (Fig.\,\ref{fig:380-h2o}). These narrow high-velocity satellite lines of the 380\,GHz maser are up to about $-2000$\,km\,s$^{-1}$ away from the central velocity, and they are likely originated from maser spots on an accretion disk surrounding the SMBH, similar to what has been observed in local galaxies \citep{2023ApJ...948..134P}. If this is true, their velocities are more than twice as large compared to the highest velocity maser spots of the 22\,GHz \hto\ line from a local survey of the \hto\ megamasers disks \citep{2011ApJ...727...20K}, indicating that the accretion disk, in which the 380\,GHz \hto\ masers are located, is very close to the central AGN or that the SMBH of \apm\ is much more massive than the local sources from \citet{2011ApJ...727...20K}. Taking scaling relations from local megamaser disks \citep[e.g.,][]{2017ApJ...834...52G}, such a high velocity indicates a mass of $\sim$\,10$^{10}$\,$M_\odot$ of the SMBH in APM\,08279+5255, which is consistent with estimates using other methods \citep[e.g.,][]{2013A&A...557A..91T}. However, we caution that the high-velocity satellite lines here are very tentative detections, and thus we have been carrying out ongoing follow-up observations with the Green Bank Telescope (GBT) to confirm if these features are real. A detailed analysis of the maser line, together with the follow-up GBT data, will be presented in a separate paper.

\subsubsection{The 448\,GHz \htot423330}

The ortho-\htot423330 line at 448.001\,GHz is predominantly excited through the absorption of 79 and 132\,$\mu$m photons in the transitions $4_{23}$\,$\leftarrow$\,$3_{12}$ and $4_{23}$\,$\leftarrow$\,$4_{14}$ (Fig.\,\ref{fig:h2o-diag}). The influence of collisional excitation on the 448\,GHz \hto\ line is generally deemed negligible due to the high energy levels involved ($E_\mathrm{up}/k_\mathrm{B}$\,>\,400\,K) and the high critical densities of these levels. The $4_{23}$ level is thus likely mostly radiatively populated. Recently, this key \hto\ line has been detected for the first time in the ISM of both local and high-redshift galaxies, where this optically thin line is found to be a good tracer of the deeply buried nuclei that have intense far-infrared fields \citep{2017A&A...601L...3P, 2020A&A...634L...3Y, 2021A&A...645A..49G}. This makes the 448\,GHz \hto\ line particularly effective in tracing the elusive population of buried AGN in environments with extreme far-infrared radiation energy densities.

From Fig.\,\ref{fig:flux-compare}, we can see that the 448\,GHz \hto\ line exhibits the most significant difference among all the water lines between \apm\ and NCv1.143, where this line is about nine times stronger in the quasar. This strongly indicates that there is a deeply buried radiation source with a large column of dust and gas in \apm\ that is nonetheless visible in the optical. Most importantly, the brighter 448\,GHz \hto\ emission suggests that the \apm\ quasar must have a significantly more intense far-infrared field buried in the nuclear region than NCv1.143. It is worth noticing that in \apm, we have also detected several transitions of the HCN-VIB line, which is believed to be the tracer of the compact obscured nuclei \citep{2015A&A...584A..42A, 2021A&A...649A.105F}, where the column density reaches beyond 10$^{24}$\,cm$^{-2}$. Comparing the ratio between the 448\,GHz \hto\ with the HCN-VIB(3--2) line, we find a comparable flux ratio of about 6.1 in \apm, which is close to the value of $7.9\pm1.6$ found in a local compact obscured nucleus of ESO\,320-G030 \citep{2021A&A...649A.105F,2021A&A...645A..49G}. Because \apm\ is visible in the soft X-ray \citep{2022A&A...662A..98B} as well as in the rest-frame UV and optical \citep[e.g.,][]{2018A&A...617A.118S}, is it unclear how \apm\ is linked to the typical compact obscured nuclei in the local Universe featured by strong HCN-VIB emission but mostly no X-ray detection  \citep{2021A&A...649A.105F}. Future high-spatial-resolution data are needed to have a clearer picture of the ISM structure in order to answer this question.

\subsubsection{The 475\,GHz \htot533440\ line}

The para-\htot533440\ at rest frame 474.689\,GHz (475\,GHz hereafter) has a very high upper energy level of $E_\mathrm{up}$\,=\,725\,K, which is even higher than the commonly observed $J_\mathrm{up}$\,=\,5 \hto\ line transition \t523514\ at 1411\,GHz  ($E_\mathrm{up}$\,=\,642\,K) in local galaxies \citep{2013ApJ...771L..24Y}. Observations of evolved stars found the 475\,GHz \hto\ lines are masers associated with fast outflows \citep{2008A&A...477..185M}.

In dusty warm dense regions of galaxies, \hto\ molecules can also absorb 53\,$\mu$m far-infrared photons through $5_{33}$\,$\leftarrow$\,$4_{22}$ (Fig.\,\ref{fig:h2o-diag}). Such absorption features have been detected in NGC\,4418 \citep{2012A&A...541A...4G}. In this case, the high \hto\ column density can efficiently populate the $4_{22}$ level and thus sustain this far-infrared pumping process. Such a process may potentially provide a route for the formation of \hto\ maser. Nevertheless, the observed line profile of \htot533440\ in \apm\ is similar to those of other dense gas tracers, suggesting the \hto\ emission is from similar regions of the dense gas tracers that are tracing the bulk motion of the gas at large scales. This is contradictory to a maser origin of \hto\ emission, where commonly, a specific physical condition has to be met (\citealt{2016MNRAS.456..374G} found a typical excitation condition with a kinetic temperature of $\sim$\,1000\,K), leading to very localized emitting regions (thus very narrow linewidths), such as accretion disks surrounding SMBHs.

\subsubsection{Model of the H$_2$O emission lines and its interpretation}
\label{sect:h2o-model}

\begin{figure*}[htbp]
\centering
\includegraphics[scale=0.44]{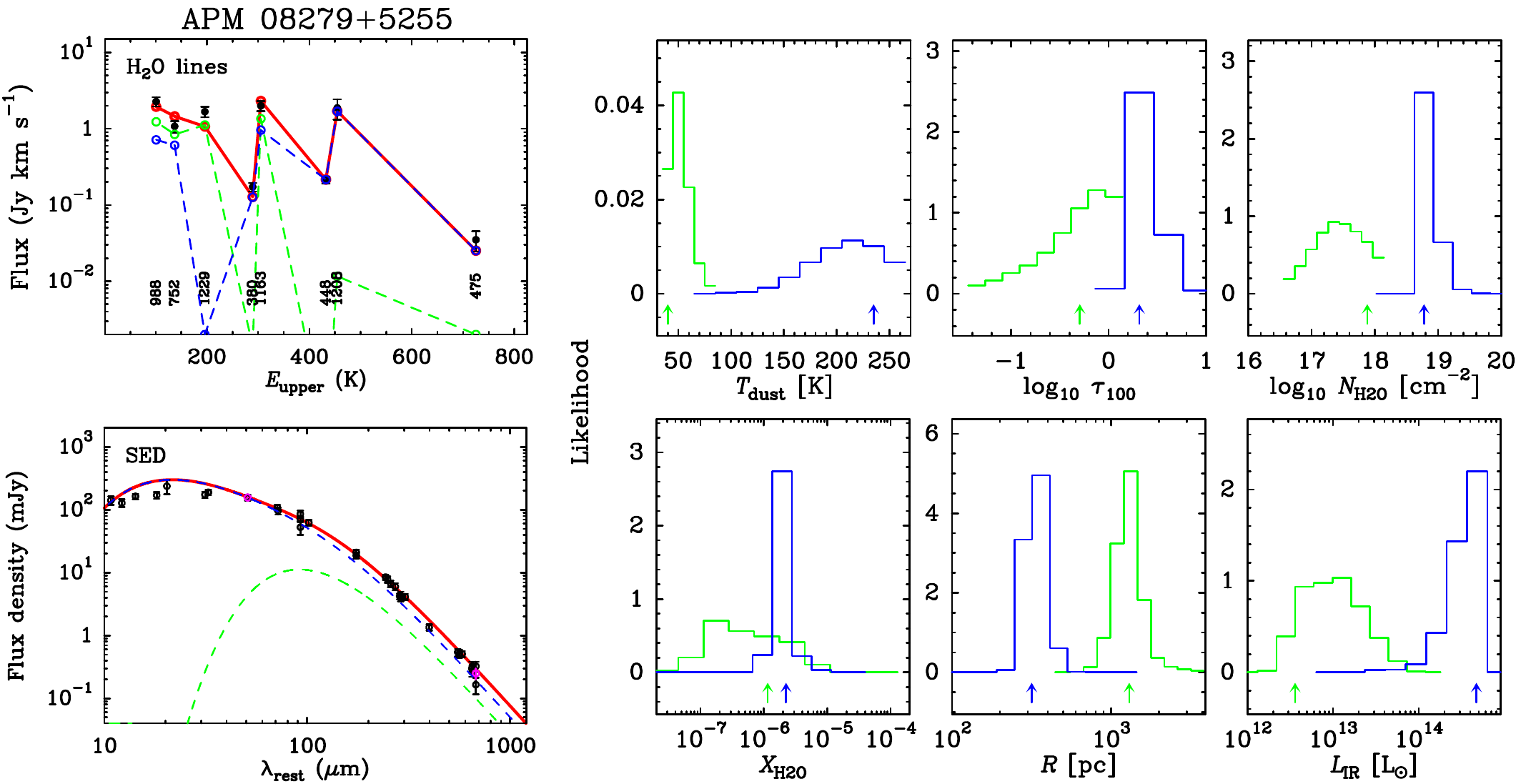}
\vspace{-0.2cm}
\caption
{Best-fit model and the likelihood of the parameters from the model of the \hto\ lines and the dust continuum in \apm. {\it Top left:} Observed lensing-corrected (black data points) versus modeled \hto\ integrated flux  (green and blue are the extended and compact components, respectively; red is the total). \textit{Bottom-left}: Lensing-corrected modeled dust SED from the two components (same colors as in the top panel). The photometric data are taken from \citet{2019ApJ...876...48L}. {\it Right panels:} Likelihood distributions of the key parameters of the \hto\ model, with the same colors as in the left panels. The arrows indicate the maximum likelihood values (Table\,\ref{tab:h2oparam}). 
}
\label{fig:h2o-model-apm}
\end{figure*}
\begin{figure*}[htbp]
\centering
\includegraphics[scale=0.44]{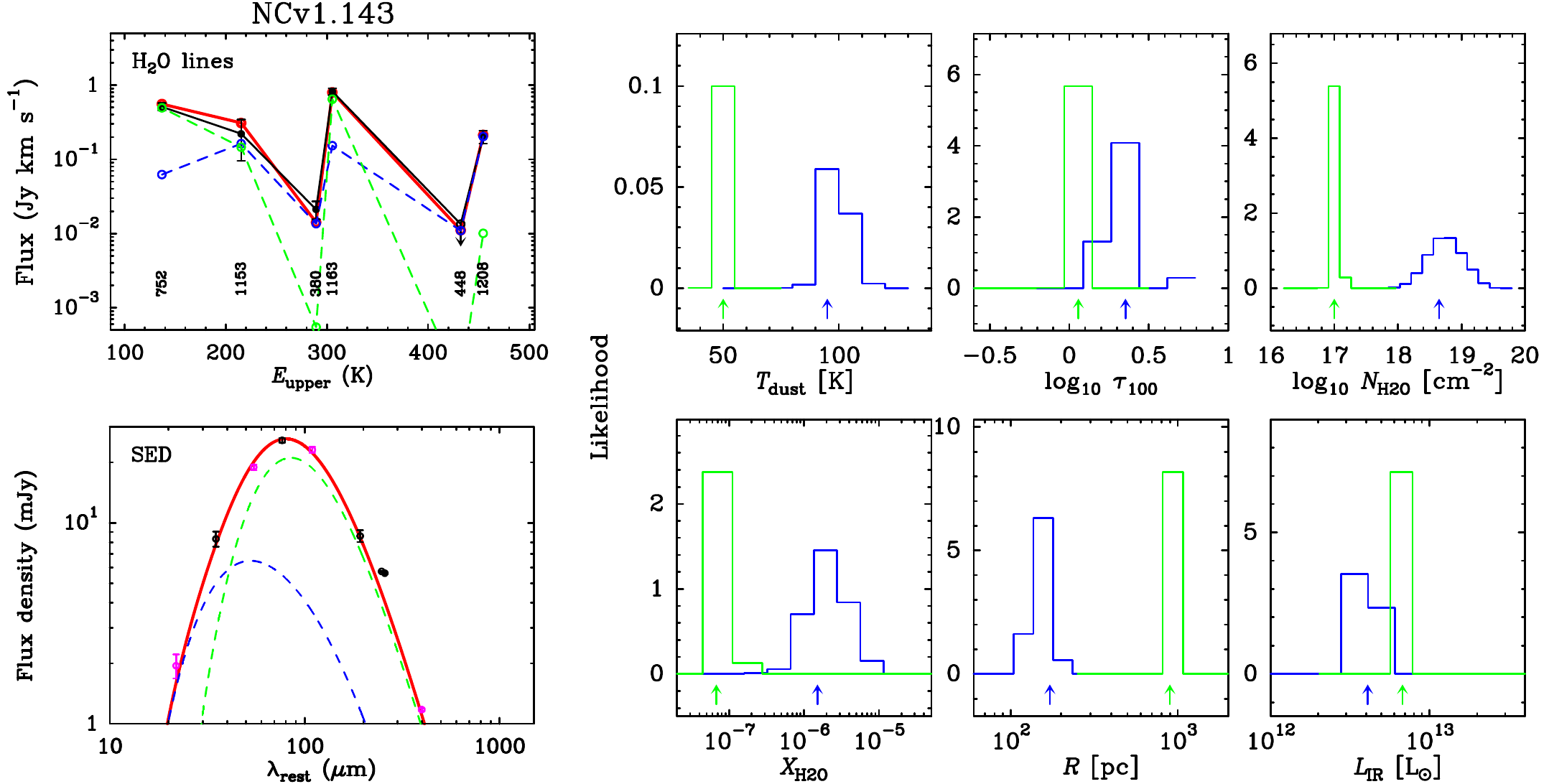}
\vspace{-0.2cm}
\caption
{Same as Fig.\,\ref{fig:h2o-model-apm} but for NCv1.143. The photometric data for the dust SED are from \citet{2013ApJ...779...25B} and \citet{2017A&A...608A.144Y}.
}
\label{fig:h2o-model-nc}
\end{figure*}

Given that in both \apm\ and NCv1.143, there are detections of other \hto\ lines from the literature \citep{2011ApJ...738L...6L, 2011ApJ...741L..38V, 2016A&A...595A..80Y, 2017PhDT........21Y}, it is worth combining all the detected \hto\ lines together with the dust continuum to constrain the excitation conditions. For \apm, there are eight \hto\ lines from $J_\mathrm{up}$\,=\,2 to 5 included, while for NCv1.143, six \hto\ lines are included (Table\,\ref{table:line-flux}, Figs.\,\ref{fig:flux-compare} and 
\ref{fig:h2o-diag}). Here, we assume that the \hto\ submillimeter excitation is dominated by the absorption of far-infrared dust-emitted photons, with additional contribution from collisional excitations. Then, we used the radiative transfer model carried out by \citet{2014A&A...561A..27G} and \citet{2021A&A...645A..49G} to constrain the physical properties of the dusty ISM. Here we adopt a magnification of $\mu$\,$\sim$\,4 \citep{2009ApJ...690..463R} for \apm\ and $\mu$\,$\sim$\,9 for NCv1.143 (Appendix\,\ref{appen:lens-model}), and all the following \hto\ analysis are performed after lensing correction.

The model assumes a number $N_\mathrm{C}$ of independent components, which are spherically symmetric sources with uniform physical properties, namely: the dust temperature $T_{\mathrm{dust}}$, the continuum optical depth at 100\,$\mu$m along a radial path $\tau_{100}$, the column density of \hto\ along a radial path $N_{\mathrm{H_{2}O}}$, the H$_2$ density $n_{\mathrm{H_{2}}}$, the velocity dispersion $\Delta V$, and the gas temperature $T_{\mathrm{gas}}$. The model components are classified into groups according to their physical parameters, each group covering a regular grid in the parameter space ($T_{\mathrm{dust}}$, $\tau_{100}$, $N_{\mathrm{H_{2}O}}$, $n_{\mathrm{H_{2}}}$). However, it should be noted that the conditions of the gas that the \hto\ lines trace can be different from the CO-traced conditions. We keep fixed $\Delta V$\,=\,100\,km\,s$^{-1}$ and $T_{\mathrm{gas}}$\,=\,150\,K (which is the typical condition and the models are insensitivity to the choice of $T_{\mathrm{gas}}$\,=\,150\,K, \citealt{2021A&A...645A..49G}). For both \apm\ and NCv1.143, $N_\mathrm{C}=2$ is required to obtain a satisfactory fit to the data, as shown below.

\setlength{\tabcolsep}{1.45em}
\begin{table*}[!htbp]
\small
\centering
\renewcommand{\arraystretch}{1.05}
\caption{Physical parameters derived from \hto\ modeling of \apm\ and NCv1.143.}
\begin{tabular}{lcccccccc}
\toprule
\multicolumn{9}{c}{{\bf \apm}} \\
\midrule
                                                & & \multicolumn{3}{c}{Compact component}           & & \multicolumn{3}{c}{Extended component}          \\  
 Parameter                                      & & Median & Range$^{\rm{a}}$ & Fiducial$^{\rm{b}}$ & & Median & Range$^{\rm{a}}$ & Fiducial$^{\rm{b}}$ \\
 \cmidrule{3-5}\cmidrule{7-9} 
 $T_{\mathrm{dust}}$\,(K)                       & & $ 211$ & $   148,   258$  & $235$               & & $  51$ & $    37,    70$  & $40$                \\
 $\tau_{100}$                                   & & $ 2.3$ & $   1.5,   5.2$  & $2.1$               & & $ 0.5$ & $   0.1,   1.3$  & $0.5$               \\
 log$_{10}$\,$N_{\mathrm{H_2O}}$\,(cm$^{-2}$)   & & $18.8$ & $  18.6,  19.2$  & $18.8$              & & $17.4$ & $  16.8,  18.0$  & $17.9$              \\
 log$_{10}$\,$n_{\mathrm{H_2}}$\,(cm$^{-3}$)    & & $ 4.7$ & $   4.1,   5.3$  & $5.2$               & & $ 4.6$ & $   4.0,   5.3$  & $4.2$               \\
 $R$\,(pc)                                      & & $ 336$ & $   253,   440$  & $317$               & & $1280$ & $   899,  1875$  & $1303$              \\
 log$_{10}$\,$L_{\mathrm{IR}}$\,(L$_{\odot}$)   & & $14.6$ & $  14.1,  14.8$  & $14.7$              & & $13.0$ & $  12.4,  13.6$  & $12.6$              \\
 log$_{10}$\,$X_{\mathrm{H_2O}}$                & & $-5.7$ & $  -6.0,  -5.5$  & $-5.7$              & & $-6.4$ & $  -7.2,  -5.4$  & $-5.9$              \\
 log$_{10}$\,$M_{\mathrm{gas}}$\,(M$_{\odot}$)                 & & $10.1$ & $   9.8,  10.5$  & $10.1$              & & $10.6$ & $   9.9,  11.0$  & $10.7$              \\
 \noalign{\smallskip}
 \toprule
 \multicolumn{9}{c}{{\bf NCv1.143}} \\
 \midrule
 \noalign{\smallskip}
 $T_{\mathrm{dust}}$\,(K)                       & & $  98$ & $    91,   109$ &  $95$                & & $  50$ & $    46,    55$  & $50$                \\
 $\tau_{100}$                                   & & $ 2.1$ & $   1.3,   4.2$ &  $2.3$               & & $ 1.1$ & $   1.0,   1.4$  & $1.1$               \\
 log$_{10}$\,$N_{\mathrm{H_2O}}$\,(cm$^{-2}$)   & & $18.7$ & $  18.3,  19.2$ &  $18.6$              & & $17.0$ & $  16.9,  17.1$  & $17.0$              \\
 log$_{10}$\,$n_{\mathrm{H_2}}$\,(cm$^{-3}$)    & & $ 4.6$ & $   4.0,   5.3$ &  $4.2$               & & $ 4.7$ & $   4.1,   5.3$  & $4.7$               \\
 $R$\,(pc)                                      & & $ 151$ & $   111,   190$ &  $172$               & & $ 925$ & $     815,1050$  & $898$               \\
 log$_{10}$\,$L_{\mathrm{IR}}$\,(L$_{\odot}$)   & & $12.6$ & $  12.4,  12.8$ &  $12.6$              & & $12.8$ & $  12.8,  12.9$  & $12.8$              \\
 log$_{10}$\,$X_{\mathrm{H_2O}}$                & & $-5.7$ & $  -6.2,  -5.3$ &  $-5.8$              & & $-7.2$ & $  -7.4,  -7.0$  & $-7.2$              \\
 log$_{10}$\,$M_{\mathrm{gas}}$\,(M$_{\odot}$)      & & $ 9.4$ & $   9.0,   9.7$ &  $9.6$               & & $10.7$ & $  10.6,  10.8$  & $10.7$              \\
            \noalign{\smallskip}
\bottomrule
\end{tabular}
   \begin{tablenotes}[flushleft]
   \small
	 \item\textbf{Note:} $^{\mathrm{a}}$: 90\% confidence intervals;  $^{\mathrm{b}}$: Values for the fiducial best-fit model, selected for detailed
  comparison with data, while the uncertainties of the derived parameters are well-characterized in Figs.\,\ref{fig:h2o-model-apm} and \ref{fig:h2o-model-nc}.  
   \end{tablenotes}
    \label{tab:h2oparam}
\end{table*}
\normalsize

Radiative pumping by far-infrared radiation means that \hto\ is sensitive to the dust SED, and thus, our goal is to fit the \hto\ SLED and the far-infrared SED simultaneously. For this purpose, we use as input data of our minimization method the fluxes of the \hto\ lines (and the upper limit of the \hto\,448\,GHz line in NCv1.143) and a number $N_{\mathrm{cont}}$ of continuum flux densities. For \apm\ $N_{\mathrm{cont}}=2$ (at $\lambda_{\mathrm{rest}}\approx51$ and 677\,$\mu$m), while we use $N_{\mathrm{cont}}=6$ for NCv1.143. These continuum data are highlighted in magenta in the SEDs in Figs.\,\ref{fig:h2o-model-apm} and \ref{fig:h2o-model-nc}. In our method, $\chi^2$ is minimized for all possible combinations among the $N_\mathrm{C}$ components, giving for each one the squared radius $R^2$ of each component. The best-fit (fiducial) model is the combination that yields the minimum $\chi^2$, and the regular grid enables the estimation of the likelihood of each parameter. The derived parameters -- $R$, the \hto\ abundance relative to H nuclei $X_{\mathrm{H_{2}O}}$, $L_{\mathrm{IR}}$, and $M_{\mathrm{gas}}$ of each component -- are also inferred. More details are given in \citet{2021A&A...645A..49G}.

We first attempted to fit the \hto\ and continuum emission with a single-model component ($N_\mathrm{C}=1$), but results were unreliable, with a best-reduced $\chi^2$ value $\chi_\mathrm{red}^2$\,$\approx$\,4. This was indeed expected because the low-lying \hto\ $J_{\mathrm{up}}$\,=\,2--3 lines are expected to arise in more extended regions than the $J_{\mathrm{up}}$\,=\,4 lines that trace buried regions \citep[e.g.,][]{2014A&A...567A..91G,2017A&A...601L...3P}. A better fit was found with $N_\mathrm{C}$\,=\,2 components, with $\chi_\mathrm{red}^2$\,$\approx$\,0.9; in Figs.\,\ref{fig:h2o-model-apm} and \ref{fig:h2o-model-nc}, the arrows indicate the best-fit values and solid histograms show their likelihood distributions. 

In both galaxies, we find that the $J_{\mathrm{up}}\geq4$ lines are formed in a  warm ``compact'' component ($R\sim330$\,pc and $\sim150$\,pc in \apm\ and NCv1.143, respectively), which we identify with the nuclear region (nuclear core). The nuclear cores are moderately optically thick in the far-infrared with  $\tau_{100}$\,$\sim$\,$2$, with warm dust temperatures of over 100\,K (Figs.\,\ref{fig:h2o-model-apm} and Fig.\,\ref{fig:h2o-model-nc}). In NCv1.143, this core has an intrinsic luminosity of $L_{\mathrm{IR}}$\,$\sim$\,$4\times10^{12}$\,$L_{\odot}$, resulting in an extreme infrared luminosity surface density $\Sigma_\mathrm{IR}\equiv L_{\mathrm{IR}}/(4\pi\,R^2)$\,=\,$1.4\times10^{13}$\,\lsun\,kpc$^{-2}$. This translates into a surface SFR value of $\mathit{SFR}/(\pi\,R^2)\sim$\,$6.2\times10^{3}$\,\msun\,yr$^{-1}$\,kpc$^{-2}$ if the contribution to $L_{\mathrm{IR}}$ by a possible obscured AGN is negligible since no strong evidence of the presence of an AGN has been found in NCv1.143 \citep{2016A&A...595A..80Y}. In \apm, however, the compact nuclear core has an extreme luminosity of $L_{\mathrm{IR}}$\,$\sim$\,$4\times10^{14}$\,$L_{\odot}$ that results in $\Sigma_\mathrm{IR}$\,=\,$3\times10^{14}$\,\lsun\,kpc$^{-2}$. Obviously, this can only be explained by a significant contribution of radiation from the AGN.

On the other hand, the $J_{\mathrm{up}}$\,$\leq$\,3 lines in both galaxies, pumped by absorption of dust-emitted 101 and 75\,$\mu$m photons, are formed in a more extended $R$\,$\sim$\,1\,kpc starburst region with moderate $T_{\mathrm{dust}}$\,$\sim$\,50\,K, most likely the circumnuclear disk of the host. In NCv1.143, this size is comparable to the projected half-light effective radius traced by the \htot211202, \htot321312\ and CO(10--9) line emission as presented in Appendix\,\ref{appen:lens-model} (Fig.\,\ref{h2o:fig:high-res:nc143-contours}), and also to the size of the averaged 870\,$\mu$m dust continuum of submillimeter galaxies \citep[SMGs, $R$\,$\sim$\,1\,kpc;][]{2019MNRAS.490.4956G}. The circumnuclear disk in NCv1.143 dominates the luminosity with $L_{\mathrm{IR}}$\,$\sim$\,$6.5\times10^{12}$\,$L_{\odot}$ and $\Sigma_\mathrm{IR}=6\times10^{11}$\,\lsun\,kpc$^{-2}$ (corresponding to $\sim$\,230\,\msun\,yr$^{-1}$\,kpc$^{-2}$).

We note that excluding the continuum flux density at 51\,$\mu$m in the fit of \apm\ yields a very similar best-fit model, with the far-infrared and mid-infrared continuum emission well fitted. The estimate of very warm $T_{\mathrm{dust}}$ of the nuclear region is mainly driven by the \htot533440\ 475\,GHz line. This means that the \hto\ lines are indeed tracing the shape of the SED, and high columns of \hto\ are present in the inner $\sim0.5$\,kpc in \apm\ provided that these nuclear regions (still) have large reservoirs of gas of $\sim10^{10}$\,\msun. 

It is also worth noting the similarities and differences between the two sources. Despite their different SEDs, the extended circumnuclear disks have similar characteristics in \apm\ and NCv1.143 (see Table\,\ref{tab:h2oparam}). For the compact core, most of the parameters in Table\,\ref{tab:h2oparam} are also similar between \apm\ and NCv1.143. However, the dust temperature and infrared-luminosity of the core in \apm\ are significantly higher than those in NCv1.143. As shown in Figs.\,\ref{fig:h2o-model-apm} and \ref{fig:h2o-model-nc}, the compact nuclear core in \apm\ is dominating the total infrared luminosity of the entire galaxy, while in NCv1.143, the extended circumnuclear disk has a much larger contribution to the total infrared luminosity. Therefore, as diagnosed by the \hto\ lines, the main difference between the two sources is a much more prominent and powerful nuclear core component, indicating the ``turn-on'' phase of the extreme quasar in \apm, which may be a short-lived phenomenon.

\subsection{H$_3$O$^+$ lines}
\label{sect:H3O+}

As a symmetric top molecule, \httop\ plays an important role in the oxygen chemistry network that is related to the \hto\ formation in ion-neutral reactions \citep{2016ARA&A..54..181G}. First detections of the \httop\ lines were achieved in two local starburst galaxy, M\,82 and Arp\,220, from the transition of p-\httop(\tp32+22-) at 364.797\,GHz, with $E_\mathrm{up}/k_\mathrm{B}$\,=\,139\,K \citep[][see also \citealt{2021ApJ...923..240S}]{2008A&A...477L...5V}. \httop\ has also been detected in the Cloverleaf quasar \citep{2022EPJWC.26500024G}. Another study of five local galaxies utilized the abundance ratios of \httop\ with \hto\ and HCO$^+$ and found the chemistry in XDRs can explain some of the high column density ratios of \httop\ over \hto\ and HCO$^+$ in some sources \citep{2011A&A...527A..69A}.

Both the \httop(\tp32+22-) and \httop(\tp30+20-) lines are robustly detected in \apm. These are the first high-redshift detections of any \httop\ lines. While in NCv1.143, these two lines are below the noise level; therefore, we do not discuss NCv1.143 in detail here until further robust detections are reached. The LTE-derived column density of \httop\ is about $N_\mathrm{H_3O^+}$\,$\sim$\,$10^{15}$\,cm$^{-2}$ after lensing correction. This value is comparable with NGC\,253, NGC\,1068, and NGC\,6240, where the  XDR and/or a high CR ionization rate in the photodissociation-dominated region (PDR) can explain the column density \citep{2011A&A...527A..69A}.

The abundance ratio between \httop\ and \hto\ can also be used as a probe of ionization states. Using the \hto\ column density derived from Sect.\,\ref{sect:h2o-model} of $2.5\times 10^{17}$\,cm$^{-2}$, we find a relative abundance ratio of \httop/\hto\,$\sim$\,$3\times10^{-3}$, which can be explained by both the PDR and the XDR \citep{2009ApJ...690..463R}. Indeed, \httop\ lines have also been detected in the pure-starburst NGC\,253 \citep{2021A&A...656A..46M, 2022ApJ...931...89H}. While the aforementioned diagnostics are inconclusive, the most evident abundance ratio that suggests the XDR plays an important role comes from \httop/HCO$^+$, where in \apm\ we find a very large abundance ratio of $\sim$\,40, which is difficult to explain by PDRs \citep{2007A&A...461..793M}. This suggests that the XDR may significantly influence the abundances of  HCO$^+$ and \httop\ abundances in \apm, though the possibility of \httop\ enhancement by CR cannot be entirely ruled out.

\subsection{Typical high dipole moment tracers --- HCN, HCO$^+$, HNC, CS and C$^{34}$S --- and non-LTE modeling}
\label{sect:lvg}

\begin{figure*}[htbp]
\centering
\includegraphics[scale=0.26]{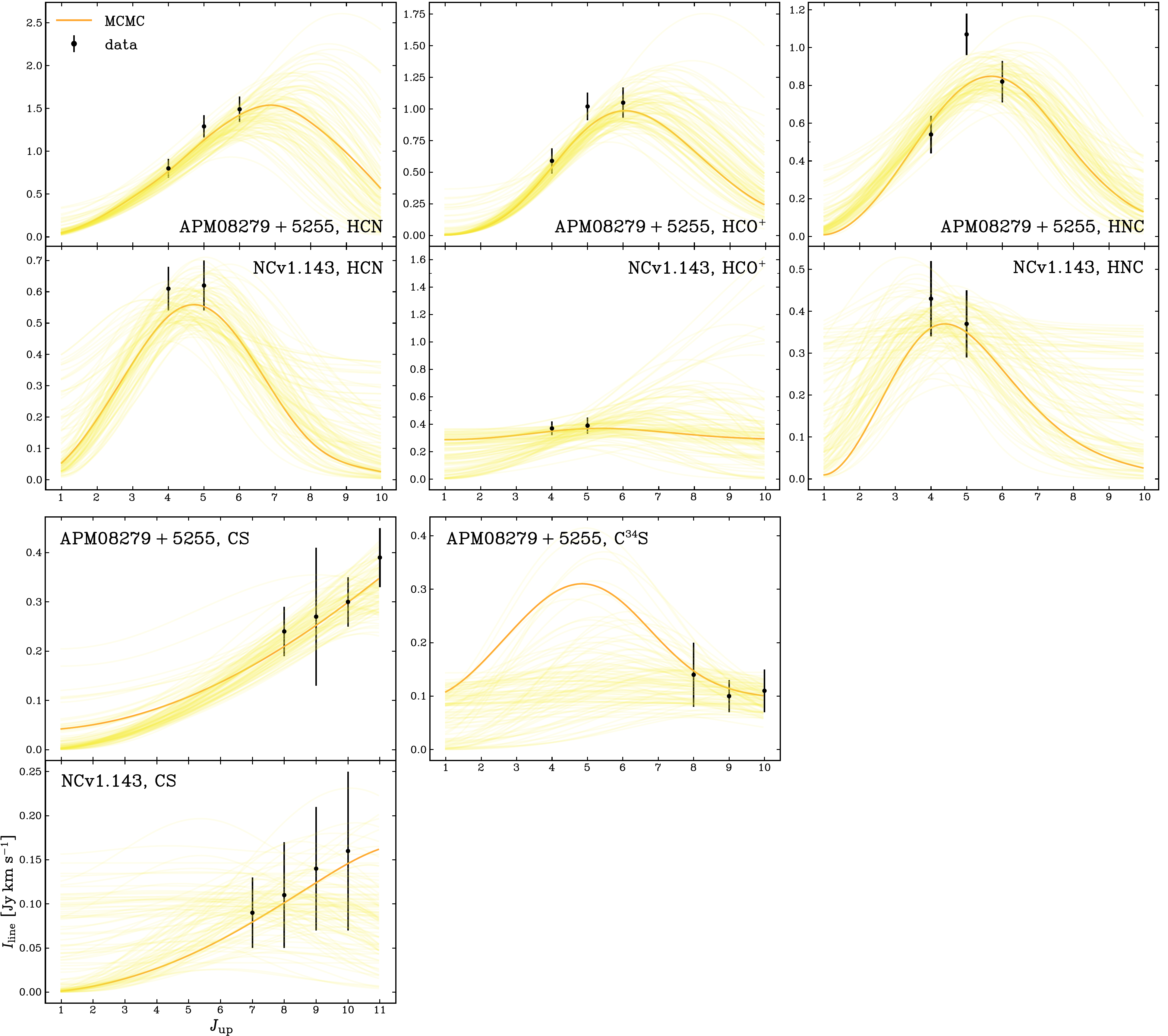}
\vspace{0.1cm} 
\caption
{Fitting of the dense gas SLEDs using LVG models via the MCMC technique. Observed fluxes and their errors are indicated with black symbols. Orange lines show the best fit (Table\,\ref{table:LVG-fit}), and light-yellow lines indicate 100 examples within the 15.87\% and 84.14\% quartiles of the posteriors. }
\label{fig:lvg-sled}
\end{figure*}

Because of their large dipole moment, the consensus is that the lines of HCN, HCO$^+$, HNC, and CS are probably among the best tracers of dense gas where stars are forming. This is demonstrated by the linear correlation between the line luminosity of the dense gas tracers, especially HCN, and \lir, which holds tightly over ten orders of magnitude, ranging all the way from dense cores of Giant Molecular Clouds to ultra-luminous infrared galaxies \citep[e.g.,][]{2004ApJ...606..271G, 2005ApJ...635L.173W, 2007ApJ...660L..93G, 2014ApJ...784L..31Z, 2015ApJ...810..140C, 2019ApJ...880..127J, 2022ApJ...936...58Z}. Unlike the super-linear correlation found between the SFR and the total molecular gas traced by \co10, the tight relation between the dense gas and the SFR suggests that the dense gas links more tightly with star formation rather than \co10\ \citep{2004ApJ...606..271G}. Therefore, the dense gas tracers such as HCN, HCO$^{+}$, HNC, and CS serve as crucial probes for understanding the most intense star formation in the SMGs. Nevertheless, the detection of the standard dense gas tracers remains meager at high redshifts because the emission lines of these dense gas tracers are very weak compared to CO. HCN is only reported in a handful of individual luminous high-redshift galaxies and only detected in strongly lensed sources \citep{2004ApJ...614L..97V, 2005ApJ...618..586C, 2007ApJ...660L..93G, 2007ApJ...671L..13R, 2011ApJ...726...50R,2013ARA&A..51..105C,2017ApJ...850..170O,2018A&A...620A.115B, 2021A&A...645A..45C, 2022A&A...667A..70R, 2023arXiv230802886R}. These lines provide the essential signature of massive starbursts, and it has been suggested that the dense gas fraction, as derived from the comparison between HCN and low-$J$ CO lines, is higher in the high-redshift dusty star-forming galaxies \citep{2017ApJ...850..170O}, though  \citet{2022A&A...667A..70R} argue that there is no difference in dense gas fractions.

\begin{figure*}[htbp]
\includegraphics[scale=0.287]{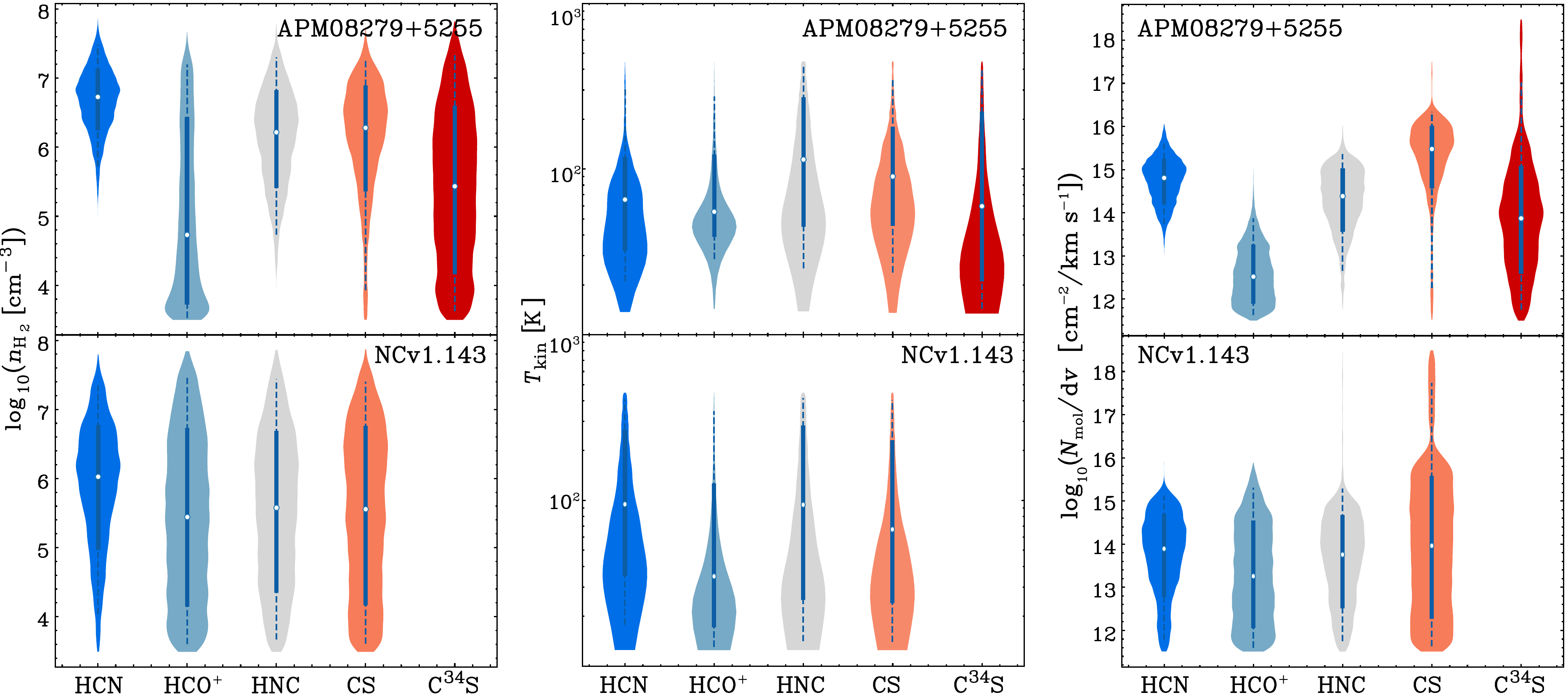}
\vspace{-0.2cm}
\caption
{Violin plots for the LVG-MCMC-derived posterior distributions of $n_{\rm H_2}$, $T_\mathrm{kin}$, and $N_\mathrm{mol}/{\rm d}V$ in \apm\ and NCv1.143 (upper and lower panels, respectively). The white dot represents the median values, while the solid blue and dashed blue lines represent $\pm$\,1$\sigma$ and $\pm$\,2$\sigma$, respectively. The violin shapes represent the posterior distributions of each parameter as presented in the corner plots (Figs.\,\ref{fig:LVG-corner-1} and \ref{fig:LVG-corner-2}).
}
\label{fig:LVG-violin}
\end{figure*}

In our NOEMA line survey, we have acquired the largest amount of multiple transitions of several dense gas tracers to date, including HCN, HCO$^+$, HNC, CS, C$^{34}$S, NO, and N$_2$H$^+$ simultaneously in single objects. This enables us to understand the gas excitation conditions by directly looking into each species; thus, we are able to perform excitation analysis via the LVG method \citep[e.g.,][]{1974ApJ...189..441G} for \apm, where at least three transitions are detected per molecule. For NCv1.143, using some strict priors based on the previous studies of this source, we can also get loose constraints of the excitation conditions (the lack of data constraints here is well reflected by the posteriors of the parameters, Fig.\,\ref{fig:LVG-violin} and Appendix\,\ref{sect:appendix:LVG}), where only two transitions are detected for most of the dense gas tracers. For the purpose of comparing \apm\ and NCv1.143, we picked HCN, HCO$^+$, HNC, CS, and C$^{34}$S for a detailed non-LTE analysis for the ISM conditions (N$_2$H$^+$ is excluded in the LVG analysis due to the lack of collisional rates data).

Following the same implementation as described in \cite{2017A&A...608A.144Y}, we use a one-dimensional (1D) non-LTE radiative transfer code \texttt{RADEX} \citep{2007A&A...468..627V}, assuming an escape probability of $\beta = (1-\mathrm{e}^{-\tau}) / \tau$ from an expanding sphere geometry \citep{sobolev60} to model the observed flux. At the same time, we also adopted the MCMC code \texttt{emcee} as we fit the integrated line flux of the SLED for HCN, HCO$^+$, HNC, and CS for both \apm\ and NCv1.143, as well as C$^{34}$S for \apm. Here, we do not simultaneously fit all the tracers with a single condition because they do not trace the same gas conditions \citep{2005pcim.book.....T}. The collisional rates are taken from the LAMDA database \citep{2005A&A...432..369S}. After deploying 100 walkers with 1000 interactions after 100 burn-in steps, we derive the posterior distributions of the kinetic temperature of the molecular gas ($T_\mathrm{kin}$), the volume density of H$_2$ ($n_\mathrm{H_2}$), and the column density of each molecule unit velocity difference ($N_\mathrm{mol}/{\rm d}V$), and the solid angle ($\Omega_\mathrm{app}$) of the source where the lensing magnification factor is included.

{\centering
\setlength{\tabcolsep}{0.57em}
\begin{table}[!htbp]
\renewcommand{\arraystretch}{1.5}
\small
\caption{Physical parameters derived from the LVG models.}
\begin{tabular}{lcccccc}
\toprule
Species          &  \multicolumn{2}{c}{log($n_{\rm H_2}$)}    &  \multicolumn{2}{c}{log$(T_{\rm kin})$}   &  \multicolumn{2}{c}{log($N$/d$V$)}          \\
                 &  \multicolumn{2}{c}{log({$\rm cm^{-3}$})}  &  \multicolumn{2}{c}{log(K)}                &  \multicolumn{2}{c}{log(cm$^{-2}$/\kms)}   \\
                 &     APM             &     NC              &        APM          &       NC            &      APM              &   NC                 \\
\midrule                                                                                                                                               
CO               & 3.6$^{+0.8}_{-1.1}$ & 3.0$^{+0.4}_{-0.5}$ & 2.3$^{+0.2}_{-0.3}$ & 2.8$^{+0.2}_{-0.3}$ &  18.2$^{+1.0}_{-0.6}$ & 17.3$^{+0.4}_{-0.4}$ \\ 
HCN              & 6.7$^{+0.4}_{-0.5}$ & 6.0$^{+0.8}_{-1.1}$ & 1.8$^{+0.3}_{-0.3}$ & 2.0$^{+0.5}_{-0.4}$ &  14.8$^{+0.5}_{-0.6}$ & 13.9$^{+0.8}_{-1.1}$ \\
HCO$^+$          & 4.7$^{+1.7}_{-1.0}$ & 5.5$^{+1.3}_{-1.3}$ & 1.7$^{+0.4}_{-0.2}$ & 1.5$^{+0.6}_{-0.3}$ &  12.5$^{+0.6}_{-0.4}$ & 13.3$^{+1.3}_{-1.2}$ \\
HNC              & 6.2$^{+0.6}_{-0.8}$ & 5.6$^{+1.1}_{-1.2}$ & 2.1$^{+0.4}_{-0.4}$ & 2.0$^{+0.5}_{-0.6}$ &  14.4$^{+0.7}_{-0.8}$ & 13.8$^{+1.0}_{-1.3}$ \\
CS               & 6.3$^{+0.6}_{-0.9}$ & 5.6$^{+1.2}_{-1.4}$ & 2.0$^{+0.3}_{-0.3}$ & 1.8$^{+0.5}_{-0.5}$ &  15.5$^{+0.5}_{-0.9}$ & 14.0$^{+1.6}_{-1.7}$ \\
C$^{34}$S        & 5.4$^{+1.1}_{-1.2}$ &     --              & 1.8$^{+0.6}_{-0.5}$ &       --            &  13.9$^{+1.2}_{-1.3}$ &  --                  \\
\bottomrule  
\end{tabular}
   \begin{tablenotes}[flushleft]
   \small
	 \item\textbf{Note:} 
        LVG-MCMC-derived parameters; the median values and the $\pm$\,1\,$\sigma$ uncertainties are noted. The results where only two lines were detected in NCv1.143 are also included despite their large uncertainties, which are well captured by the posterior distributions shown in the corner plots (Figs.\,\ref{fig:LVG-corner-1} and \ref{fig:LVG-corner-2}). 
        For comparison, we also include the LVG model results of CO, using the line fluxes from \citet{2007A&A...467..955W} and \citet{2017A&A...608A.144Y}, which trace much less extreme physical conditions compared to the dense gas tracers.
   \end{tablenotes}
   \label{table:LVG-fit}
\end{table}
\normalsize}

Table\,\ref{table:LVG-fit} lists the results from the LVG-MCMC analysis, while the fitted SLEDs are shown in Fig.\,\ref{fig:lvg-sled}. The LVG models can well reproduce the observed integrated fluxes. However, it is worth noticing that the $J_\mathrm{up}$\,=\,5 HNC line is about 40\% brighter than the LVG model (over 1 $\sigma$ level). It is difficult to explain such a flux excess of the $J_\mathrm{up}$\,=\,5 HNC line. One possible cause is the blending by the CN(4--3) line. Nevertheless, our current data limit the ability to assess this possibility thoroughly, and we do not see a similar overestimate of the HNC(5--4) flux in NCv1.143. Thus, the source of the flux excess of the HNC(5--4) line remains elusive. 

In order to facilitate a more effective comparison of the results from our LVG-MCMC analysis, we present the posterior distributions of the two sources using the violin plots in Fig.\,\ref{fig:LVG-violin}. It becomes apparent that all the $n_\mathrm{H_2}$ values derived from HCN, HCO$^+$, HNC, CS, and C$^{34}$S significantly exceed that of the CO from previous studies (\citealt{2007A&A...467..955W} and \citealt{2017A&A...608A.144Y}, $n_\mathrm{H_2}$\,$\sim$\,$10^3$--$10^{4}$\,cm$^{-3}$), falling within the range of $10^{5}$--$10^{7}$\,cm$^{-3}$. This is not surprising as these mid-$J$ lines of the dense gas tracers probe more extreme conditions of the high-density ISM than the CO lines. Within \apm, the gas density traced by HCO$^+$ is markedly lower compared to that traced by all other molecules. As for the kinetic temperature, $T_\mathrm{kin}$, all tracers align to suggest a similar value of approximately 100\,K. When examining the column density values derived from the LVG analysis, \apm\ displays well-defined values, with most of the species falling within the $10^{14}$--$10^{15}$\,cm$^{-2}/\kms$ range. Notably, we observe that the column density of HCO$^+$ is significantly below other species, particularly in \apm. 

When compared with the LVG-derived column densities and assuming a small d$V$, the abundances of \apm\ and NCv1.143 align with the \texttt{MADCUBA} LTE-derived values listed in Table\,\ref{table:madcuba-fit}. The only exception is the CO abundance of NCv1.143, where the LVG-derived value is one order of magnitude smaller in NCv1.143. The LVG-derived values confirm the trend we found in Fig.\,\ref{fig:abundances-compare-2} that the relative abundance of the dense gas tracers seems all boosted in the quasar where AGN is dominated. In the next section, we take a closer look at how different ISM environments might affect the abundance of the dense gas tracers in the two sources.

\subsection{AGN diagnostics with HCN, HCO$^+$, HNC and CS}
\label{sec.HCNHCOpCSdiagnostic}

Here, we place the observed line flux ratios from high dipole moment tracers from our line surveys into what has been reported within the local Universe. A number of molecular line ratios have been proposed as tracers of the powering sources and/or physical conditions within the obscured central regions of bright galaxies.

The ratio between the HCN and HCO$^+$ is a classic molecular discriminator of nuclear regions dominated by either a starburst or an AGN \citep[e.g.,][]{Kohno2001, 2008ApJ...677..262K, 2023ApJ...950...75I}. Observations of a statistically significant sample with low spatial resolutions have shown that an enhancement of the HCN/HCO$^+$ can occur in galaxies hosting an AGN \citep{Privon2015, Izumi2016, Imanishi2016}. However, \citet{2020ApJ...893..149P} found no trends between HCN/HCO$^+$ and the AGN fraction from hard X-ray observations. High-resolution imaging of nearby AGN galaxies has shown this enhancement to be located in the circumnuclear disk and not directly toward the AGN position \citep{2014A&A...570A..28V, Martin2015}, which is claimed to be the result of high-temperature chemistry due to mechanical heating in the surrounding of the AGN \citep{Izumi2013, Harada2010}. All these local studies show the complex nature of the HCN/HCO$^+$ diagnostics.

\citet{Izumi2016} proposed a diagnostic diagram based on the ratios of HCN/HCO$^+$(4--3) and HCN(4--3)/CS(7--6), which we cover in our observations. We derive HCN/HCO$^+$ ratios of 1.4$\pm$0.2 and 1.6$\pm$0.2 toward \apm\ and NCv1.143, respectively. Similar ratios are obtained with the $J$\,=\,5--4 and \mbox{6--5}. The HCN/CS ratio is measured to be between 3 and 4 \footnote{Ratios around 3.3 are derived with HCN(4--3)/CS(8--7) and HCN\mbox{(6--5)}/CS(8--7) toward \apm.} and 6.7$\pm$3.3, for \apm\ and NCv1.143, respectively.

The ratios derived for the quasar \apm\ are not distinctive from what is observed toward starburst-dominated galaxies. This is not uncommon and matches what is observed in other observations of AGN at a low spatial resolution, which shows a wide range of ratios \citep{Privon2015, Izumi2016, Imanishi2016}. Conversely, the ratios observed toward the starburst NCv1.143 are very uncertain and might lie within the range where only galaxies with an AGN are found.

Although we could expect these ratios to be affected by continuum absorption or self-absorption from intervening material, the analysis from \citet{Imanishi2016} showed that line flux ratios are only moderately affected by such effects. Since we do not find any AGN footprint in NCv1.143 \citep{2016A&A...595A..80Y}, especially from the absence of detection in radio and in the mid-infrared bands of the Wide-field Infrared Survey Explorer \citep[WISE;][]{2010AJ....140.1868W}, a buried AGN is unlikely to reside in NCv1.143. However, the sensitivity and the angular resolution of our data on the CS, HCN, and HCO$^+$ lines are limited to reaching conclusions on whether they actually point to the presence of such an AGN, although we do see general trends of the abundance difference between AGN- and starburst-dominated sources as discussed in Sect.\,\ref{section:abundance-LTE}, and NCv1.143 is aligned with the non-AGN-dominated abundances.

The abundance ratios we derived for \apm\ and NCv1.143, expressed as $N$(HCN)/$N$(CS), are given in log base ten as $-0.7^{+1.3}_{-1.1}$ and $-0.1^{+2.5}_{-2.7}$, respectively. These correspond to the values of $N$(HCN)/$N$(CS) ratios of about 0.2 and 0.8 for \apm\ and NCv1.143, respectively, with large uncertainties. Compared with a detailed case study of the chemical networks in NGC\,1068 \citep{2022A&A...667A.131B}, where AGN tends to enhance $N$(HCN)/$N$(CS), our values do not show such clear evidence of the enhancement. This is not surprising considering the large uncertainties of the abundance ratios and low spatial resolution of our measurements. The abundance ratios of $N$(HCN)/$N$(HCO$^+$) for \apm\ and NCv1.143 are $2.3^{+1.1}_{-1.3}$ and $0.6^{+2.0}_{-2.4}$ in log base ten, which might show a significant abundance enhance of HCN over HCO$^+$. This is consistent with what has been found in the AGN-dominated nuclear regions of NGC\,1068, at 40\,pc to 100\,pc scales. If such similarities are valid, it also indicates that the spatial distribution of HCN and HCO$^+$ are similar, as the case in NGC\,1068 \citep{2022A&A...667A.131B}. Thus, their abundance ratio is a better representation of a consistent inference of their ISM condition compared to the abundance ratios of $N$(HCN)/$N$(CS).

\subsection{CCH}
\label{sect:cch}

The ethynyl radical, CCH, is abundant in the ISM and is known to be enhanced in PDRs, both from galactic and extragalactic observations, although ionizing processes such as CR ionizing could also be responsible for the CCH production \citep[][see the discussion and references therein]{Martin2014, Holdship2021}. 

In this work, we observed line ratios of CCH over HCN, HCO$^+$, and HNC using their $J$\,=\,5--4 transitions of 0.29/0.37/0.46 and 0.29/0.46/0.48, toward \apm\ and NCv1.143, respectively. These ratios fall within the ratios presented by \citet{Martin2014} over a small sample of galaxies with a variety of environments. We do not see any particular flux enhancement of the CCH lines in our targets. However, when we compare the relative abundances from the LTE analysis directly, we find that CCH is more abundant in \apm\ than in NCv1.143. This is consistent with an enhancement of CCH abundance in the region dominated by AGN and shocks \citep{2023ApJ...955...27N}.

\subsection{CH lines}
\label{sect:ch}

The methydilyne radical, CH, is a major interstellar molecule both in the diffuse and dense ISM, as well as in XDRs \citep[e.g.,][]{2016ARA&A..54..181G}, typically tracing densities $n_\mathrm{H}$ of 100--5000\,cm$^{-3}$ and temperatures of 15-100\,K \citep{2006ARA&A..44..367S}. It was one of the very first interstellar molecules detected through its visible bands in the optical and later through its $\Lambda$-doublet lines in the ground state in radio frequencies. CH is one of the key probes of the diffuse ISM. Its absorption lines have been detected in the foreground galaxy at $z$\,=\,0.89 toward the quasar PKS\,1830-211 \citep{2014A&A...566A.112M, 2023A&A...674A.101M}. However, its study in the ISM of local galaxies is limited by the absence of millimeter lines and the complete atmospheric absorption of its first submillimeter transition at $\sim$\,530\,GHz \citep{2021A&A...650A.133J}.

The lowest spin-rotational (560\,$\mu$m) transition of CH has six hyperfine components grouped at 532.8 and 536.8\,GHz. They were detected in four nearby galaxies, NGC\,1068, Arp\,220, M\,82, and NGC\,253, using the {\it Herschel} \citep{2011ApJ...743...94R, 2014ApJ...788..147R}, who found that the CH lines were a factor of $\sim$\,20 brighter than the adjacent HCN and HCO$^+$ lines.  

While this frequency range was not covered for NCv1.143, the two groups of hyperfine components of the CH(4--3) transition were detected with S/N\,>\,4 in \apm\ (Table\,\ref{table:line-flux}). This is the first clear detection of CH in an individual high-redshift galaxy, although \citet{2014ApJ...785..149S} presented a marginal 2.5-$\sigma$ detection of these lines in the stacked spectrum of the dusty star-forming galaxies selected by the South Pole Telescope survey.

As shown in Table\,\ref{table:ch-ratios}, the ratio of CH lines to adjacent $J$\,=\,6--5 lines of HCN and HCO$^+$ is more than an order of magnitude lower in \apm\ than that in the local galaxies observed by \citet{2014ApJ...788..147R}. This holds true even for NGC\,1068 and Arp\,220, where the ratios are lower by a factor of approximately 20. The origin of such a substantial difference is challenging to comprehend without a detailed comparison to other high-redshift starbursts and AGN galaxies. Additionally, a higher angular resolution of \apm\ would be beneficial in better understanding the origin of the CH emission, shedding light on the cause of the CH ``deficit.'' As the maximum factor of differential lensing reaches only up to 3 \citep{2009ApJ...690..463R} in \apm, it is unlikely the explanation. Interestingly, however, the average CH/HCO$^+$ ratio in the stacked spectrum of SPT-discovered dusty galaxies appears intermediate, with an approximate value of $0.5\pm0.2$ \citep{2014ApJ...785..149S}, a factor of six times smaller than the local values. In contrast, the CH/HCO$^+$ flux ratios reach as high as $\sim$\,4--5 in the stacked spectrum of the {\it Herschel} lensed sources \citep{2023MNRAS.521.5508H}. In the absence of alternative explanations, the primary cause of the weakness of the CH lines in \apm\ might be due to the destruction of CH by the strong quasar UV field, which is akin to the weak \ci\ emission in \apm\ compared with NCv1.143.

{\centering
\setlength{\tabcolsep}{0.26em}
\begin{table}[!htbp]
\renewcommand{\arraystretch}{1.1}
\footnotesize
\caption{Flux ratios between CH versus HCN and HCO$^+$.}
\begin{tabular}{lcccc}
\toprule
 {Source}   & {CH$_{532}$/HCN} & {CH$_{532}$/HCO$^+$} & {CH$_{536}$/HCN} & {CH$_{536}$/HCO$^+$} \\ 
\midrule
{NGC\,1068} & 3.0         & 3.1             & 3.4         & 3.6  \\
{NGC\,253}  & 1.9         & 3.0             & 1.5         & 2.5  \\
{Arp\,220}  & 4.3         & --              & 4.5         & --   \\     
{APM\,08279}& 0.16        & 0.23            & 0.16        & 0.23 \\\bottomrule  
\end{tabular}
   \begin{tablenotes}[flushleft]
   \small
	 \item\textbf{Note:} 
        Comparison of CH(532\,GHz) and CH(532\,GHz) line ratios to HCN(6--5) and HCO$^+$(6--5) in \apm\ and local starburst (NGC\,253 and Arp\,220) and AGN (NGC\,1068) galaxies from \citet{2014ApJ...788..147R}.
   \end{tablenotes}
   \label{table:ch-ratios}
\end{table}
\normalsize}

Taking the LTE-derived column densities, we derive a relative abundance ratio between CO and CH in \apm\ of CH/CO\,$\sim$\,$8\times10^{-5}$. Unlike the typical Milky Way condition where the diffuse gas is dominating the CH emission, in typical fully shielded molecular regions, the CH/CO abundance ratio is expected to be about 10$^{-4}$ \citep{2008ApJ...687.1075S}, which is consistent with our finding in \apm, suggesting the CH lines are unlikely to arise from diffuse gas in \apm. Such an abundance ratio is higher than the value found in Arp\,220, M\,82, and NGC\,253 \citep{2014ApJ...788..147R} by at least a factor of 2. However, the abundance ratio of \apm\ is very close to that of NGC\,1068, where the XDR is likely driving the high abundance of CH. This is also consistent with our previous finding of the abundance comparison between AGN and non-AGN sources (Sect.\,\ref{section:abundance-LTE}), as well as the explanation of the abundance ratio of \httop/HCO$^+$ (Sect.\,\ref{sect:H3O+}). If CH/CO ratios are indeed a robust XDR tracer as proposed in \citet{2014ApJ...788..147R}, future observations of the CH lines in starburst galaxies will be revealing, and we will be able to compare the abundance ratios between AGN and non-AGN sources to test the assumption. If the test is successful, the CH/CO ratio can be further used as a diagnostic tool for high-redshift ISM observations to detect XDR imprints.

\subsection{CO isotopologs: $^{13}$CO and C$^{18}$O }
\label{sect:isotopologs}

CO isotopologs are conceivably promising tracers of stellar IMF in high-redshift dust-enshrouded starburst galaxies \citep[e.g.,][]{2013MNRAS.436.2793D, 2017MNRAS.470..401R, 2018Natur.558..260Z}. In those galaxies, the flux ratio of $^{13}$CO/C$^{18}$O may be assumed to be close to the isotope ratio $^{13}$C/$^{18}$O if both lines of $^{13}$CO and C$^{18}$O are optically thin (however, we caution that in very high-density gas, the $^{13}$CO lines may have a moderate optical depth; see Appendix~\ref{app:taus}). The isotopes, $^{13}$C and $^{18}$O, predominantly originate from low- to intermediate-mass stars (less than 8\,\msun) and massive stars (greater than 8\,\msun), respectively. Therefore, the flux ratio $^{13}$CO/C$^{18}$O, which represents their abundance ratio, can serve as a dust-insensitive stellar IMF probe. \citet{2018Natur.558..260Z} demonstrated that low flux ratios $^{13}$CO/C$^{18}$O, approaching unity, found in a sample of starburst galaxies can be interpreted as evidence of a top-heavy stellar IMF. This implication can be further substantiated by high flux ratio values of CO/$^{13}$CO, specifically those exceeding 20.

\setlength{\tabcolsep}{1.73em}
\begin{table}[!htbp]
\renewcommand{\arraystretch}{1.15}
\footnotesize
\caption{CO isotopolog ratios.}
\begin{tabular}{ccc}
\toprule
Line ratios                                                                  & \multicolumn{2}{c}{Isotopologue line ratio}      \\    
                                                                             &      APM               &     NC                  \\    
\midrule   
\llap{$^{13}$}${\mathrm{CO}}(4\text{--}3)$/${\mathrm{C^{18}O}}(4\text{--}3)$ & $1.10\pm0.4$           &  $\la$ 1                \\
\llap{$^{13}$}${\mathrm{CO}}(3\text{--}2)$/${\mathrm{C^{18}O}}(3\text{--}2)$ &    --                  &  $0.66\pm0.27$          \\
\llap{$^{12}$}${\mathrm{CO}}(4\text{--}3)$/${\mathrm{^{13}CO}}(4\text{--}3)$ & $17\pm5.3$             &  $\ga$ 26               \\
\llap{$^{12}$}${\mathrm{CO}}(3\text{--}2)$/${\mathrm{^{13}CO}}(3\text{--}2)$ &    --                  &  $28\pm9$               \\
                                                                     
\bottomrule  
\end{tabular}
   \label{tab:line-isotop}
      \begin{tablenotes}[flushleft]
   \small
	 \item\textbf{Note:} 
        APM stands for \apm\ and NC for NCv1.143.
   \end{tablenotes}
\end{table}
\normalsize

\citet{2018Natur.558..260Z} present deep observations of the three isotopologs --- CO, $^{13}$CO and C$^{18}$O --- toward four high-redshift lensed galaxies, including three starbursts and one quasar. They found flux ratios $^{13}$CO/C$^{18}$O and CO/$^{13}$CO ranging from 1.0 to 1.6 and from 20 to 25, respectively, which can be interpreted as evidence for a top-heavy stellar IMF. In the spectral surveys reported here, we detected all three CO isotopologs in NCv1.143 in the $J$\,=\,3--2 and 4--3 transitions and, in \apm\ in the $J$\,=\,4--3 transition (Table\,\ref{table:line-flux}). However, the $^{13}$CO(4--3) line is blended with the CS(9--8) line, which is blueshifted with respect to $^{13}$CO(4--3) by only 26 \,km/s. Therefore, the value given for the flux of the blended line in Table\,\ref{table:line-flux} should be corrected for the contribution of the CS(9--8) line. The flux of this line may be estimated by interpolation between the observed values of the CS(8--7) and CS(10--9) lines, assuming a smooth excitation SLED shape. We also acknowledge the substantial uncertainties of the integrated fluxes of the CS lines in NCv1.143 (Table\,\ref{table:line-flux}), which we take into consideration when computing the line ratios conservatively as upper and lower limits. These ratios listed in Table\,\ref{tab:line-isotop} are consistent for the 3--2 and 4--3 lines, and their values for both sources --- $^{13}$CO/C$^{18}$O\,$\la$\,1 and CO/$^{13}$CO\,$\ga$\,20 --- are comparable to the values reported by \citet{2018Natur.558..260Z}. The similar isotopolog ratios can be interpreted as evidence for a top-heavy stellar IMF in both \apm\ and NCv1.143, with implications for parameters derived using common assumptions about the IMF, SFR, and stellar mass.

\subsection{Rotational lines of the vibrationally excited HCN}
\label{sect:hcn-vib}

From our NOEMA line survey, we have detected for the first time emission lines from the vibrationally excited state $v_2$\,=\,1$f$ of HCN (HCN-VIB hereafter) at high redshifts. In the \apm\ quasar, all three transitions of HCN-VIB, \mbox{6--5}, \mbox{5--4}, and 4--3, are detected (Table\,\ref{table:line-flux}). 

These HCN-VIB lines have very high energy levels that exceed 1050\,K \citep{2015A&A...584A..42A}, suggesting that our detection of the bright HCN-VIB lines are unlikely caused only by collisional excitation because of the extreme physical conditions needed \citep{1986ApJ...300L..19Z}. In local galaxies, observations of the HCN-VIB lines found that the most common process powering these lines involve radiative pumping from the mid-infrared (such as 14\,$\mu$m), where HCN-VIB is absorbing the mid-infrared in the most obscured regions where $N_\mathrm{H}$ reach beyond 10$^{24}$--10$^{25}$\,cm$^{-2}$, then releasing the energy via its emission at submillimeter and millimeter bands through the HCN-VIB lines, forming the so-called compact obscured nuclei \citep[e.g.,][]{2010ApJ...725L.228S, 2013AJ....146...91I, 2015A&A...584A..42A, 2019ApJ...882..153G, 2019A&A...623A..29F, 2019A&A...627A.147A, 2021A&A...649A.105F}. This makes the HCN-VIB lines a powerful tool for probing the presence of extremely obscured nuclei.

Based on the observed integrated flux of the HCN-VIB\mbox{(4--3)} line listed in Table\,\ref{table:line-flux}, we estimate line luminosities of $L_\mathrm{HCN\text{-}VIB(4\text{--}3)}=(2.5\pm1.0)\times10^7$\,\lsun\ for APM\,08275+5255. This yields a luminosity ratio of $L_\mathrm{HCN\text{-}VIB(4\text{--}3)}/L_\mathrm{IR}=(2.9\pm1.2)\times10^{-7}$. This value is at the high end of the ratios found in compact obscured nuclei and an order of magnitude above the threshold of $10^{-8}$ considered to enter the criteria of a compact obscured nucleus \citep{2021A&A...649A.105F}. Nevertheless, the HCN-VIB line luminosity could be overestimated if the emitting region is much more compact than the dust continuum and is located at the caustic lines, though we argue that the differential lensing is insignificant (Sect.\,\ref{section:lensing}). While, for NCv1.143, a 3\,$\sigma$ upper limit of the HCN-VIB lines of 0.13\,Jy\,\kms\ results in a limit to its luminosity of $L_\mathrm{HCN\text{-}VIB(4\text{--}3)}<1.1\times10^7$\,\lsun, about a factor of 2--3 below the detected line luminosity in \apm. Such a big difference in the HCN-VIB line between \apm\ and NCv1.143 highlights the fact that the nuclear region of the \apm\ quasar resides in a much more intense infrared radiation field than NCv1.143. This is consistent with the conclusions we derived from our \hto\ excitation model in Sect.\,\ref{sect:h2o-model}. Nevertheless, the upper limit of the ratio $L_\mathrm{HCN\text{-}VIB(4\text{--}3)}/L_\mathrm{IR}<9\times10^{-7}$ in NCv1.143 is still compatible with a compact obscured nucleus, though it is much less powerful than that of \apm. Although a better measurement is needed for NCv1.143, this could be consistent with a buried nuclear starburst or AGN as speculated in Sect.\,\ref{sec.HCNHCOpCSdiagnostic} should the compact obscured nuclei be unequivocally associated with buried AGN. If such a deeply obscured AGN exists in NCv1.143, its influence on the bulk of the ISM is still limited, consistent with most of the diagnostics from other lines presented in this work.

We also note that the surface luminosity of the HCN-VIB line (converting to $J$\,=\,3--2 using local line ratios between 4--3 and 3--2) is about 50\,\lsun\,pc$^{-2}$. Such a value is about five times stronger than the brightest compact obscured nuclei in the local Universe, such as NGC\,4418 (surface brightness of HCN-VIB is about 9.1\,\lsun\,pc$^{-2}$, \citealt{2021A&A...649A.105F}), making \apm\ the brightest HCN-VIB source ever discovered. Yet, unlike most of the compact obscured nuclei, where there have been no X-ray detections \citep{2021A&A...649A.105F}, \apm\ is rather unique. As argued in \citet{2016ApJ...825...44I}, the HCN-VIB line can be boosted by AGN; thus, the unusual brightness of the HCN-VIB in \apm\ is likely powered by the central AGN.

\subsection{Other lines: CN, NO, N$_2$H$^+$, and c-C$_3$H$_2$}
\label{sect:other-lines}
In \apm, we have detected the CN(4--3) emission line, confirming the first detection reported by \citet{2007A&A...462L..45G} with a significantly better S/N. While in NCv1.143, we have detected both the 3--2 and 4--3 transitions of the CN line. CN can be a dense gas tracer with a few factors lower critical density than HCN \citep{2002A&A...381..783A}. In PDRs, CN is primarily formed in the ionization fronts where HCN is dissociated. CN radical can also be formed from CH and N via neutral-neutral reactions \citep{2013RvMP...85.1021T}. On the other hand, in hot-temperature environments, CN can also form HCN directly by interacting with H$_2$ \citep{2010ApJ...721.1570H}. Comparing \apm\ and NCv1.143, we find that the luminosity ratio between CN and HCN is significantly lower in \apm. Comparing directly the abundance ratios (Figs.\,\ref{fig:abundance-compare} and Fig.\,\ref{fig:abundances-compare-2}), we do see the same trend of a smaller ratio of CN/HCN in \apm\ than in NCv1.143. Such a deficiency of CN with regard to HCN in the quasar \apm\ indicates that rather than the ionization process, hot-temperature chemistry powered by the AGN might dominate the CN formation/destruction \citep{2013ApJ...765..108H} in \apm.

Nitric oxide NO lines have been detected in \apm, the first high-redshift detection of this molecule. NO can form from the interaction of N\,$+$\,OH and NH\,$+$\,O, can be dissociated to N$^+$ and NO$^+$, and can also form CN by interacting with C; it is thus an important part of the network of nitrogen-bearing species \citep[e.g.,][]{1989ApJ...340..869N}. NO can also trace dense gas as its critical densities can reach above 10$^5$\,cm$^{-3}$. The derived column density of NO in \apm\ is higher than the values found in local starburst galaxies such as M\,82 \citep{2011A&A...535A..84A} and NGC\,253 \citep{2021A&A...656A..46M}, which can indicate a high abundance of nitrogen and oxygen. Future detailed analysis of the nitrogen chemical-reaction network can help us understand the chemistry.

In \apm, we have also detected the 4--3 and \mbox{5--4} transitions of the classic dense gas tracer, diazenylium N$_2$H$^+$, confirming the first tentative detection of the N$_2$H$^+$(5--4) line report by \citet{2017A&A...608A..30F}. Also, we report a tentative detection of the cyclopropenylidene c-C$_3$H$_2$ line at 485.7\,GHz for the first time at high redshifts, which needs further confirmation.

\section{Chemical analysis of the dense molecular gas}
\label{subsection:Chemical-model}

The unique richness of observational data for both galaxies presented here makes them ideal systems for measuring the heating and cooling processes as well as determining what powers the energy of their ISM via chemical modeling. Using all available lines simultaneously to perform such modeling, however, is an unprecedented and rather challenging task that deserves its own separate and detailed study. 

For the purposes of the present work, we focus on the column densities of four species (HCN, HCO$^+$, HNC, and CS), which are dense gas tracers and thus can tell much about the conditions prevailing in the molecular gas of both galaxies. In particular, the contribution from the stellar far-ultraviolet (FUV) photons can be neglected since the FUV radiation will be severely extinguished in the dense regions with volume densities predicted by the LVG models (Table\,\ref{table:LVG-fit}). For instance, a density of $n_{\rm H}=10^5\,{\rm cm}^{-3}$ is likely to have an effective visual extinction of $A_{\rm V}\simeq30\,{\rm mag}$ \citep[][]{2023MNRAS.519..729B}, and such visual extinction will extinguish the effect of the FUV field even as strong as 10$^5$\,$\chi_0$ \footnote{$\chi_0$ is the FUV field normalized to the \citet{1978ApJS...36..595D} field, and it equals to about 1.7$G_0$, where $G_0$ is the \citet{1968BAN....19..421H} field.}, which is at least one order of magnitude higher than the typical value found in dusty high-redshift starburst galaxies \citep[e.g.,][]{2018A&A...620A..61C,2019ApJ...876..112R}. At such high column densities, the chemistry is no longer driven by FUV radiation but rather by CR, which can be quantified by the CR ionization per H-nuclei ratio ($\zeta_{\rm CR}/n_{\rm H}$) that describes the ionization degree. We therefore focus only on $\zeta_{\rm CR}/n_{\rm H}$ hereafter, which plays an important role in triggering chemical networks at high column densities \citep[e.g.,][]{Bialy15}.
 
For the purposes of this chemical modeling, we use the \texttt{3D-PDR}\footnote{\url{https://uclchem.github.io/3dpdr/}} code, which is an astrochemical code treating photodissociation regions \citep{2012MNRAS.427.2100B}. Using various heating and cooling functions and the LVG approximation \citep{sobolev60,poelman05}, the code performs iterations over thermal balance and terminates once the total heating is approximately equal to the total cooling. In this work, the full UMIST2012 \citep{2013A&A...550A..36M} chemical network is used, consisting of 215 species and approximately 3000 reactions. The initial abundances (normalized to total hydrogen) adopted here are [Mg]=3.981$\times10^{-5}$, [C$^+$]=10$^{-4}$, [S$^+$]=3.236$\times10^{-5}$, [N]=6.76$\times10^{-5}$, [O]=3$\times10^{-4}$ [He]=0.1. A metallicity of $Z=1\,{\rm Z}_{\odot}$ \citep[e.g.,][]{2022MNRAS.517..962D} and a microturbulent velocity of $v_{\rm turb}=1\,{\rm km}\,{\rm s}^{-1}$ are further assumed. 

In addition, we adopted the following expression of optical depths, which best describes macroturbulent molecular clouds \citep{1999ApJ...516..114P}:
\begin{eqnarray}
\label{eqn:tau}
    \tau_{ij}=\frac{A_{ij}c^3}{8\pi\nu_{ij}}\left(\frac{dV}{dr}\right)^{-1}_{\rm VIR}\left[\frac{n_jg_i-n_ig_j}{g_j}\right],
\end{eqnarray}
where
\begin{eqnarray}
\left(\frac{dV}{dr}\right)_{\rm VIR}=0.65\sqrt{\alpha}\left[\frac{n_{\rm H}}{10^3\,{\rm cm}^{-3}}\right]^{1/2}\,{\rm km}\,{\rm s^{-1}}\,{\rm pc^{-1}}.
\end{eqnarray}
In the above equations, $A_{ij}$ is the Einstein coefficient of levels $i$ and $j$, $\nu_{ij}$ is the corresponding line frequency, $n_i$ and $n_j$ the level populations, and $g_i$, $g_j$ their statistical weights. The parameter $\alpha$ depends on the density profile \citep{1999ApJ...516..114P}. Here, we have considered $0.65\sqrt{\alpha}=1$ \citep[see also][]{2015ApJ...803...37B}. Equation~\ref{eqn:tau} is used based on the assumption that in macroturbulent clouds, the optical depth is considered as a local property rather than building along the line-of-sight as treated for local clouds.

\begin{figure}
    \centering
    \includegraphics[width=0.45\textwidth]{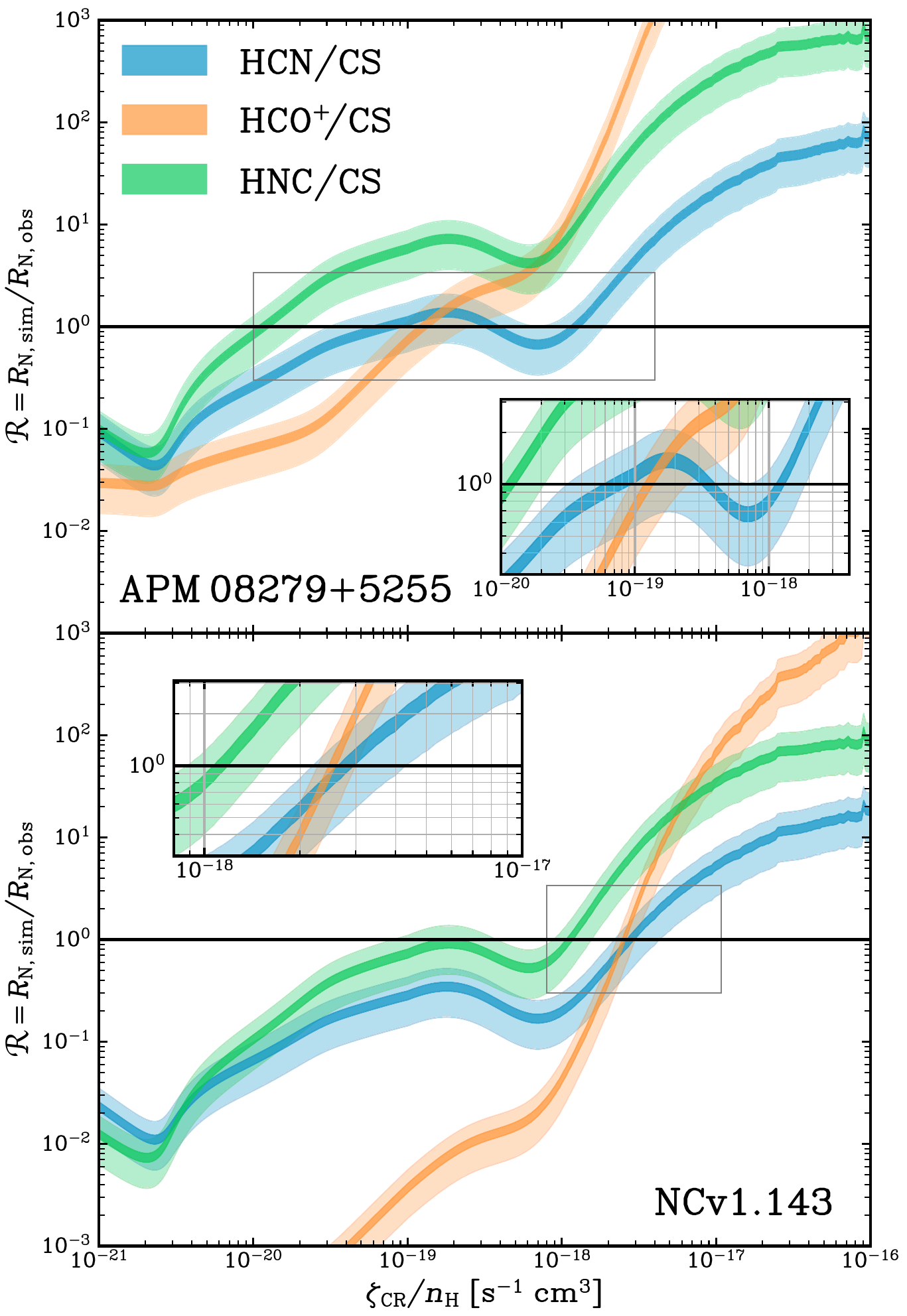} 
    \caption{Chemical modeling using \texttt{3D-DPR} to identify the best range of the CR ionization parameter $\zeta_{\rm CR}/n_{\rm H}$ ($x$-axis) based on the observed column density ratios. The column densities of HCN (blue), HCO$^+$ (orange), and HNC (green) are normalized with CS. The $y$-axis shows the ratio of the modeled column density ratio to the observed one. Unity marks agreement between the models and observations. The upper panel shows the results for APM~08279+5255, and the lower panel shows those for NCv1.143. The insets are zoom-in regions of the gray rectangular area in each panel. The thickness of each line represents the uncertainty (light color is for 50\% uncertainty, and dark color is for 10\% uncertainty).}
    \label{fig:zetan}
\end{figure}

Absolute column densities strongly rely on the assumptions made on the size of the emitting source, which also introduces uncertainty on the opacity of the emission. In order to minimize this uncertainty and the matching to models, we base our analysis on the normalized column density. In this case, we normalize it to the column density of CS as a reference, which will be less affected by optical thickness than, for example, CO. Here, we assume that the spatial distribution among HCN, HCO$^+$, HNC, and CS is similar; however, due to the limit of our current data, this is not certain. Figure\,\ref{fig:zetan} shows the results of our chemical modeling in which we plot for both sources the ratio ${\cal R}=R_{\rm N,sim}/R_{\rm N,obs}$ versus $\zeta_{\rm CR}/n_{\rm H}$, where ${\cal R}$ is the ratio between the modeled column densities ratios and the observed ones from Table~\ref{table:LVG-fit}. We adopt two uncertainties for this comparison, a 10\% (dark color) and a 50\% (light color) of the mean reported in this table, for the purpose of demonstrating how the measurement uncertainties impact our analysis. However, we acknowledge that our estimations of CR ionization rates have large uncertainties.  

\setlength{\tabcolsep}{0.72em}
\begin{table}[!htbp]
\renewcommand{\arraystretch}{1.15}
\footnotesize
\caption{\texttt{3D-PDR} model results for \apm\ and NCv1.143. The $\zeta_{\rm CR}$ value of the third column was calculated using the $n_{\rm H}$ density of CS from the LVG results shown in Table~\ref{table:LVG-fit}.}
\begin{tabular}{rrcc}
\toprule
                            &    Ratio      &  $\zeta_{\rm CR}/n_{\rm H}$ (cm$^3$\,s$^{-1}$) & $\zeta_{\rm CR}$ (s$^{-1}$) \\ 
\midrule
\multirow{3}{*}{APM\,08279} &    HCN/CS     & $1\text{--}15\times10^{-19}$ & $0.2\text{--}3.0\times10^{-12}$ \\
                            &    HCO$^+$/CS & $2\times10^{-19}$ & $0.4\times10^{-13}$ \\
                            &    HNC/CS     & $2\times10^{-20}$ & $0.4\times10^{-14}$ \\
\bottomrule   
\multirow{3}{*}{NCv1.143}   &    HCN/CS     & $3\times10^{-18}$ & $1.2\times10^{-12}$ \\
                            &    HCO$^+$/CS & $3\times10^{-18}$ & $1.2\times10^{-12}$ \\
                            &    HNC/CS     & $2\times10^{-18}$ & $0.8\times10^{-12}$ \\
\bottomrule  
\end{tabular}
   \label{tab:NCv_results}
\end{table}
\normalsize 

Tables~\ref{tab:NCv_results} present the modeled results for NCv1.143 and \apm, respectively. It appears that for both cases, no single solution for the column densities considered is found. Using the $n_{\rm H}$ number density of CS from Table~\ref{table:LVG-fit} it can be seen that the estimated range of CR ionization rates of $\zeta_{\rm CR}\sim(4-40)\times10^{-14}\,{\rm s}^{-1}$ for \apm\ (although the high $\zeta_{\rm CR}/n_{\rm H}$ solution of the HCN/CS ratio suggests a higher value of $\zeta_{\rm CR}\sim10^{-12}\,{\rm s}^{-1}$), and of $\zeta_{\rm CR}\simeq(8-12)\times10^{-13}\,{\rm s}^{-1}$ for the NCv1.143 galaxy. In both galaxies, the HNC/CS ratio shows systematically lower values of $\zeta_{\rm CR}$, while the results from the HCN/CS and HCO$^+$/CS ratios generally agree with each other.
 
Similar high $\zeta_{\rm CR}$ values, as those found here, have been previously reported in systems from the Galactic Center to high-redshift galaxies --- Bremsstrahlung radiation \citep{Yusef2013} and H$_3^+$ observations \citep{Goto2014,LePetit2016} in the Galactic Center suggest a CR ionization rate $\zeta_{\rm CR}\simeq(1-100)\times10^{-15}\,{\rm s}^{-1}$, which is 2--4 orders of magnitude higher than the Milky Way average. The nearby and well-studied NGC~253 starburst is suggested to have a $\zeta_{\rm CR}\gtrsim10^{-14}\,{\rm s}^{-1}$ derived from the observed HCO$^+$/HOC$^+$ emission ratio \citep{Harada2021}. Similarly, \citet{Holdship2021} report an ionization rate $\zeta_{\rm CR}\sim10^{-11}\,{\rm s}^{-1}$ derived from chemical models of C$_2$H. However, in a subsequent paper and using H$_3$O$^+$ and SO observations, \citet{Holdship2022} suggest a lower value of $\zeta_{\rm CR}\simeq(1-80)\times10^{-14}\,{\rm s}^{-1}$ for NGC~253. Very high $\zeta_{\rm CR}$ values have also been derived for other star-forming galaxies. OH$^+$, H$_2$O$^+$ and H$_3$O$^+$ observations in Arp~220 find a $\zeta_{\rm CR}/n_{\rm H}\simeq(1-2)\times10^{-17}\,{\rm cm}^3\,{\rm s}^{-1}$, which leads to a $\zeta_{\rm CR}\gtrsim10^{-13}\,{\rm s}^{-1}$ for an estimated $n_{\rm H}\sim10^4\,{\rm cm}^{-3}$ \citep{2013A&A...550A..25G}. Similarly, the aforementioned lines observed for Mrk~231 also suggest a $\zeta_{\rm CR}/n_{\rm H}\simeq(1-2)\times10^{-16}\,{\rm cm}^3\,{\rm s}^{-1}$ for the outflowing component implying a $\zeta_{\rm CR}\sim(0.5-2)\times10^{-12}\,{\rm s}^{-1}$ \citep{2018ApJ...857...66G}. \citet{2018ApJ...853L..25I} studied OH$^+$ and H$_2$O$^+$ lines of the extended halos of $z\sim2.3$ galaxies and derived a $\zeta_{\rm CR}\sim(1-100)\times10^{-16}\,{\rm s}^{-1}$, where the ionization rates in the compact star-forming regions are expected to be orders of magnitude higher \citep{2018ApJ...865..127I}. \citet{2016A&A...595A.128M} also found higher than Milky Way CR ionization rate in the $z$\,=\,0.89 molecular absorbers toward PKS1830-211. All these results suggest that the dense molecular gas in these high-redshift star-forming galaxies is subjected to high CR ionization rates. As pointed out by \citet{2023MNRAS.520.5126K}, assuming star formation is the dominant source of CR, $\zeta_{\rm CR}$ scales with the gas depletion time. Therefore, such high $\zeta_{\rm CR}$ values of $\sim10^{-14}\text{--}10^{-12}\,{\rm s}^{-1}$ indicates a very short gas depletion timescale of possible only a few to a few tens of megayears, taking the prescription from \citet{2023MNRAS.520.5126K}, consistent with their intense starburst nature.

\begin{figure}
    \includegraphics[width=0.465\textwidth]{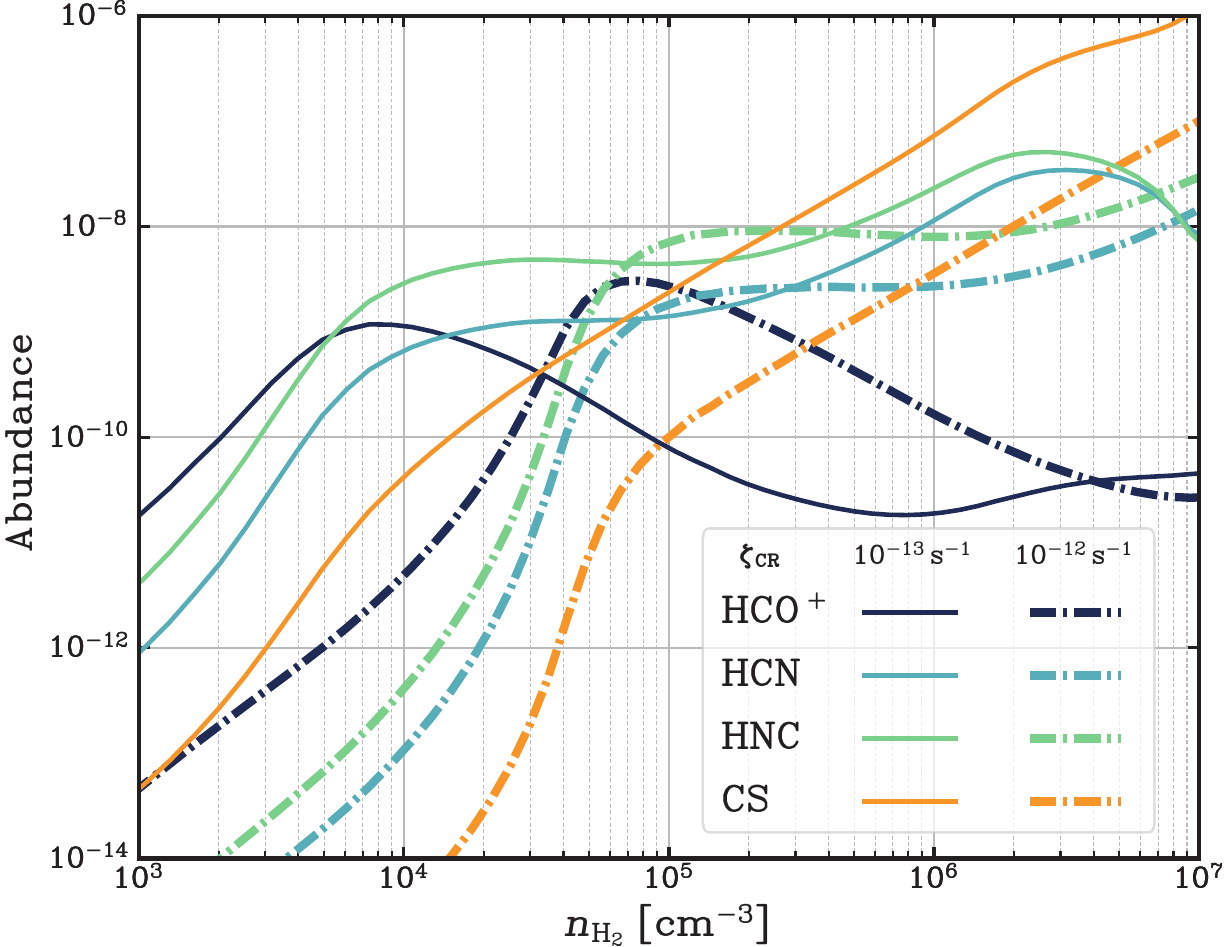}
    \caption{Abundances of HCN, HCO$^+$, HNC and CS versus $n_{\rm H_2}$, for two different CR ionization rates (solid lines for $\zeta_{\rm CR}$\,=\,10$^{-13}\,{\rm s}^{-1}$ and solid-dotted lines for 10$^{-12}\,{\rm s}^{-1}$). Our model shows that with such a high ionization rate $\zeta_{\rm CR}$\,=\,10$^{-13}$--10$^{-12}$\,${\rm s}^{-1}$, the abundances of HCO$^+$ is expected to be lower than HCN, HNC, and CS in the dense gas regions of $n_{\rm H_2}$\,>\,10$^{5}$\,cm$^{-3}$. However, with increasing CR ionization rates, at the same H$_2$ densities, the abundance of HCO$^+$ is rising for the dense gas while the abundance of CS is decreasing.}
    \label{fig:abundances_density_CR}
\end{figure}

Taking the derived CR ionization rate values in Table\,\ref{tab:NCv_results} for both galaxies, in Fig.\,\ref{fig:abundances_density_CR}, we show the variation of the abundances in HCN, HCO$^+$, HNC, and CS with gas density. For \apm, where the gas densities reach above $n_{\rm H_2}$\,$\sim$\,10$^5$\,cm$^{-3}$ and $\zeta_{\rm CR}\simeq 10^{-13}\,{\rm s}^{-1}$, we found that our chemical model predicts a low HCO$^+$ abundance compared to other species, lower by about two orders of magnitude. This is strikingly consistent with the LVG-derived column densities, as shown in Fig.\,\ref{fig:LVG-violin}, that the abundance of HCO$^+$ is about two orders of magnitude lower than that of HCN, HNC, and CS. For NCv1.143, if we assume a $\zeta_{\rm CR}\simeq 10^{-12}\,{\rm s}^{-1}$, we see a significant increase in the relative abundance of the HCO$^+$ (a detailed discussion of the chemistry of HCO$^+$ is given in Appendix\,\ref{app:hcop_chem}), and a small decrease in the CS abundance while HCN and HNC remain similar values of abundances. For NCv1.143, both a higher gas density and likely a higher $\zeta_{\rm CR}$ lead to similar relative abundances among the dense gas tracers, consistent with Fig.\,\ref{fig:LVG-violin}. It is also worth noticing that, given such a CR ionization rate, the relative abundance of HCO$^+$ drops toward the denser conditions from $n_{\rm H_2}$\,$\sim$\,10$^5$\,cm$^{-3}$, while CS abundance increases with increasing gas densities. Around the gas densities of $n_{\rm H_2}$\,$\sim$\,10$^{5\text{--}6}$\,cm$^{-3}$, the abundance ratio of HCO$^+$/CS is most sensitive to the change of the CR ionization among the four species, with the HCO$^+$ abundance rising and that of CS decreasing as $\zeta_{\rm CR}$ increases. 

We note that the discussions above assume that volumetric heating at high column densities arises from the presence of CRs only, neglecting other heating sources that may also contribute to gas heating. For instance, strong X-ray sources can equally contribute to the CR heating and also affect the chemistry in a similar way \citep[e.g.,][]{2007A&A...461..793M, Gallerani2014}, though they are alternated much more quickly than CRs and $\propto R^{-2}$. Should such heating sources be accounted for, they will reduce the values of CR ionization rates we find. We therefore note that the $\zeta_{\rm CR}$ values presented above shall be considered as upper limits. 

Nevertheless, regardless of whether the heating is caused by CRs only or with contributions from X-rays and shocks, the initial conditions of star formation will be altered under such ISM conditions, causing a boosted Jean mass and a top-heavy stellar IMF \citep{2010ApJ...720..226P, 2011MNRAS.414.1705P}. As predicted by hydrodynamical simulations, such high-temperature gas of $\sim$\,100\,K and H$_2$ density of $\gtrsim$\,10$^5$ will lead to a stellar IMF peaking at $\gtrsim$\,15\,$M_\odot$ \citep{2007MNRAS.374L..29K}. Moreover, such a top-heavy stellar IMF is consistent with the observed CO isotopolog ratios in both galaxies, which have been interpreted as a result of enrichment by massive stars (Sect.\,\ref{sect:isotopologs}). A similar scenario is presented in the lensed starburst galaxy SMM\,J2135-0102, where strong gas heating (likely by CR) leads to an enhanced C$^{18}$O abundance \citep{2013MNRAS.436.2793D}.

\section{Impact of the lensing on the analysis}
\label{section:lensing}

Given that our observations are based on globally integrated values using unresolved interferometric data, we do not have information on how each line emission is distributed spatially. In both sources, the background galaxy is strongly lensed, while the foreground galaxies are not visible at longer wavelengths and thus do not contaminate our observations. The lensing can bring potential differential lensing effects, where the magnification factor can be different for emissions coming from different parts of the lensed galaxies \citep[e.g.,][]{2011ApJ...733L..12R, 2012MNRAS.424.2429S, 2012ApJ...761...20H, 2012ApJ...753..134F, 2019A&A...624A.138Y, 2022MNRAS.510.3734D}. This can bring more uncertainties to our analysis because we assume minimum differential lensing across all the molecular lines, which is, of course, a first-order approximation, and yet the only assumption we can make until high-angular-resolution imaging of individual lines is obtained. 

For NCv1.143, to better understand how lensing magnification impacts our results, we built the lens model based on 0.3\arcsec\ resolution data of the CO and \hto, as described in Appendix\,\ref{appen:lens-model}. We find a median lensing magnification of about 12 for the dust continuum and about 8--10 for the CO(10--9), \htot211202\ and \htot321312\ emission. Because the line profiles are not strongly distorted by the magnification, and the resolved line profiles of mid/high-$J$ CO and \htot211202\ and \htot321312 are similar, we can expect that the differential lensing is negligible among the dense gas tracers, at least. This is further supported by the fact that almost all the line profiles from our line survey are similar, except for the 380\,GHz \hto\ maser line. However, we caution that some of the diffuse gas tracers in NCv1.143 can have a significantly smaller magnification factor, likely leading to an underestimation of the intrinsic fluxes.

For \apm, we do not expect significant differential lensing. As noted by \citet{2009ApJ...690..463R}, even under the most extreme scenarios, the magnification fluctuates mildly, from $\lesssim$\,5 with small variations for a source radius up to 3\,kpc, down to 3 if the source extends further past 10\,kpc. Given that such an extensive expansion is highly improbable for the molecular emissions observed, substantial differential lensing in \apm\ is unlikely.

\section{Conclusions}
\label{section:conclusion}

Using NOEMA, we conducted deep 3\,mm spectral line surveys toward a dusty quasar, \apm\ at $z$\,=\,3.911, and a dusty star-forming galaxy, NCv1.143 at $z$\,=\,3.565, with total a bandwidth coverage of about 200\,GHz in the rest frame, from around 330 to 550\,GHz. The line survey data allow us to study the high-redshift ISM with an unparalleled level of detail. Our main findings are:

\begin{itemize}

    \item Utilizing \texttt{MADCUBA} \citep{2019A&A...631A.159M} plus a matched-filtering method, we detect 38 emission lines in \apm\ and 25 emission lines in NCv1.143, from \ci, CO, $^{13}$CO, C$^{18}$O, CN, CCH, HCN, HCO$^+$, HNC, CS, C$^{34}$S, H$_2$O, H$_3$O$^+$, NO, N$_2$H$^+$, CH, and \mbox{c-C$_3$H$_2$}, as well as the rotational levels of the vibrationally excited HCN (HCN-VIB). The CH, CCH, c-C$_3$H$_2$, N$_2$H$^+$, HCN-VIB, and H$_3$O$^+$ lines are the first-ever individual high-redshift detections of any transitions of the species. The luminosities of the lines between the two sources behave differently: the SLEDs of most of the tracers are more elevated in \apm, indicating more extreme excitation conditions. The line luminosities are higher in \apm, except for the \ci(1--0) line. The most significant difference is seen in 448\,GHz \hto\ line, where \apm\ is about nine times brighter than NCv1.143 in flux.

    \item Under the LTE assumption, we derived the relative abundances of the molecules for both sources. By comparing them with some prototype local galaxies, we find that NCv1.143 is more similar to that of the CMZ of the starburst-dominated galaxy, NGC\,253, while \apm\ behaves similarly to some local galaxies where the ISM is expected to be dominated by AGN, such as the central $\sim$\,2\,kpc regions of NGC\,1068 and NGC\,4418. We have also found that the abundances of dense gas tracers are more enhanced in the AGN-dominated sources. Additionally, we find a lower CN/HCN ratio in \apm\ than in NCv1.143, suggesting that the AGN-powered hot-temperature chemistry that converts CN into HCN, is important in \apm.

    \item We performed non-LTE LVG modeling of HCN, HCO$^+$, HNC, CS, and C$^{34}$S, which are trace much denser gas than CO: in the range $n_\mathrm{H_2}$\,$\sim$\,10$^{5}$--10$^{7}$\,cm$^{-3}$. The kinetic temperatures derived from all the species are consistent with $T_\mathrm{kin}$\,$\sim$\,50--130\,K. Interestingly, the LVG-derived column densities are consistent with the LTE-derived values, and we highlight a generally low abundance of HCO$^+$ compared with other dense gas tracers.

    \item In \apm, we detect the 380\,GHz \hto\ maser lines, which contain high-velocity components that are likely linked to the maser spots located on the fast-rotating accretion disk around the central SMBH. This 380\,GHz feature is the only identified maser line in \apm\ from the whole rest-frame frequency of $\sim$\,330--550\,GHz. The flux of this line is at least three times stronger than the undetected 22\,GHz, making it a very promising submillimeter \hto\ megamaser candidate in high-redshift galaxies. Simply scaling from local galaxies, we crudely derived the SMBH mass, which is about $10^{10}$\,\msun, consistent with previous literature values.

    \item After including the two water emission lines detected here for the first time, we built a \hto\ excitation model using multiple transitions of \hto\ and the dust continuum. We find the \hto\ excitation conditions in both sources are best explained by two components, the more extended of which are about 1.3 and 0.9\,kpc in \apm\ and NCv1.143, respectively, and both have a dust temperature of $\sim$\,50\,K. The compact component of \apm\ ($R\sim320$\,pc) has a high dust temperature of about 211\,K that dominates the total infrared luminosity output,  and that of NCv1.143 ($R\sim170$\,pc) has a dust temperature of about 98\,K with a minor contribution to the infrared luminosity. Such a difference in their compact component shows that \apm\ has a highly prominent compact core that exhibits extreme conditions and dominates the total power output at the far infrared, indicating that the AGN in \apm\ is in the ``turn-on'' phase, in contrast with NCv1.143. This showcases the power of \hto\ lines as a probe of ISM structures in dusty galaxies.

    \item While the line flux ratios among HCN, HCO$^+$, and CS are inclusive of the imprint of the XDRs versus the PDRs, the abundance ratios from the LVG analysis suggest that the AGN plays an important role in regulating the abundance of these dense gas species in \apm. This is further supported by the abundance ratios between \httop\ (detected in both \apm\ and NCv1.143, the first high-redshift detections) and HCO$^+$, where we find that XDRs play an important role.

    \item We have also detected the isotopologs of CO ($^{13}$CO and C$^{18}$O) in both sources. We find similar flux ratios of $^{13}$CO/C$^{18}$O and $^{12}$CO/$^{13}$CO, as in other starburst galaxies. These ratios are interpreted as evidence for a top-heavy stellar IMF as proposed by \citet{2018Natur.558..260Z}, which can significantly impact estimations of the SFRs.

    \item In \apm, we have detected the HCN-VIB lines (rotational transition from $J_\mathrm{up}$\,=\,4 to 6) at high redshifts for the first time. These lines probe deep into the compact obscured nucleus of the quasar. In NCv.1.143, the upper limit of the HCN-VIB line does not rule out the possibility of a deeply buried AGN. Our calculation of the HCN-VIB surface brightness makes \apm\ the brightest HCN-VIB emitter discovered to date.

    \item Using a three-dimensional PDR model, we explored the properties of the ISM and the radiation fields by taking the chemistry into account. Using the LVG-derived abundance ratios, we find that both our sources are extremely dust-obscured and that FUV radiation plays a limited role because of the high $A_\mathrm{V}$. However, we do find high ionization rates in both sources at a level of 10$^{-13}$ to 10$^{-12}$\,s$^{-1}$, which is likely caused by high densities of CRs, although we do not rule out the possibility of contributions from X-rays. Such an extreme condition is likely altering the initial conditions of star formation, leading to a top-heavy stellar IMF, consistent with the enhanced C$^{18}$O/$^{13}$CO ratio observed. Additionally, our chemical model predicts a low abundance of HCO$^+$ compared to HCN, HNC, and CS in such high-density and high-CR-ionization conditions. This provides a possible explanation of the relatively low abundance of HCO$^+$ derived from our LVG modeling.

    \item We built detailed models taking chemistry and radiative transfer into account. However, we caution that our line fluxes are derived from spatially unresolved integrated values. In reality, the spatial distribution of each molecular emission can vary significantly across a single galaxy, depending on the ISM conditions in different regions. This has been observed in the pure-starburst prototype galaxy NGC\,253 \citep{2017ApJ...835..265W, 2021A&A...656A..46M}, where the CMZ has a very distinct chemical process and thus different spatial distributions of molecular emission than its galactic disk. The variation is even more pronounced when a central AGN is present, as shown in the case study of NGC\,1068 \citep{2014A&A...567A.125G}. In our line survey data, the intrinsic variations found in NGC\,253 and NGC\,1068 add more uncertainties and lead to difficulties when interpreting the data, although one can group several tracers according to the typical conditions they trace in common and assume that their emission distributions are coincident. Nevertheless, high-resolution imaging of the individual line emission is needed for a more comprehensive picture of the ISM in \apm\ and NCv1.143.

\end{itemize}

With the drastic improvement of the bandwidths and sensitivities of (sub)millimeter interferometers, at high redshifts, we can now detect not only CO, the traditional low-density gas tracer, but also the emission of a score of high-excitation and more complex molecules, including some rare isotopologs. With the richness of lines detected from these molecules, we are able to understand the physical conditions, structures, and chemical processes of the ISM at high redshifts in details never before achieved. Future high-angular-resolution line survey data will help us further break the degeneracies in the models of gas excitation and ISM chemistry.

\begin{acknowledgement}
We thank the anonymous referee for many helpful comments and suggestions.
This work is based on observations carried out under project numbers W15EP, S18DC, and W18EB with the IRAM NOEMA Interferometer. IRAM is supported by INSU/CNRS (France), MPG (Germany), and IGN (Spain). The authors are grateful to the IRAM staff for their support. 
Chentao Yang and Susanne Aalto acknowledge support from ERC Advanced Grant 789410. Chentao Yang gratefully thanks Serena Viti for the enlightening discussions during several conferences. Chentao Yang thanks Melanie Krips for her involvement in the observation proposals. Chentao Yang is in debt of the insightful discussion with Santiago Del Palacio about X-ray corona emission, as well as the wonderful discussions with Jim Braatz, Dom Pesce, Violette Impellizzeri, Christian Henkel, Elizabeth Humphreys, and Chengyu Kuo on \hto\ masers. Chentao Yang also thanks Masatoshi Imanishi and Matus Rybak for the nice discussions. 
EG-A thanks the Spanish MICINN for support under projects PID2019-105552RB-C41 and PID2022-137779OB-C41.
Kirsten Knudsen acknowledges support from the Swedish Research Council and the Knut and Alice Wallenberg Foundation. 
Rob J. Ivison acknowledges funding by the Deutsche Forschungsgemeinschaft (DFG, German Research Foundation) under Germany's Excellence Strategy -- EXC-2094 -- 390783311.
\\
 
We acknowledge the usage of the following \texttt{Python} packages: 
\texttt{Astropy} \citep{2018AJ....156..123A}, 
\texttt{Matplotlib} \citep{Hunter:2007}, 
\texttt{NumPy} \citep{harris2020array},
\texttt{SciPy} \citep{2020SciPy-NMeth},
\href{https://github.com/keflavich/pyradex}{\texttt{PyRadex}},
\texttt{Emcee} \citep{emcee}, and
\texttt{Corner} \citep{corner}. 
And the following \texttt{Julia} package: \texttt{Cosmology.jl} \citep{2019ASPC..523..147T}, and
\texttt{Measurements.jl} \citep{2016arXiv161008716G}.

\end{acknowledgement}

\small{
\bibliographystyle{aa_url}
\bibliography{ref.bib} 
}

\begin{appendix}

\onecolumn 
\normalsize	

\section{Zoomed-in view of the spectra and the fittings}
\label{app:zoom-in}
\begin{figure*}[htbp]
\centering
\includegraphics[scale=0.84]{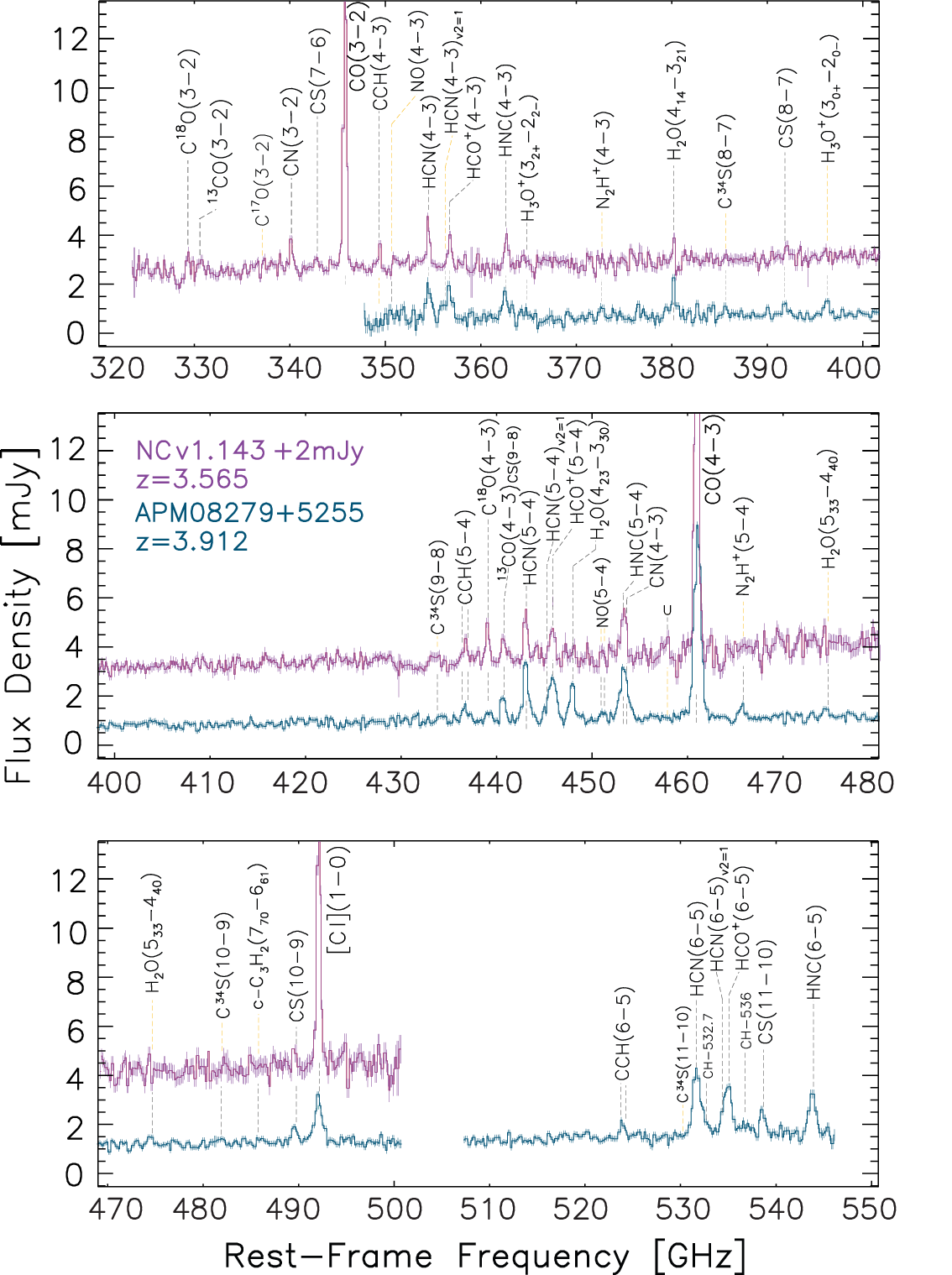}
\caption
{
Same as Fig\,\ref{fig:spec}, but zoomed in to show details 
of the line profiles.
}
 \label{fig:zoom-in_full-spec}
 \end{figure*}
 
\clearpage

\begin{figure*}[htbp]
\centering
\includegraphics[scale=0.445]{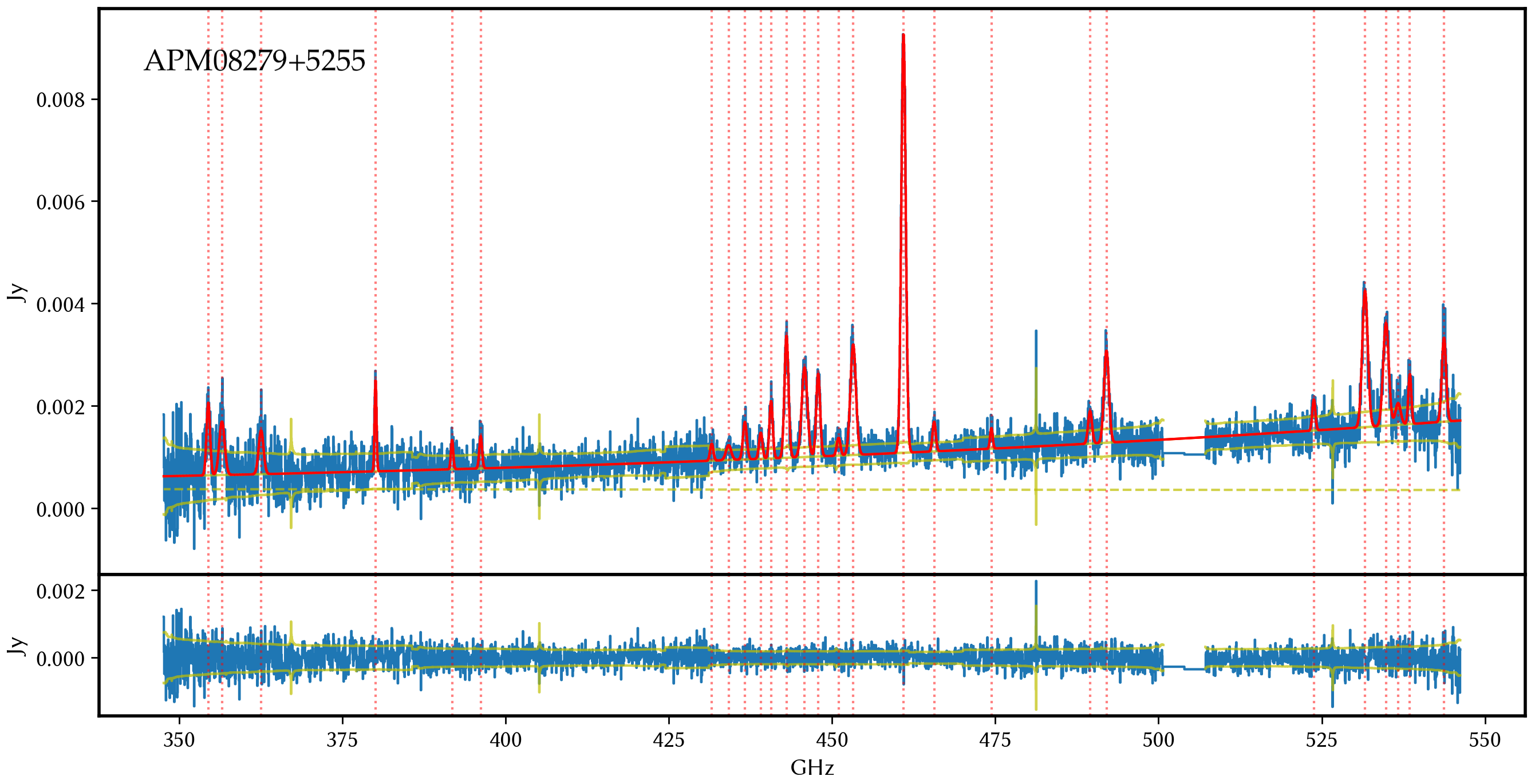}
\vspace{-0.2cm}
\caption
{
Fitting results of the spectra of \apm, with the x-axis as the rest-frame frequency and the y-axis as the observed flux density. The top panel shows the observed spectrum in blue with the uncertainties in the green line. The red solid lines indicate the fit to the entire spectrum, and the dashed red lines are the position of the identified lines. The lower panel shows the residual.
}
\label{fig:flux-fitting-1}
\end{figure*}

\begin{figure*}[htbp]
\centering
\includegraphics[scale=0.435]{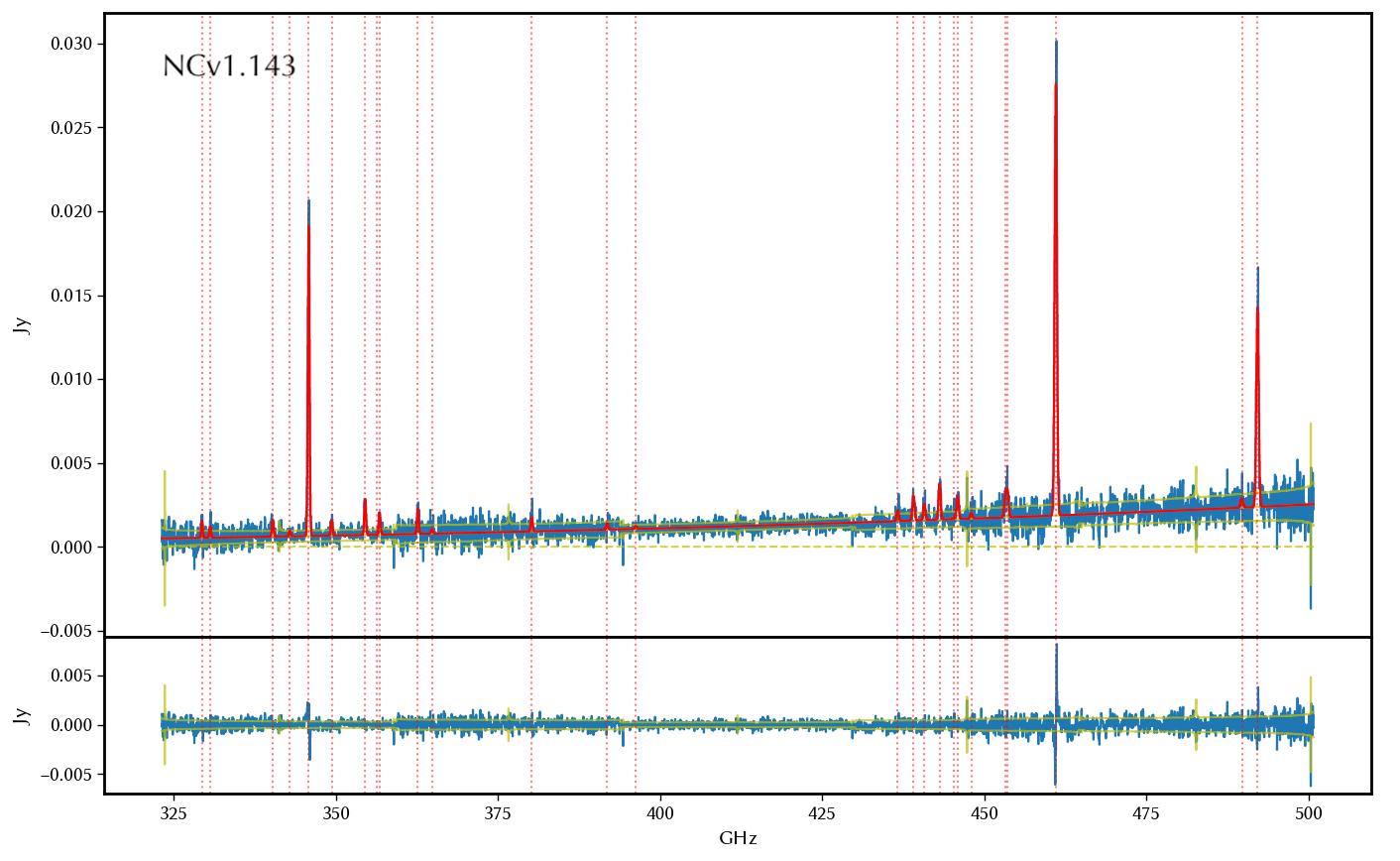}
\vspace{-0.2cm}
\caption
{
Same as Fig.\,\ref{fig:flux-fitting-1} but for NCv1.143.
}
\label{fig:flux-fitting-2}
\end{figure*}

\clearpage

\section{Posterior distribution of the LVG parameters}
\label{sect:appendix:LVG}
This section shows the posterior probability distributions of molecular gas density $n_\mathrm{H_2}$, gas temperature $T_\mathrm{kin}$ and column density per velocity $N\mathrm{_{mol}/d}v$ of HCN, HCO$^+$, HNC, CS and C$^{34}$S in \apm\ and NCv1.143.

\begin{figure*}[htbp]
\centering
\includegraphics[scale=0.36]{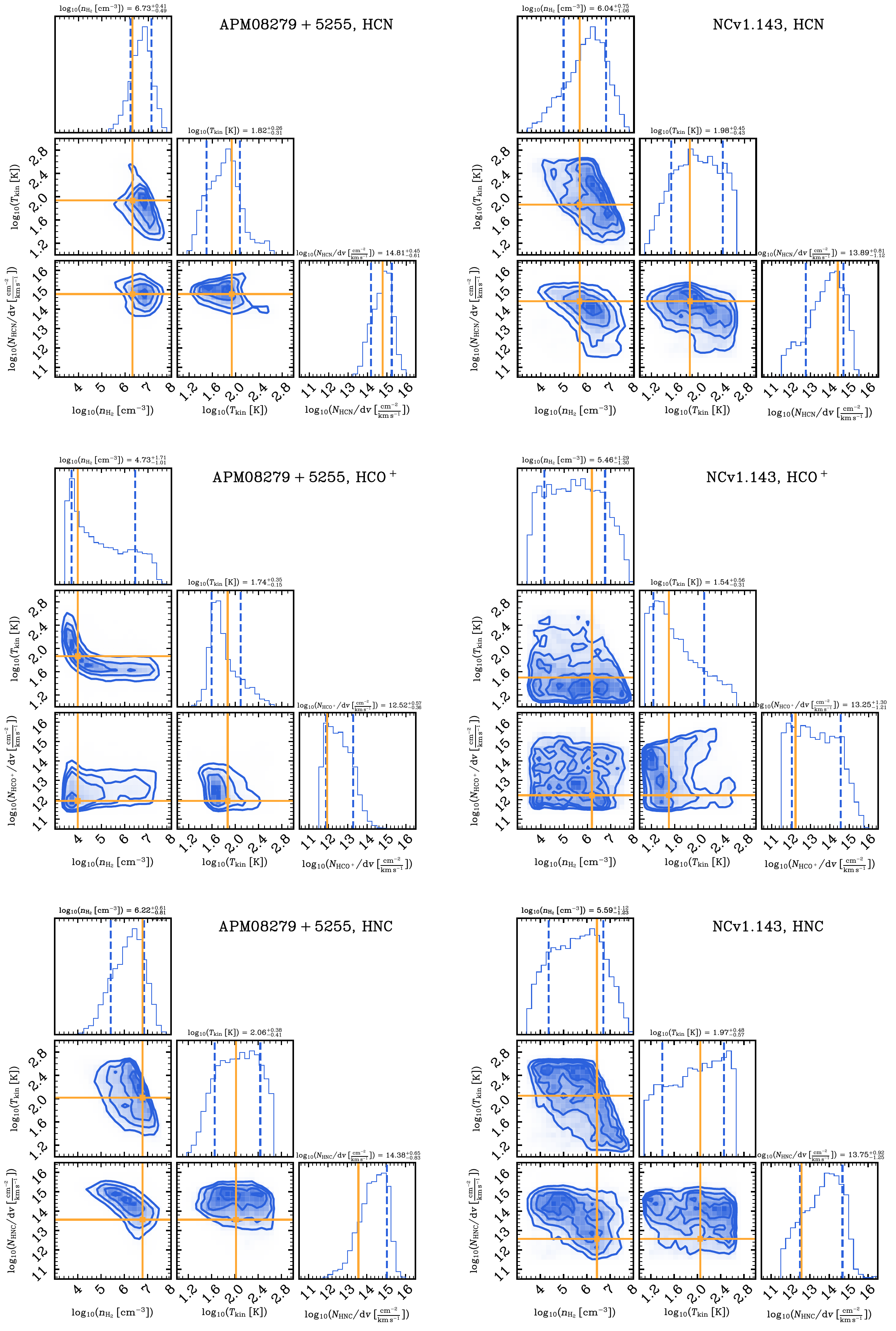}
\vspace{-0.2cm}
\caption
{Posterior probability distributions of molecular gas density ($n_\mathrm{H_2}$), gas temperature ($T_\mathrm{kin}$) and column density per velocity ($N\mathrm{_{mol}/d}v$) of each species in \apm\ and NCv1.143, with the maximum posterior possibility point in the parameter space shown in orange lines and points. The contours are in steps of 0.5\,$\sigma$ starting from the center.
}
\label{fig:LVG-corner-1}
\end{figure*}

\newpage

\begin{figure*}[htbp]
\centering
\includegraphics[scale=0.36]{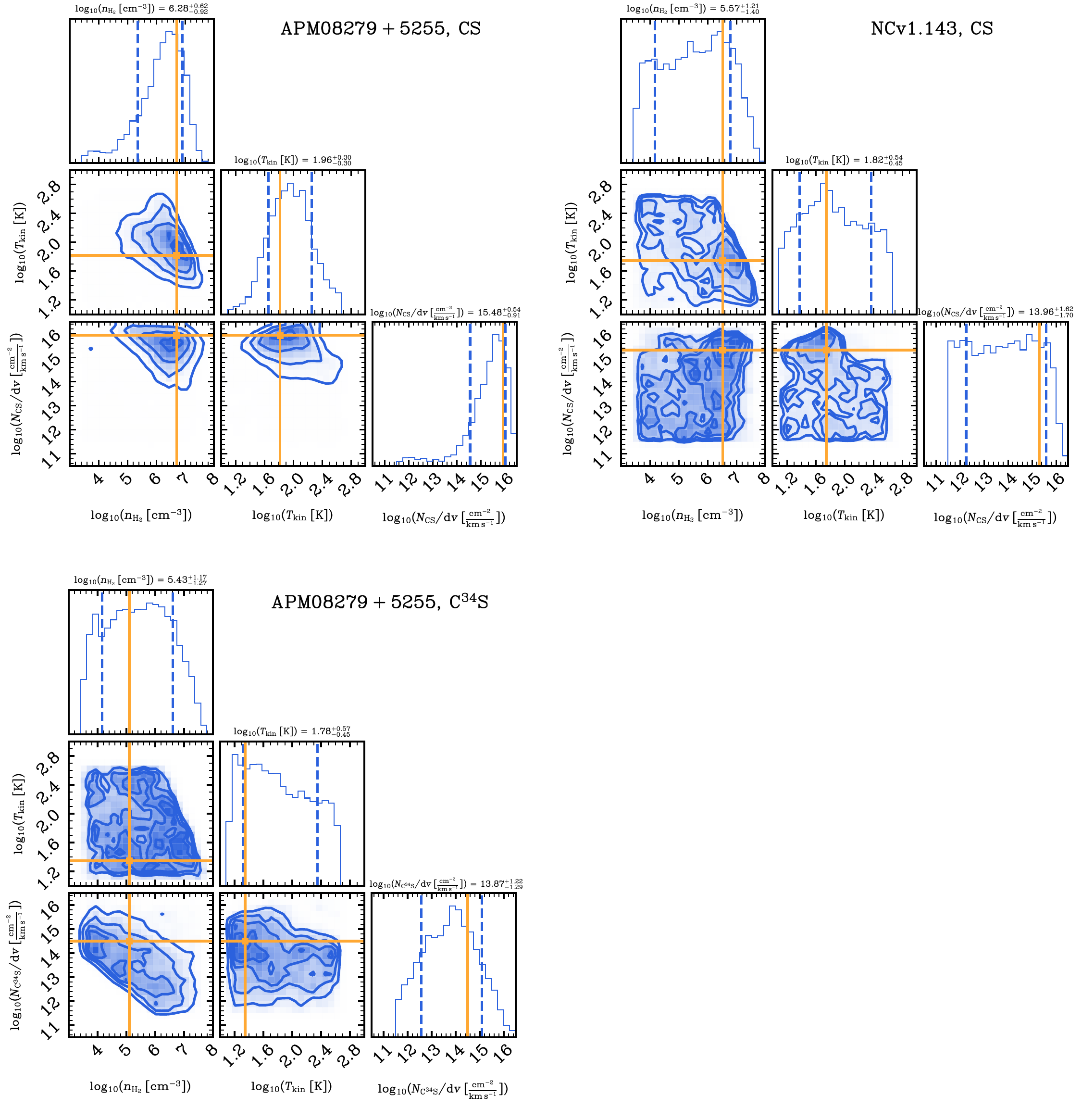}
\vspace{-0.2cm}
\caption
{Same as Fig.\,\ref{fig:LVG-corner-1} but for the CS and C$^{34}$S lines.
}
\label{fig:LVG-corner-2}
\end{figure*}

\clearpage
\twocolumn
 
\section{Lens model}
\label{appen:lens-model}

Our lens model for NCv1.143 was based on high-spatial-resolution data gathered from NOEMA under project W15EP, which has improved sensitivity and spatial resolution compared with the Submillimeter Array (SMA) observations from \citet{2013ApJ...779...25B}. Additionally, the NOEMA data also covers the multiple molecular line emissions, including two \hto, one \htop, and the CO(10--9) lines. 

NOEMA observations were conducted under excellent atmospheric conditions (seeing of 0.3\arcsec--1.0\arcsec and PWV$\leq$1 mm) from February to March 2016, using A-configuration (baselines are from 93 to 760\,m). The nominal angular resolution reached is approximately 0.5"$\times$0.3" at 1.2 mm and 0.7\arcsec$\times$0.4\arcsec\ at 1.8 mm. Seven antennas were used with a total on-sources time of 5.7 hours for the entire series of observations. Standard NOEMA calibrators, including 3C279, 3C273, MWC349, and 0923+392, were used for phase and bandpass calibration. Calibration, imaging, \texttt{CLEAN}ing, and spectra extraction were executed within the \texttt{GILDAS} packages \texttt{CLIC} and \texttt{MAPPING}.

 \begin{figure*}[htbp]
 \centering
 \includegraphics[scale=0.24]{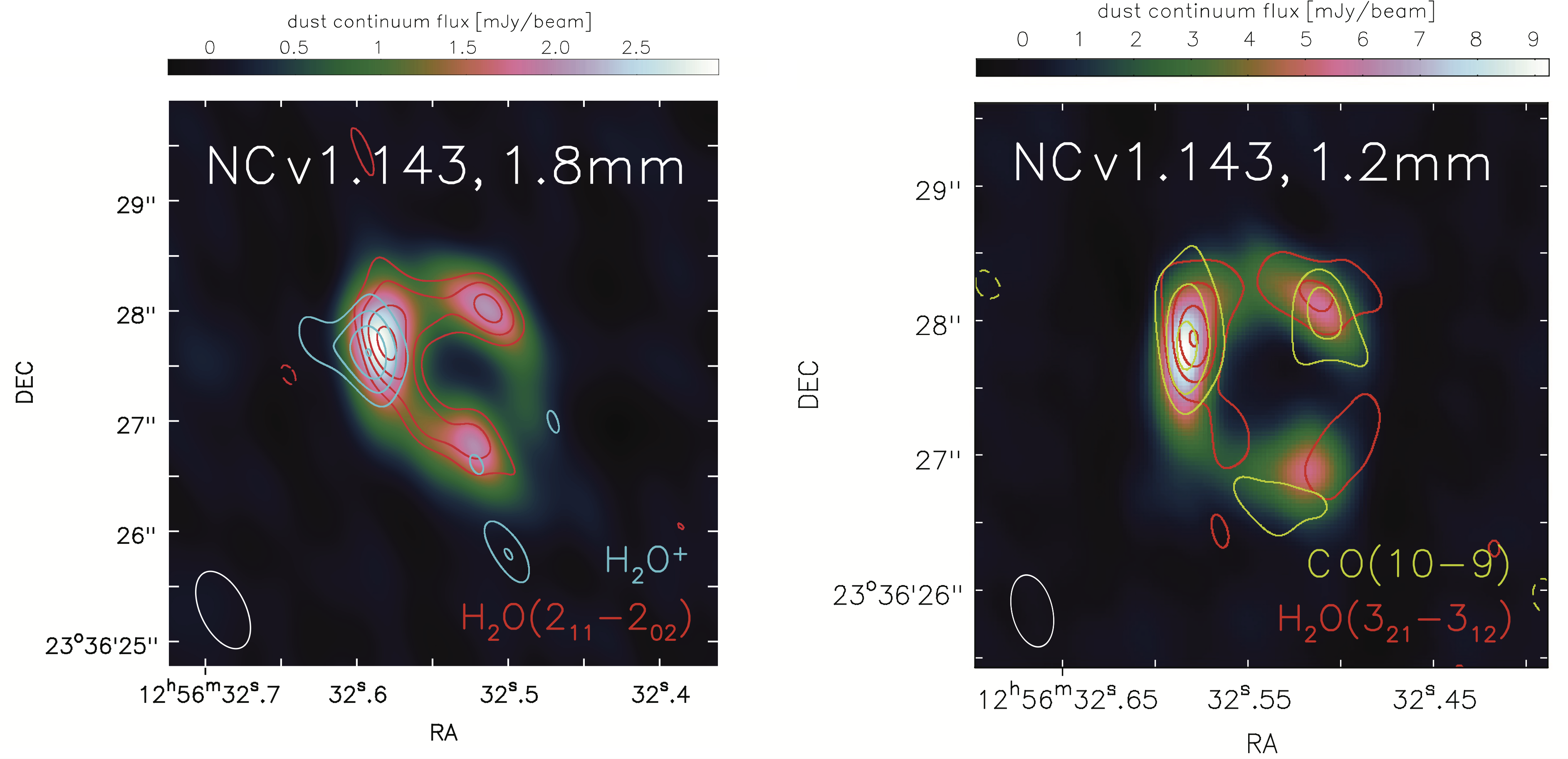}
 \caption{
          NOEMA images of NCv1.143. Color images are the $\sim$1.8\,mm and $\sim$1.2\,mm dust continuum. Yellow, red, and blue contours are the CO(10--9), \hto, and \htop\ lines.  
          The synthesized beams are 0.76"$\times$0.41" and 0.54"$\times$0.30" for the 
          1.8 and 1.2\,mm bands, respectively.
          The contour levels are: in steps of 1\,$\sigma$ started 
          from $\pm$\,3\,$\sigma$ for the \htot211202\ and the \htop\ lines;
          and in steps of 3\,$\sigma$ started from $\pm$\,3\,$\sigma$
          for the \htot321312\ and the CO(10--9) lines.
         }
 \label{h2o:fig:high-res:nc143-image}
 \end{figure*}

In Fig.\,\ref{h2o:fig:high-res:nc143-image} we present the 1.8\,mm and 1.2\,mm dust continuum of NCv1.143. These images reveal dust continuum emissions at approximately 390\,$\mu$m and 260\,$\mu$m in the rest frame. We also show the line emission of the CO(10--9), \htot211202, and \htot321312\ lines and the sum of the emissions from the four \htop\ lines. The images show arc-like emission components, which are typically found in galaxy-galaxy lensing systems.

 \begin{figure}[htbp]
  \centering
    \includegraphics[width=0.42\textwidth]{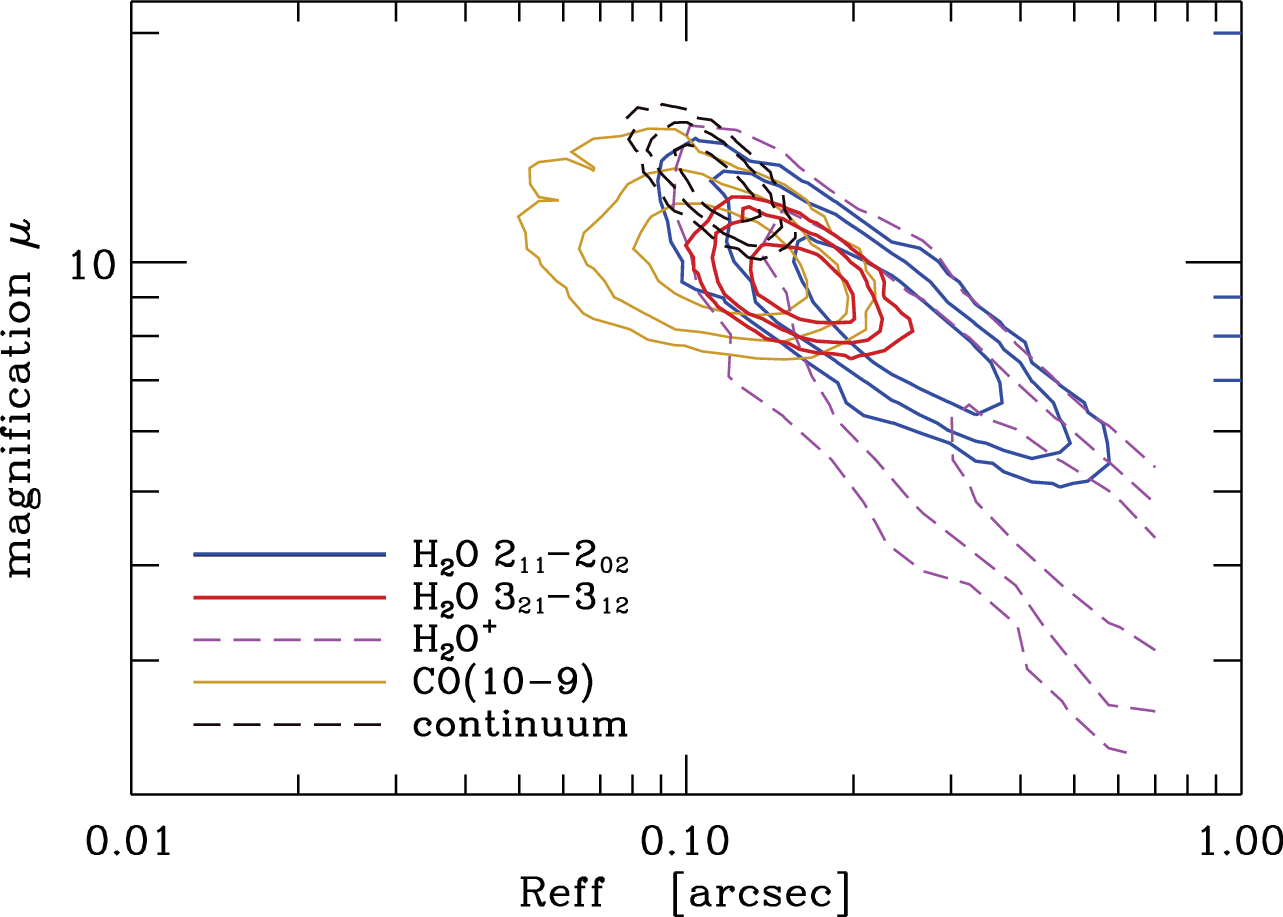}
  \caption{Posterior possibility distributions of the $R_\mathrm{eff}$
           and $\mu$ from the MCMC sampling.
           The contours are from 1\,$\sigma$ to 3\,$\sigma$ from inside out.
           The solid blue line, solid red line, dashed purple line, solid
           yellow line, and dashed black line indicate the 
           results for \htot211202, \htot321312, \htop, CO(10--9),  
           and the dust continuum, respectively.
		   }
 \label{h2o:fig:high-res:nc143-contours}
 \end{figure}

Utilizing a simple lens model such as a singular isothermal ellipsoid (SIE), as used in NCv1.143 and outlined in \citet{1994A&A...284..285K}, we can reproduce lensed signatures and determine deflector properties based on mass profile assumptions. Following the procedures outlined by \cite{2013ApJ...779...25B} and \citet{2011ApJ...738..125G}, we used the SIE model and the \texttt{GRAVLENS} software to analyze the source plane emission and the NOEMA \texttt{CLEAN}ed images. These techniques yield the parameters of both the deflector's SIE profiles and the lensed source S\'ersic profiles.

Our images of NCv1.143, with a resolution of 0.3"--0.4", surpass the resolution of the SMA images from \cite{2013ApJ...779...25B} and represent the highest spatial resolution data of NCv1.143 to date. In addition to dust continuum images, we obtained data cubes, including line emission images at various velocity channels. We assumed that the deflector comprises a single SIE mass distribution, allowing the center to vary within $0.5"$ of the optical center of the foreground galaxies. The emitted light was approximated by an elliptical exponential light profile, as supported by \cite{2013ApJ...779...25B}. The modeling was accomplished with the \texttt{sl\_fit} code, using an MCMC method to explore the parameter space.

Using the integrated continuum dust emission for mass distribution and dust emission constraining (simultaneously fitting the 1.2\,mm and 2\,mm continuum images), the best-fit model was then applied to retrieve intrinsic gas properties. We believe this approach adequately captures the key information of the data. The results of our lensing model are detailed in Table\,\ref{tab:lensmodel-NC143}. The positions of line emissions are reported for the channel closest to the frequency center, and magnification is integrated over the entire line.

The model aligns well with the observed continuum images (Fig.\,\ref{tab:lensmodel-NC143}), allowing us to accurately reconstruct the image of the dust continuum emission in the source plane. The average magnification factor of the dust continuum, $\mu_\mathrm{2,mm}$, was found to be 12.2$\pm$1.2, and its radius of extension, $R_\mathrm{eff}$, was determined to be (0.8$\pm$0.1)\,kpc. These results are consistent with, but more precise than, the $\mu_\mathrm{880}$ = 11.3$\pm$1.7 value reported by \citet{2013ApJ...779...25B}.

\begin{table*}[htbp]
\small
\caption{Lens modeling results for NCv1.143.}\label{tab:lensmodel-NC143}
\begin{center}
\begin{tabular}{ccccc} 
\toprule
\multicolumn{5}{c}{SIE mass component}\\
$x_{\rm def}$ & $y_{\rm def}$ & $q_{\rm def}$ & PA$_{\rm def}$ & $R_{\rm Ein}$ \\
arcsec & arcsec &  & deg (East of North) &   arcsec \\  
\midrule
0 &  0  & $0.65 \pm 0.03 $ &  $-62 \pm  2$ & $ 0.69 \pm 0.01 $ \\
\bottomrule
\end{tabular}
 
\begin{tabular}{cc}
        &  \\
\end{tabular}
 
\begin{tabular}{rcccccc} 
\toprule
\multicolumn{7}{c}{Reconstructed source}\\ 
& $x_{\rm src}$ & $y_{\rm src}$ & $q_{\rm src}$ & PA$_{\rm src}$ & $R_{\rm eff}$ & $\mu$ \\
&  arcsec & arcsec &  & deg &   arcsec &  \\
\midrule
continuum &  $+0.000 \pm 0.004 $ &  $-0.093 \pm 0.008 $  & $0.76 \pm 0.07 $ &  $-37 \pm  16$ & $0.113 \pm 0.017$ &  $ 12.2 \pm 1.2$ \\
CO(10--9) & $-0.029 \pm 0.016 $ &  $-0.125 \pm 0.016$  & $0.40\pm 0.19 $ &   $-38 \pm 17$ &  $0.111 \pm 0.032 $  &  $10.1 \pm 1.3$ \\
\htot211202 & $+0.006 \pm 0.026 $ &   $-0.123 \pm 0.040 $ & $0.63\pm 0.17$ &   $\;\;25 \pm 30$ &   $0.23 \pm 0.10 $  &  $8.3 \pm 1.8$  \\
\htot321312 &  $+0.102 \pm 0.019 $ &   $-0.106 \pm 0.016 $ & $0.69\pm 0.13$ &   $-17 \pm 60$ &   $0.158 \pm 0.030 $  &  $9.4 \pm 1.0$  \\
H$_2$O$^+$ &  $-0.20 \pm 0.12 $ &   $-0.06 \pm 0.09 $ & $0.54\pm 0.18$ &   $\;\;43 \pm 60$ &   $0.38 \pm 0.20 $  &  $5.0 \pm 2.0$  \\
\bottomrule
\end{tabular}
\end{center}
\end{table*}
\normalsize

\begin{figure*}[htbp]
\centering
\includegraphics[scale=0.7]{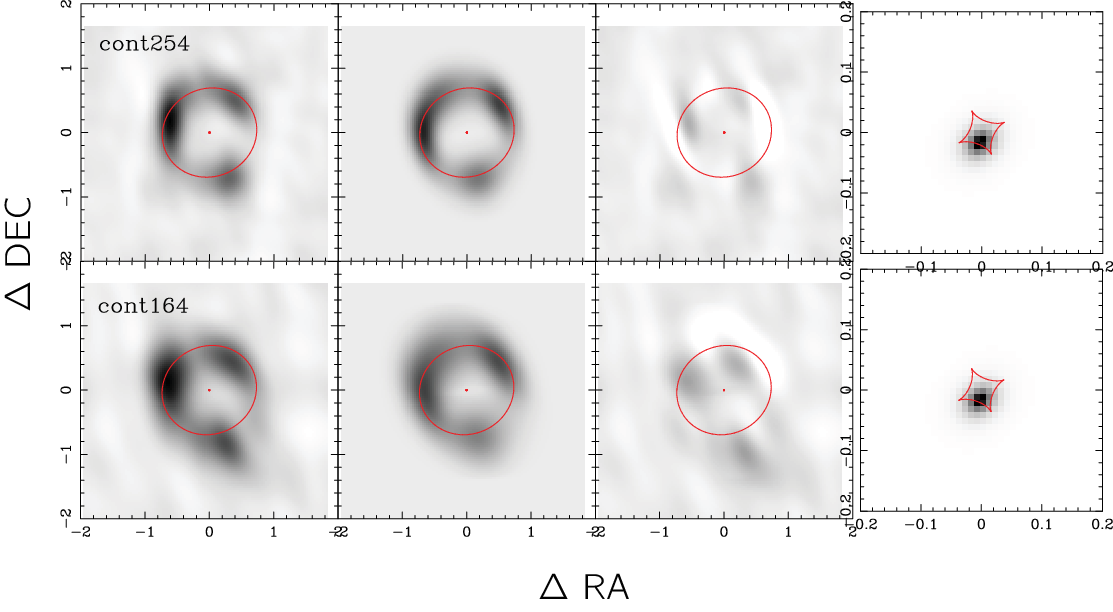}
\caption{
          Lens model of the 254\,GHz (top row) and 164\,GHz (bottom row) dust continuum emission 
          in NCv1.143. {\it From left to right}: 
          (1) Observed image of the 2\,mm continuum. 
          (2) Model images with the model parameters of Table\,\ref{tab:lensmodel-NC143}. 
          (3) Residuals from the difference between observed and model images. 
          (4) Reconstructed image of the dust continuum emission 
          in the source plane.          
          The critical curves and caustics are shown in red lines in the first three
          columns and the fourth columns, respectively. The position of the 
          deflector is indicated by red points.
          }
 \label{h2o:fig:high-res:resid-nc143-ctm}
 \end{figure*}

 \begin{figure*}[htbp]
 \centering
 \includegraphics[scale=0.5]{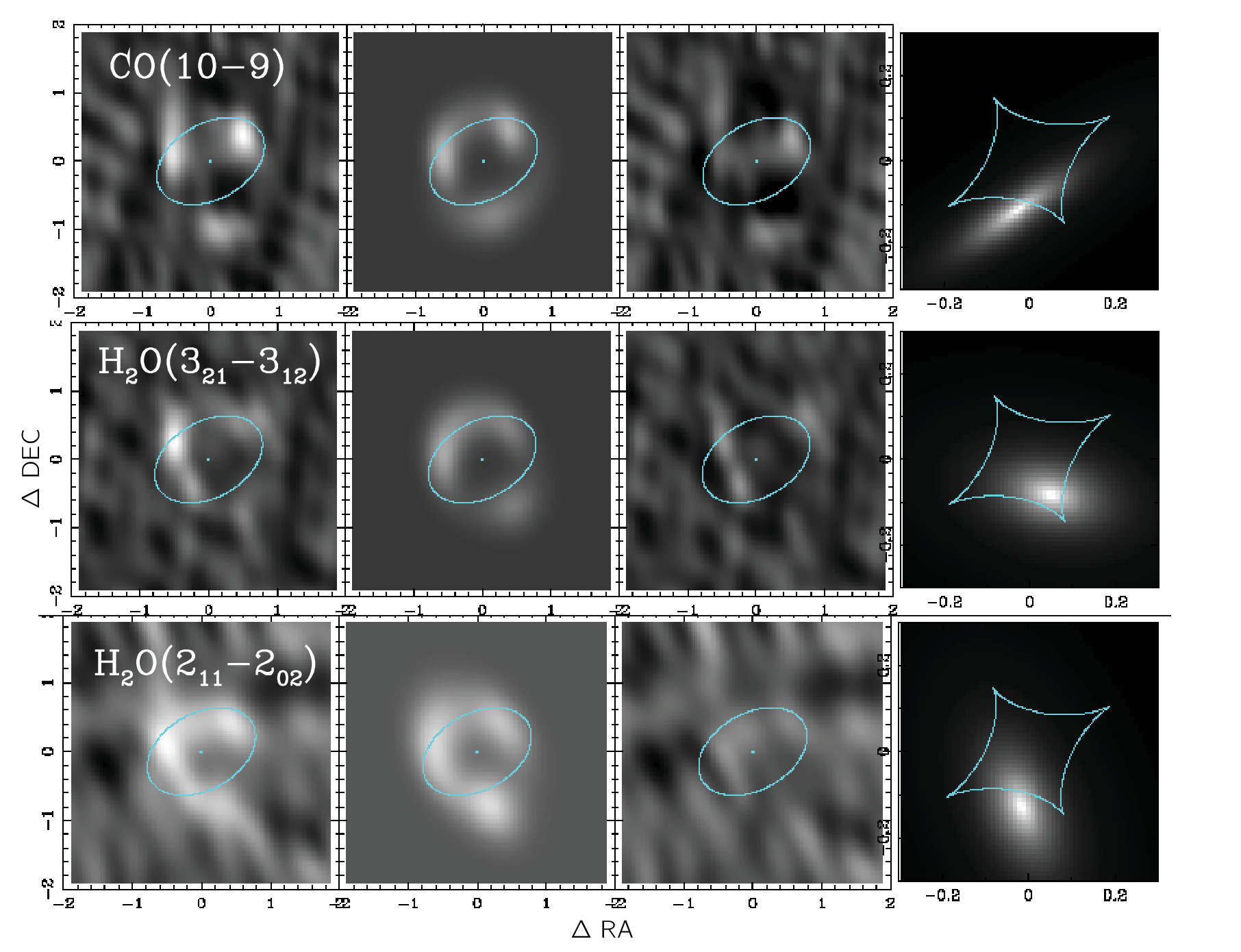}
 \caption{
          {\it From top to bottom, each row}: Lens model of 
          the CO(10--9), \htot321312\ and \htot211202\ lines in NCv1.143. 
          {\it From left to right}: 
          (1) Observed \texttt{CLEAN}ed images of the emission line. 
          (2) Model images with the model parameters of Table\,\ref{tab:lensmodel-NC143}. 
          (3) Residuals from the difference between observed and model images. 
          (4) Reconstructed images of the emission line in the source plane.   
          The critical curves and caustics are shown in cyan lines in the first three
          columns and the fourth column, respectively. The position of the 
          deflector is indicated by the cyan points.
          One should note that the position of the background source in the 
          source plane is not well-constrained. Thus, the differences
          seen in the positions of the three lines are not significant, considering 
          the uncertainties.
          }
 \label{h2o:fig:high-res:resid-nc143-J2_J3}
 \end{figure*}

\clearpage

\section{Optical depths}
\label{app:taus}

We performed a PDR calculation to explore how the optical depth of $^{13}$CO(4--3) and C$^{18}$O(4--3) builds along with the cloud depth. We constructed a density distribution resulting from the $A_{\rm V,\,eff}-n_{\rm H}$ relation presented in \citet{2019MNRAS.485.3097B} and \cite{2023MNRAS.519..729B}. This empirical relation connects the local visual extinction with the most probable local number density.
The calculation was performed using an FUV intensity of $\chi/\chi_0=10$ \citep[normalized to the spectral shape of][]{1978ApJS...36..595D}, a CR ionization rate of $\zeta_{\rm CR}=10^{-13}\,{\rm s}^{-1}$ (as derived from Sect.~\ref{subsection:Chemical-model}), and a metallicity of $1\,{\rm Z}_{\odot}$. The microturbulent velocity controlling the turbulent heating was taken to be $v_{\rm turb}=10\,{\rm km}\,{\rm s}^{-1}$, and the velocity dispersion was taken to be $v_{\rm disp}=180\,{\rm km}\,{\rm s}^{-1}$ to be in accordance to the observed linewidth. Furthermore, we assumed a $^{12}$C/$^{13}$C isotope ratio of 30 and $^{12}$CO/C$^{18}$O of 200 \citep[e.g.,][]{2019A&A...624A.125M, 2019A&A...629A...6T}.  

\begin{figure}[htbp]
    \centering
    \includegraphics[width=0.45\textwidth]{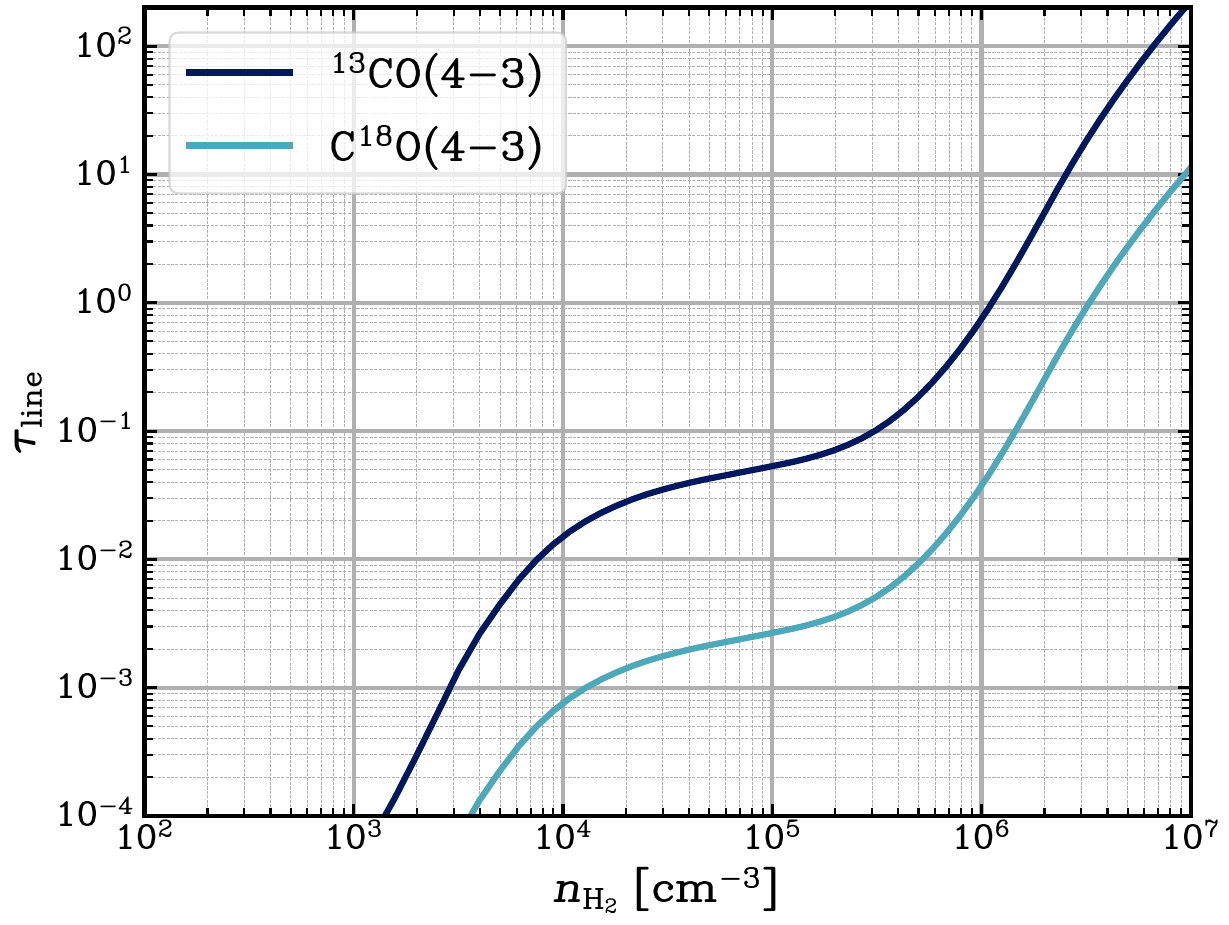}
    \caption{\texttt{3D-PDR} calculations using a density distribution representing the $A_{\rm V,eff}-n_{\rm H}$ relationship \citep{2019MNRAS.485.3097B}. The figure shows the optical depth of $^{13}$CO(4--3) and C$^{18}$O(4--3) lines as a function of the  H$_2$ volume density. We find that both lines remain optically thin for $n_{\rm H_2}\lesssim10^6\,{\rm cm}^{-3}$), while the $^{13}$CO(4--3) starts to have moderate optical depth with increasing H$_2$ densities.}
    \label{fig:taus}
\end{figure}

Figure~\ref{fig:taus} shows the optical depths of the $^{13}$CO(4--3) and C$^{18}$O(4--3) lines as a function of the H$_2$ density, $n_{\rm H_2}$. For values of $n_{\rm H_2}\lesssim10^6\,{\rm cm}^{-3}$, both lines remain optically thin. However, when $n_{\rm H_2}$ reach beyond 10$^6$\,cm$^{-3}$, the $^{13}$CO(4--3) line starts to have a moderate optical depth while C$^{18}$O(4--3) remains optical thin until the density getting close to 10$^7$\,cm$^{-3}$. Both lines start to get optically thick if $n_{\rm H_2}\gtrsim10^7\,{\rm cm}^{-3}$. In the conditions of the $n_{\rm H_2}$ between 10$^6$ and 10$^7$\,cm$^{-3}$, the optical depth of the $^{13}$CO(4--3) will reach about 10. Such an optical depth will lead to a low flux ratio of $^{13}$CO/C$^{18}$O close to unity and a very low flux ratio of $^{12}$CO/$^{13}$CO, as pointed out by \citet{2018Natur.558..260Z}. However, the observed $^{12}$CO/$^{13}$CO ratios in both our sources (Table\,\ref{tab:line-isotop}) are about 20--30, inconsistent with a optically thick scenario of the $^{13}$CO.

\section{Detailed chemical modeling of HCO$^+$}
\label{app:hcop_chem}

\begin{figure*}[htbp]
    \centering
    \includegraphics[width=0.407\textwidth]{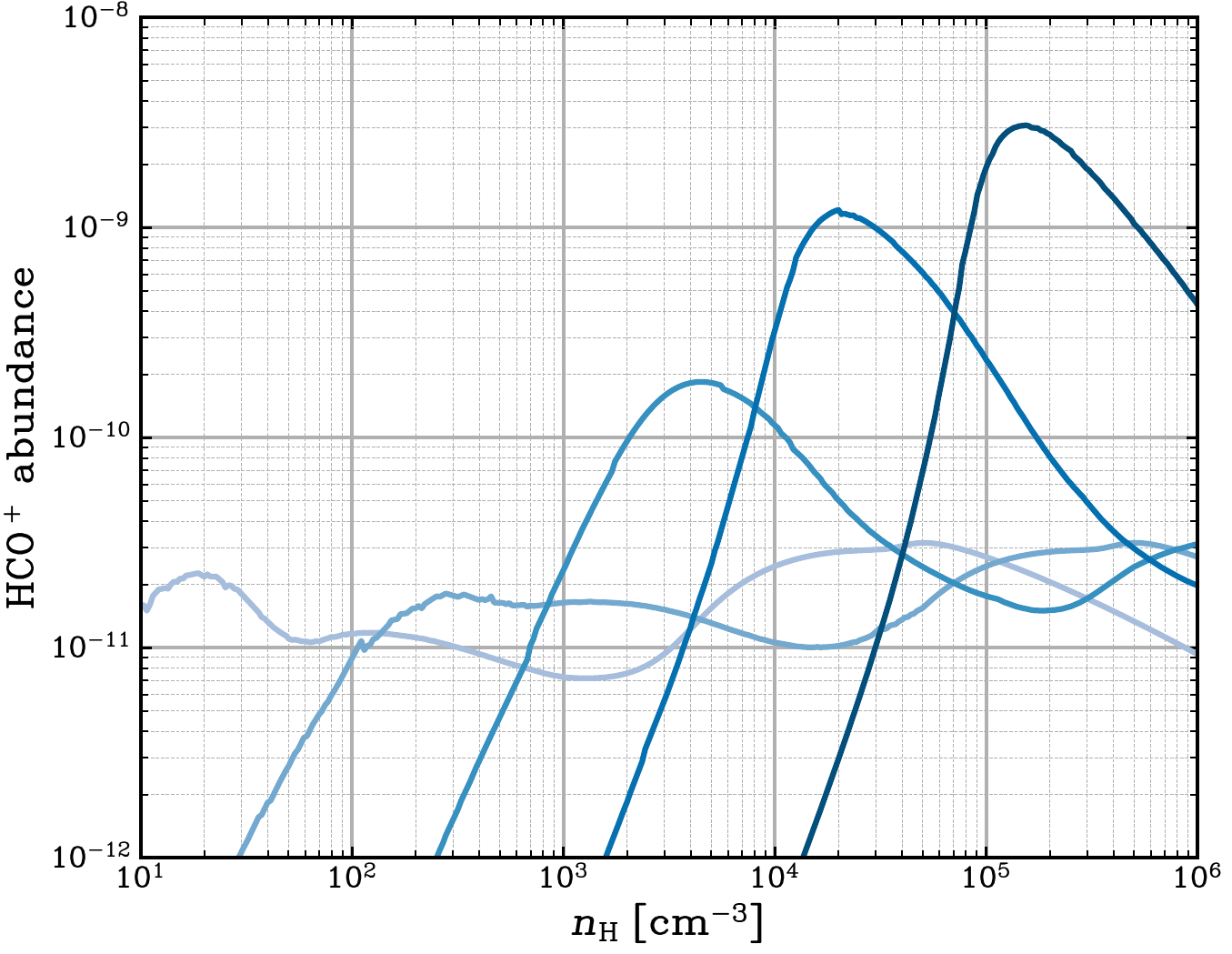}
    \includegraphics[width=0.4\textwidth]{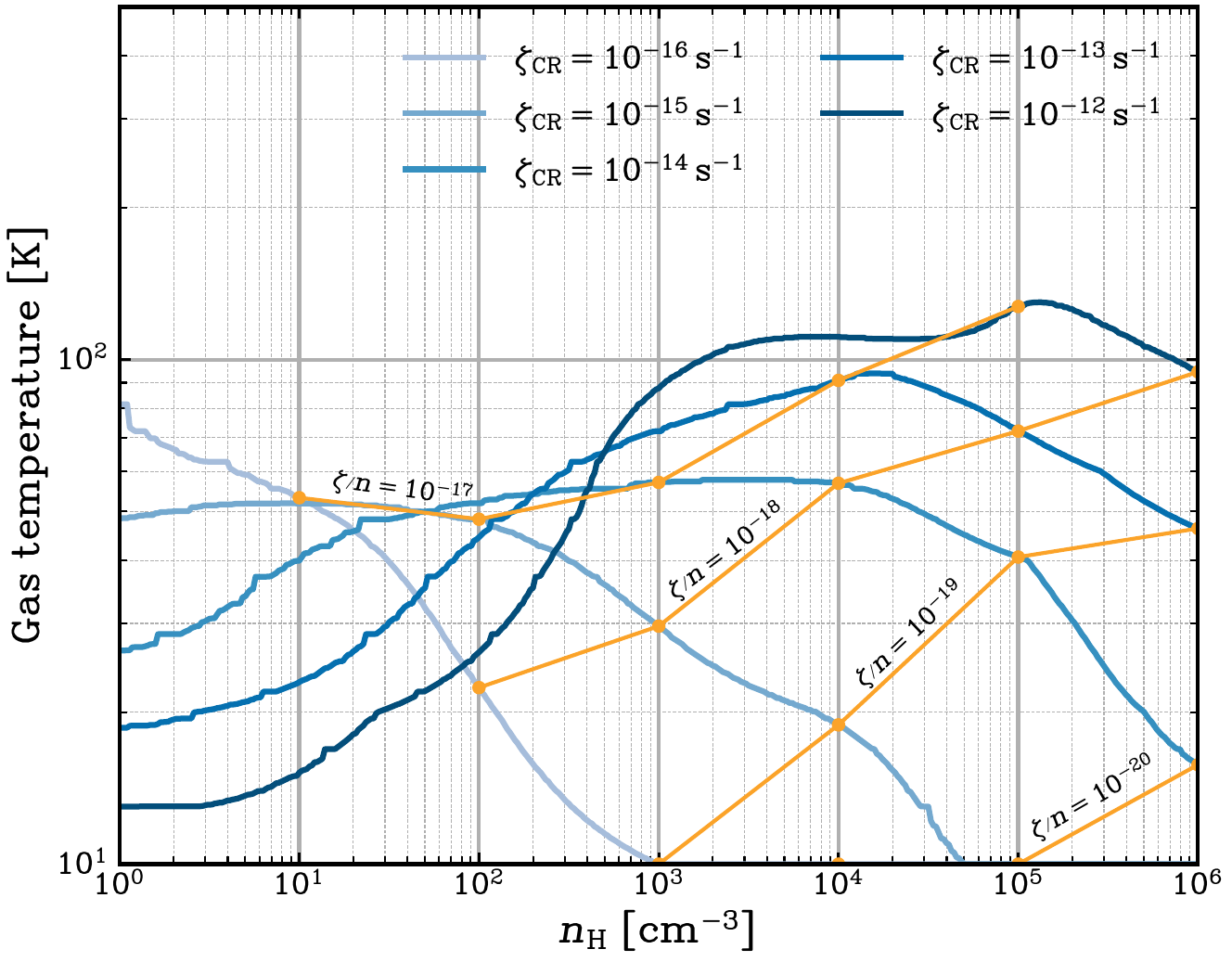}
    \caption{Chemical models for HCO$^+$.
    {\em Left panel}: HCO$^+$ abundance versus the local total H-nucleus number density, $n_{\rm H}$, for different CR ionization rates. In all cases, the abundance of HCO$^+$ peaks for densities $8\times10^{3}\lesssim n_{\rm H}\lesssim3\times10^5\,{\rm cm}^{-3}$. {\em Right panel}: Temperature profile, with the iso-($\zeta/n$) lines overplotted in orange, for different CR ionization rates ($\zeta_{\rm CR}=10^{-16}-10^{-12}\,{\rm s}^{-1}$, with the same color code as the left panel).}
    \label{fig:hcop}
\end{figure*}

The left panel of Fig.~\ref{fig:hcop} displays the HCO$^+$ abundance versus the local number density, $n_{\rm H}$, for five different CR ionization rates ($\zeta_{\rm CR}=10^{-16}-10^{-12}\,{\rm s}^{-1}$). The FUV radiation is neglected in these models, as mentioned in the main text, and the macroturbulent optical depth expression (Eq.~\ref{eqn:tau}) is adopted. Overall, the abundance of HCO$^+$ peaks at densities in the range of $n_{\rm H}\simeq10^4$ to a few $10^5\,{\rm cm}^{-3}$, with the effect to be better seen when $\zeta_{\rm CR}\gtrsim10^{-13}\,{\rm s}^{-1}$. In the low end of the above $\zeta_{\rm CR}$ values, HCO$^+$ is formed mainly via the reactions 
\begin{align}
    &\rm H_2 + HOC^+ \rightarrow \rm  HCO^+ + H_2 \tag{R1} \label{R1} \\
    &\rm H_2 + CO^+ \rightarrow \rm HCO^+ + H \tag{R2} \label{R2}
\end{align}
\ref{R1} and \ref{R2} together form approximately 80\% of the total HCO$^+$ abundance, with \ref{R2} becoming more effective for densities $n_{\rm H}\lesssim 10^2\,{\rm cm}^{-3}$. For $n_{\rm H}\gtrsim10^3\,{\rm cm}^{-3}$, the reaction
\begin{align}
    &\rm H_3^+ + CO \rightarrow \rm HCO^+ + H_2 \tag{R3} \label{R3}
\end{align}
dominates the formation route of HCO$^+$. For very high $\zeta_{\rm CR}$ values, HCO$^+$ abundance builds rapidly when $n_{\rm H}\gtrsim$ a few $10^3\,{\rm cm}^{-3}$, below which is in practice negligible (Fig.~\ref{fig:hcop}). Contrary to the aforementioned case, \ref{R2} remains, here, the main reaction that is able to form more than 50\% of the total HCO$^+$ abundance. In all cases, HCO$^+$ is destroyed with an efficiency of almost 100\%  due to its reaction with free electrons, leading to the formation of CO and H.  

The right panel of Fig.~\ref{fig:hcop} shows the temperature profile of HCO$^+$. The temperature profile varies as a function of the CR ionization rate since it inserts heating. We performed various runs with different CR ionization rates, and we have overplotted iso-($\zeta$/$n$) lines, the range of which corresponds to the solutions we found for the two galaxies. Assuming the typical density for the dense gas, we find that their gas at high column densities is relatively warm ($\sim$\,50--100\,K), consistent with the LVG-derived gas temperatures. 

For low densities, an interesting behavior occurs in which the gas temperature decreases as the CR ionization ray increases. This is counter-intuitive, given that CRs heat up the gas. The decrease in gas temperature occurs because the production of C$^+$ abundance increases, making the emission line of [C{\small \,II}] increase. The latter is a strong coolant, which in turn decreases the overall gas temperature at low densities \citep{2017ApJ...839...90B}.

\end{appendix}

\end{CJK*}
\end{document}